\pdfoutput=1  

\documentclass[aps,twocolumn,superscriptaddress,floatfix,preprintnumbers,nofootinbib,eqsecnum]{revtex4-2}

\usepackage{amsmath, float, graphicx,placeins,amssymb,multirow,pgfplots,pgfplotstable}
\usepackage{tikz,hyperref}
\usepackage{CJKutf8}
\usepackage{cjhebrew}
\usetikzlibrary{shapes.geometric,arrows}

\begin{document}
\title{
Extracting Dark-Matter Velocities from Halo Masses:\\ A Reconstruction Conjecture   
}

\def\andname{\hspace*{-0.5em}} 

\author{Keith R. Dienes}
\email{dienes@email.arizona.edu}
\affiliation{Department of Physics, University of Arizona, Tucson, AZ 85721 USA}
\affiliation{Department of Physics, University of Maryland, College Park, MD 20742 USA}
\author{Fei Huang}
\email{huangf4@uci.edu}
\affiliation{CAS Key Laboratory of Theoretical Physics, Institute of Theoretical Physics,
    Chinese Academy of Sciences, Beijing 100190, China}
\affiliation{Department of Physics and Astronomy, University of California, Irvine, CA 92697, USA}
\author{Jeff Kost}
\email{J.D.Kost@sussex.ac.uk}
\affiliation{Department of Physics and Astronomy, University of Sussex, Brighton BN1 9QH, UK}
\author{Kevin Manogue}
\email{manoguek@lafayette.edu}
\affiliation{Department of Physics, Lafayette College, Easton, PA 18042 USA}
\author{Brooks Thomas}
\email{thomasbd@lafayette.edu}
\affiliation{Department of Physics, Lafayette College, Easton, PA 18042 USA}

\begin{abstract}
Increasing attention has recently focused on non-traditional dark-matter 
production mechanisms which result in primordial dark-matter velocity 
distributions with highly non-thermal shapes.  In this paper, we undertake 
an assessment of how the detailed shape of a general dark-matter velocity 
distribution impacts structure formation in the non-linear regime.  In 
particular, we investigate the impact on the halo-mass and subhalo-mass 
functions, as well as on astrophysical observables such as satellite and 
cluster-number counts.  We find that many of the standard expectations no 
longer hold in situations in which this velocity distribution takes a 
highly non-trivial, even multi-modal shape.  For example, we find that the 
nominal free-streaming scale alone becomes insufficient to characterize the 
effect of free-streaming on structure formation.
In addition, we propose a simple one-line conjecture which can be used to 
``reconstruct'' the primordial dark-matter velocity distribution directly from the 
shape of the halo-mass function. Although our conjecture is completely heuristic, 
we show that it successfully reproduces the salient features of the underlying 
dark-matter velocity distribution even for non-trivial distributions which are 
highly non-thermal and/or multi-modal, such as might occur for non-minimal dark
sectors.  Moreover, since our approach relies only on the halo-mass function, 
our conjecture provides a method of probing dark-matter properties even for 
scenarios in which the dark and visible sectors interact only gravitationally.
\end{abstract}

\maketitle


\newcommand{\PRE}[1]{{#1}} 
\newcommand{\ul}{\underline}
\newcommand{\del}{\partial}
\newcommand{\nbox}{{\,\lower0.9pt\vbox{\hrule \hbox{\vrule height 0.2 cm
\hskip 0.2 cm \vrule height 0.2 cm}\hrule}\,}}

\newcommand{\postscript}[2]{\setlength{\epsfxsize}{#2\hsize}
   \centerline{\epsfbox{#1}}}
\newcommand{\gweak}{g_{\text{weak}}}
\newcommand{\mweak}{m_{\text{weak}}}
\newcommand{\mplanck}{M_{\text{Pl}}}
\newcommand{\mstar}{M_{*}}
\newcommand{\sigmaan}{\sigma_{\text{an}}}
\newcommand{\sigmatot}{\sigma_{\text{tot}}}
\newcommand{\sigmaSI}{\sigma_{\rm SI}}
\newcommand{\sigmaSD}{\sigma_{\rm SD}}
\newcommand{\OmegaM}{\Omega_{\text{M}}}
\newcommand{\OmegaDM}{\Omega_{\text{DM}}}
\newcommand{\ipb}{\text{pb}^{-1}}
\newcommand{\ifb}{\text{fb}^{-1}}
\newcommand{\iab}{\text{ab}^{-1}}
\newcommand{\ev}{\text{eV}}
\newcommand{\kev}{\text{keV}}
\newcommand{\mev}{\text{MeV}}
\newcommand{\gev}{\text{GeV}}
\newcommand{\tev}{\text{TeV}}
\newcommand{\pb}{\text{pb}}
\newcommand{\mb}{\text{mb}}
\newcommand{\cm}{\text{cm}}
\newcommand{\m}{\text{m}}
\newcommand{\km}{\text{km}}
\newcommand{\kg}{\text{kg}}
\newcommand{\g}{\text{g}}
\newcommand{\s}{\text{s}}
\newcommand{\yr}{\text{yr}}
\newcommand{\Mpc}{\text{Mpc}}
\newcommand{\etal}{{\em et al.}}
\newcommand{\eg}{{\em e.g.}}
\newcommand{\ie}{{\em i.e.}}
\newcommand{\ibid}{{\em ibid.}}
\newcommand{\Eqref}[1]{Equation~(\ref{#1})}
\newcommand{\secref}[1]{Sec.~\ref{sec:#1}}
\newcommand{\secsref}[2]{Secs.~\ref{sec:#1} and \ref{sec:#2}}
\newcommand{\Secref}[1]{Section~\ref{sec:#1}}
\newcommand{\appref}[1]{App.~\ref{sec:#1}}
\newcommand{\figref}[1]{Fig.~\ref{fig:#1}}
\newcommand{\figsref}[2]{Figs.~\ref{fig:#1} and \ref{fig:#2}}
\newcommand{\Figref}[1]{Figure~\ref{fig:#1}}
\newcommand{\tableref}[1]{Table~\ref{table:#1}}
\newcommand{\tablesref}[2]{Tables~\ref{table:#1} and \ref{table:#2}}
\newcommand{\Dsle}[1]{\slash\hskip -0.28 cm #1}
\newcommand{\met}{{\Dsle E_T}}
\newcommand{\mpt}{\not{\! p_T}}
\newcommand{\Dslp}[1]{\slash\hskip -0.23 cm #1}
\newcommand{\Dsl}[1]{\slash\hskip -0.20 cm #1}

\newcommand{\mB}{m_{B^1}}
\newcommand{\mq}{m_{q^1}}
\newcommand{\mf}{m_{f^1}}
\newcommand{\mKK}{m_{\rm KK}}
\newcommand{\WIMP}{\text{WIMP}}
\newcommand{\SWIMP}{\text{SWIMP}}
\newcommand{\NLSP}{\text{NLSP}}
\newcommand{\LSP}{\text{LSP}}
\newcommand{\mWIMP}{m_{\WIMP}}
\newcommand{\mSWIMP}{m_{\SWIMP}}
\newcommand{\mNLSP}{m_{\NLSP}}
\newcommand{\mchi}{m_{\chi}}
\newcommand{\mgravitino}{m_{\gravitino}}
\newcommand{\mmed}{M_{\text{med}}}
\newcommand{\gravitino}{\tilde{G}}
\newcommand{\Bino}{\tilde{B}}
\newcommand{\photino}{\tilde{\gamma}}
\newcommand{\stau}{\tilde{\tau}}
\newcommand{\slepton}{\tilde{l}}
\newcommand{\snu}{\tilde{\nu}}
\newcommand{\squark}{\tilde{q}}
\newcommand{\mgaugino}{M_{1/2}}
\newcommand{\epsEM}{\varepsilon_{\text{EM}}}
\newcommand{\mmess}{M_{\text{mess}}}
\newcommand{\lmess}{\Lambda}
\newcommand{\nmess}{N_{\text{m}}}
\newcommand{\signmu}{\text{sign}(\mu)}
\newcommand{\Omegachi}{\Omega_{\chi}}
\newcommand{\lambdafs}{\lambda_{\text{FS}}}
\newcommand{\be}{\begin{equation}}
\newcommand{\ee}{\end{equation}}
\newcommand{\bea}{\begin{eqnarray}}
\newcommand{\eea}{\end{eqnarray}}
\newcommand{\beq}{\begin{equation}}
\newcommand{\eeq}{\end{equation}}
\newcommand{\beqn}{\begin{eqnarray}}
\newcommand{\eeqn}{\end{eqnarray}}
\newcommand{\baln}{\begin{align}}
\newcommand{\ealn}{\end{align}}
\newcommand{\lsim}{\lower.7ex\hbox{$\;\stackrel{\textstyle<}{\sim}\;$}}
\newcommand{\gsim}{\lower.7ex\hbox{$\;\stackrel{\textstyle>}{\sim}\;$}}
\newcommand{\il}[1]{\mbox{$#1$}} 

\newcommand{\ssection}[1]{{\em #1.\ }}
\newcommand{\rem}[1]{\textbf{#1}}

\def\ie{{\it i.e.}\/}
\def\eg{{\it e.g.}\/}
\def\etc{{\it etc}.\/}
\def\calN{{\cal N}}

\def\mptwo{{m_{\pi^0}^2}}
\def\mp{{m_{\pi^0}}}
\def\sqtsn{\sqrt{s_n}}
\def\sqtsn{\sqrt{s_n}}
\def\sqtsn{\sqrt{s_n}}
\def\sqts0{\sqrt{s_0}}
\def\Dsqts{\Delta(\sqrt{s})}
\def\Omegatot{\Omega_{\mathrm{tot}}}
\def\rhotot{\rho_{\mathrm{tot}}}
\def\rhocrit{\rho_{\mathrm{crit}}}
\def\OmegaDM{\Omega_{\mathrm{DM}}}
\def\OmegaDMbar{\overline{\Omega}_{\mathrm{DM}}}
\def\weff{w_{\mathrm{eff}}}
\def\tLS{t_{\mathrm{LS}}}
\def\aLS{a_{\mathrm{LS}}}
\def\zLS{z_{\mathrm{LS}}}
\def\tnow{t_{\mathrm{now}}}
\def\znow{z_{\mathrm{now}}}
\def\tMRE{t_{\mathrm{MRE}}}
\def\tast{t_{\ast}}
\def\Ndof{N_{\mathrm{d.o.f.}}\/}
\def\LambdaIR{\Lambda_{\rm IR}}
\def\LambdaUV{\Lambda_{\rm UV}}
\def\shin{{\large\textcjheb{+s}}}
\def\smallshin{{\large\textcjheb{+s}}}   

\section{Introduction}


It is now evident that the majority of matter in our universe is dark, in
the sense that it interacts at most weakly with the fields of the Standard 
Model (SM).~  Nevertheless, despite an impressive array of experimental 
efforts dedicated to probing the particle properties of this dark matter, 
its fundamental nature remains a mystery.  It is still possible that a 
conclusive discovery --- at the LHC, at 
one of the many direct-detection experiments currently in operation or under
construction, at one of the telescopes sensitive to indirect-detection signatures of 
dark-matter annihilation or decay, or at one of the experiments dedicated to
the detection of axions and/or axion-like particles --- will revolutionize our
understanding of dark matter within the next few years.  However, such a breakthrough
is far from assured.  Indeed, most experimental strategies for probing the particle 
properties of the dark matter rely on the assumption that the dark matter has 
appreciable non-gravitational interactions with the particles of the SM,  
but there is no guarantee that the dark matter possesses such interactions.  
For this reason, it is crucial to explore other possible methods for probing 
the particle properties of the dark matter --- methods which do not rely on 
its interactions with SM particles. 

One characteristic of the dark matter which could reveal information about 
both its particle properties and its production mechanism in the early 
universe is its primordial velocity distribution $f(v,t)$.  This distribution 
is conventionally described in terms of {\il{f(v)\equiv f(v,\tnow)}}, \ie, the
distribution obtained by redshifting $f(v,t)$ from 
some early time $t$ to the present time $\tnow$, while ignoring effects such as 
virialization.  This dark-matter velocity distribution plays a crucial role in 
determining the structure of the present-day universe.  In fact, many important 
quantities follow from the form of $f(v)$, such as the non-linear matter power 
spectrum and the halo-mass function $dn/d\log M$, where $n$ is the number density 
of matter halos with mass $M$.

The form which $f(v)$ takes in any given dark-matter scenario is primarily 
determined by the early-universe dynamics through which the dark matter is 
produced.  In warm-dark-matter (WDM) scenarios, wherein the dark matter
freezes out from the radiation bath while still relativistic, 
the dark-matter velocity distribution takes a simple, unimodal shape.

By contrast, in other dark-matter scenarios,
the production of dark matter is far more complicated than it is in 
WDM scenarios and includes contributions
from multiple production channels that leave markedly different kinematic 
imprints on $f(v)$.  For example, freeze-in production and 
production from the out-of-equilibrium decays of unstable particles can both 
contribute non-negligibly to the overall dark-matter
abundance~\cite{Konig:2016dzg,Du:2021jcj,Decant:2021mhj}, or multiple
decay pathways for the same unstable particle species can contribute to 
that abundance --- pathways that can involve the direct decays of this species 
into dark-matter particles~\cite{Heeck:2017xbu} or dark-matter production 
via extended decay chains~\cite{Dienes:2020bmn}.  The dark-matter velocity 
distributions which arise in such scenarios are generally far more 
complicated than those which arise in WDM scenarios and are indeed often 
multi-modal.  Moreover, features in the dark-matter velocity distribution 
encode information about the underlying particle-physics processes which gave 
rise to them.  For example, features generated by the thermal freeze-out
of a relativistic particle species encode information about 
the mass of the species.  Likewise, a feature generated by the decay of a 
heavy, unstable particle encodes information about the decay width of that
particle and the relationship between its mass and the mass of the particles
into which it decays~\cite{Dienes:2020bmn}.  This information can be 
correlated with the results of particle-physics experiments which are
capable of probing these quantities more directly. 
It is therefore important to assess how the detailed shape
of $f(v)$ affects the formation of structure within both the linear
and non-linear regimes.

A systematic study of how the detailed shape of $f(v)$ affects structure 
within the linear regime was performed in Ref.~\cite{Dienes:2020bmn}.  
The relationship between the dark-matter phase-space distribution and the 
linear matter power spectrum $P(k)$ was investigated numerically, and
it was shown that the power spectra associated with complicated 
$f(v)$ distributions deviate from those associated with simple, 
unimodal distributions of the sort that arise in WDM models in 
a quantifiable way.  On the basis of these results, 
an empirical procedure was formulated by means of which the shape of $f(v)$ 
can be reconstructed solely from information contained within $P(k)$.

By contrast, the impact of $f(v)$ on structure formation within the non-linear 
regime is far less straightforward to assess.  The spectrum of primordial 
density perturbations initially established during the epoch of cosmic 
inflation evolves with time according to the Einstein-Boltzmann equations ---
equations which depend both on $f(v)$ and on other aspects of the background
cosmology.  While these perturbations are sufficiently small at early times 
that their evolution may be reliably modeled using a linearized-gravity 
approach, this approach remains valid only until non-linear feedback becomes
significant and perturbation theory becomes less reliable.  As a result, 
the time-evolution of the density perturbations is complicated (and not 
even invertible), and one must adopt a different strategy for understanding 
the mass density at late times.  Such strategies typically involve approaching 
the problem numerically, using $N$-body or hydrodynamic simulations.  Such
simulations are computationally expensive and therefore impractical
to perform when surveying broad classes of dark-matter models.

In this paper, we make a first foray into assessing how the detailed shape of 
the dark-matter velocity distribution impacts structure formation in the 
non-linear regime.  In doing so, we use the analytic approach originally pioneered by 
Press and Schechter~\cite{Press:1973iz} and subsequently refined in a 
number of ways by others~\cite{Bardeen:1985tr,Bond:1990iw,Sheth:1999mn,Sheth:1999su}.
We investigate how the halo-mass and subhalo-mass functions obtained
for complicated, multi-modal $f(v)$ distributions differ from those obtained 
for simple, unimodal distributions with the same na\"{i}ve free-streaming
scale.  On the basis of these results, we then
propose a simple technique for extracting information about the 
primordial dark-matter velocity distribution from the spatial distribution
of dark matter within the present-day universe.  In particular, we posit an empirical 
conjecture for reconstructing $f(v)$ directly from the shape of $dn/d\log M$.

We note that while the halo- and subhalo-mass functions are of course challenging 
(or perhaps even impossible)
to measure directly, there has recently been considerable interest --- and 
progress --- in the development of methods for probing and constraining their 
properties on the basis of observational data in a model-independent 
way~\cite{Castro:2016jmw,Dong:2019mch,Sonnenfeld:2019,Li:2019zvm,Cueli:2021dai}.
It is therefore interesting and timely to consider how information about the 
underlying cosmology might potentially be extracted from these functions.
Along these lines, we demonstrate within the context of an illustrative model 
that our reconstruction conjecture is quite robust.  Indeed our 
reconstruction conjecture is capable of reproducing the salient features of 
$f(v)$, even in situations in which this velocity distribution is non-thermal 
and even multi-modal.

This paper is organized as follows.  
In Sect.~\ref{sec:VelsMPS}, we review the manner in which the free-streaming 
effects associated with the primordial dark-matter velocity distribution 
modify the shape of the linear matter power spectrum $P(k)$.  We then proceed 
to demonstrate that complicated, multi-modal dark-matter velocity distributions 
can give rise to matter power spectra which cannot be realized within the context 
of warm-dark-matter scenarios.  In Sect.~\ref{sec:HaloMassFunction}, we 
review how modifications of $P(k)$ in turn lead to modifications of the 
halo-mass function within the context of the extended Press-Schechter formalism.
In Sect.~\ref{sec:ResultsHMF}, we make use of this formalism in order to 
investigate the ways in which the halo-mass function is influenced by the 
shape of $f(v)$ and highlight the ways in which the results obtained
for complicated, multi-modal dark-matter velocity distributions differ 
from those obtained for unimodal distributions such as those associated with 
warm dark matter.
In Sect.~\ref{sec:Observables}, we investigate how the 
detailed shape of $f(v)$ impacts observables such as cluster-number counts and
the number of satellites within the halo of a typical Milky-Way-sized galaxy.
In Sect.~\ref{sec:ConjAll}, we formulate our conjecture which allows us to work 
backwards and reconstruct the underlying dark-matter velocity distribution 
from the shape of the halo-mass function.  We then apply this reconstruction 
conjecture in the context of an illustrative model --- a model which is capable
of producing highly non-thermal and even multi-modal dark-matter velocity
distributions.  In Sect.~\ref{sec:Conclusions}, we conclude with a summary
of our results  and discuss possible directions for future work.


\section{From Dark-Matter Velocity Distributions to Matter Power Spectra\label{sec:VelsMPS}}


Our ultimate goal in this paper is to investigate the relationship between the 
detailed shape of the primordial dark-matter velocity distribution $f(v)$ and 
quantities such as the halo-mass function $dn/d\log M$ and the subhalo-mass 
function $dN_{\rm SH}/d\log M$ and to examine ways in which the form of this
relationship could potentially be exploited in order to reveal meaningful 
information about $f(v)$. A necessary first step toward this goal is to 
investigate the relationship between $f(v)$ and the linear matter power 
spectrum $P(k)$.  A detailed investigation along these lines was performed 
in Ref.~\cite{Dienes:2020bmn}.  In this section we briefly review the results
of Ref.~\cite{Dienes:2020bmn}.  In particular, we review the 
physics behind how dark-matter velocities affect structure formation 
through free-streaming.  We also highlight how standard
approximations designed to assess the effects of free-streaming on $P(k)$ in 
WDM models fail --- sometimes spectacularly --- when applied to dark-matter 
scenarios with more complicated $f(v)$ distributions. 

We begin by reviewing the way in which the velocity distribution of dark-matter 
particles affects the development of structure in the linear regime. 
For simplicity, we focus on the case in which the dark matter consists of a 
single particle species and assume an otherwise standard background cosmology.
The velocity distribution $f(\vec{\mathbf{v}},t)$ of that species at time $t$ is 
conventionally normalized such that 
\begin{equation}
  n(t) ~=~ \frac{g_{\rm int}}{(2\pi)^3} \int d^3 v\, f(\vec{\mathbf{v}},t)~,
\end{equation} 
where $g_{\rm int}$ is the number of internal number of degrees of freedom for a particle
of that species and where $n(t)$ is its physical number density.  Within 
a Friedmann-Robertson-Walker (FRW) universe, the underlying assumption of isotropy 
implies that {\il{f(\vec{\mathbf{v}},t) = f(v,t)}} depends only on the magnitude 
{\il{v \equiv \vec{\mathbf{v}}}}.  Within such a universe, we may also define a 
corresponding comoving number density {\il{N(t)\equiv a^3 n(t)}}, where $a$ is the 
scale factor, defined according the usual convention in which {\il{a=1}} 
at {\il{t=\tnow}}.  

We note that this comoving dark-matter number density may also be written in 
the form
\begin{eqnarray}
  N(t) &~=~& \frac{g_{\rm int} a^3}{2\pi^2} \int d v\, v^2 f(v,t)  
    \nonumber \\
    &~=~& \frac{g_{\rm int}}{2\pi^2} \int d\log v\, g_v(v,t)~,
 \label{eq:ComovingNumberDens}
\end{eqnarray}
where we have defined 
\begin{equation}
  g_v(v,t) ~\equiv~ (a v)^3 f(v,t)~.
  \label{eq:gofp}
\end{equation}
The advantage of working with $g_v(v,t)$ is that this form of the dark-matter
velocity distribution transforms in a particularly straightforward way 
during any epoch within which the rates for dark-matter production, scattering,
and decay are negligible.  In particular, Eq.~(\ref{eq:ComovingNumberDens}) 
implies that the $g_v(v,t)$ distribution shifts uniformly toward smaller 
values of $\log v$ during such an epoch as a consequence of cosmological 
redshifting --- \ie, that the {\it shape}\/ of this distribution remains 
invariant~\cite{Dienes:2020bmn}.  For convenience,  we define 
{\il{g_v(v) \equiv g_v(v,\tnow) = v^3f(v)}} to refer to the corresponding 
present-day velocity distribution.  As with $f(v)$, this distribution is 
obtained by redshifting $g_v(v,t)$ from some early time $t$ to $\tnow$ 
while ignoring effects such as virialization.

The inhomogeneity of matter halos in the present-day universe
arises due to spatial variations in the density of matter in the early universe.  
Such variations can be characterized by the fractional overdensity
$\delta(\vec{\mathbf{x}},t)$, while point-to-point correlations in 
$\delta(\vec{\mathbf{x}},t)$ are given by the two-point correlation function
$\xi(\vec{\mathbf{r}},t)$.  For a universe which is homogeneous and 
isotropic on large scales, {\il{\xi(\vec{\mathbf{r}},t) = \xi(r,t)}} 
depends only on the magnitude $r$ of the displacement vector.  Given 
these assumptions, the Fourier transform of $\xi(r,t)$, which is commonly 
referred to as the matter power spectrum, may be written in the form
\begin{equation}
  P(k,t) ~\equiv~ 4\pi \int dr\, r^2 \frac{\sin(kr)}{kr} \xi(r,t)~.
\end{equation} 
In the following we shall evaluate $P(k,t)$ using linear perturbation theory 
(thereby producing the linear matter power spectrum), and we shall adopt the shorthand
{\il{P(k) \equiv P(k,\tnow)}}.  We also define the transfer function 
$T(k)$ according the the relation
\begin{equation}
  T^2(k) ~=~ \frac{P(k)}{P_{\rm CDM}(k)}~,~
  \label{eq:Tk}
\end{equation}
where $P_{\rm CDM}(k)$ is the matter power spectrum obtained for purely 
cold dark matter (CDM).

The velocity distribution of dark-matter particles in the early universe 
affects the manner in which $P(k,t)$ evolves with time.  For example, 
dark-matter particles with sufficiently large velocities can free-stream out 
of overdense regions which might have otherwise collapsed into halos, thereby
suppressing the growth of structure on small scales.  The distance scale 
below which a population of dark-matter particles with
present-day velocity $v$ is capable of suppressing small-scale structure 
in this way is set by the corresponding particle horizon 
\begin{equation}
    d_{\rm hor}(v) ~\equiv~ \int_{t_{\rm prod}}^{\tnow} \frac{dt}{a(t)}v(t)
    \label{eq:dhor}
\end{equation}
where $t_{\rm prod}$ denotes the time at which these particles were 
initially produced and $v(t)$ is the velocity of these particles at 
time $t$.  Given that
\begin{equation}
    v(t) ~=~ \frac{p/a(t)}{\sqrt{p^2/a^2(t)+m^2}} 
      ~=~ \frac{\gamma v}{\sqrt{\gamma^2 v^2+a^2(t)}}~,~
\end{equation}
where $\gamma = (1-v^2)^{-1/2}$ is the usual relativistic factor and 
where $p = \gamma mv$ is the present-day momentum, one finds that the 
horizon distance may also be expressed as  
\begin{equation}
    d_{\rm hor}(v) ~=~
    \int_{a_{\rm prod}}^1 \frac{da}{Ha^2} 
    \frac{\gamma v}{\sqrt{\gamma^2 v^2 +a^2}}~,~
    \label{eq:dhorda}
\end{equation}
where $a_{\rm prod} \equiv a(t_{\rm prod})$ and where 
{\il{H\equiv \dot{a}/a}} is the Hubble parameter.

On the one hand, large-scale-structure considerations imply that the
present-day dark-matter velocity distribution $g_v(v)$ must receive 
non-negligible support only at non-relativistic speeds $v \ll 1$.
On the other hand, in order for free-streaming effects to have 
a significant impact on small-scale structure, the dark matter 
must typically be relativistic at production.
Within this regime, one finds that $d_{\rm hor}(v)$ is insensitive to the 
value of $t_{\rm prod}$ and well approximated by
\begin{align}
    d_{\rm hor}(v) ~&=~ \int_{a_{\rm prod}}^1 \frac{da}{Ha^2} 
    \frac{v}{\sqrt{v^2 +a^2}} \nonumber\\
    ~&\approx~ \frac{v}{(a^2H)_{\rm MRE}}
    \left[2-2a_{\rm MRE}^{1/2} + 
    \log\left(\!\frac{2a_{\rm MRE}}{v}\!\right)\right]~,~
    \label{eq:dFSH}
\end{align}
up to corrections of $\mathcal{O}(v^3)$, where the subscript ``MRE'' indicates 
the value of the corresponding quantity at the time of matter-radiation
equality.  

In order to assess the impact of free-streaming on $P(k)$, it is 
useful to associate a wavenumber $k_{\rm hor}(v) \sim 1/d_{\rm hor}(v)$ 
with the particle horizon.  While $d_{\rm hor}(v)$ is an unambiguously 
defined quantity, the precise relationship between this distance scale and 
$k_{\rm hor}(v)$ depends on the conventions adopted and is defined only up 
to an overall $\mathcal{O}(1)$ multiplicative factor.  In other words,
the horizon wavenumber may be written as
\begin{equation}
  k_{\rm hor}(v) ~\equiv ~   
     \xi \left[ \int_ {a_{\rm prod}}^1  \frac{da}{ H a^2 } \, 
      \frac{\gamma v}{\sqrt{\gamma^2 v^2 + a^2 }}\right]^{-1}~,
  \label{eq:khor}
\end{equation}
where $\xi$ represents this $\mathcal{O}(1)$ factor.  Physically, 
$k_{\rm hor}(v)$ represents the wavenumber above which the 
free-streaming of a population of particles with velocity $v$ leads to a 
suppression of power in $P(k)$.

Up to this point, we have considered the free-streaming effects associated 
with a population of dark-matter particles with a uniform speed $v$.  For
such a population of particles, there exists a single, well-defined 
particle horizon $d_{\rm hor}(v)$, and thus a single horizon wavenumber 
$k_{\rm hor}(v)$.  By contrast, for a population of dark-matter particles 
with a continuous distribution of speeds $g_v(v)$, the situation is
significantly more complicated.  Indeed, for such a distribution of
particle speeds, Eq.~(\ref{eq:khor}) implies that there is a continuous
distribution of horizon wavenumbers, and thus that different
parts of the dark-matter velocity distribution contribute to the suppression
of structure above different threshold values of $k$.  
Nevertheless, even for complicated $g_v(v)$ distributions spanning a broad
range of dark-matter speeds, the impact of free-streaming on the shape 
of the linear matter power spectrum at late times may reliably be assessed 
numerically by means of Einstein-Boltzmann solvers such as the 
\texttt{CLASS} software 
package~\cite{Lesgourgues:2011re,Blas:2011rf,Lesgourgues:2011rg,Lesgourgues:2011rh}.  
In this way, under standard cosmological assumptions, a given dark-matter velocity 
distribution $g_v(v)$ gives rise to a particular form for $P(k)$. 

We also observe that for any particular choice of $\xi$, 
the function $k_{\rm hor}(v)$ in Eq.~(\ref{eq:khor}) represents a one-to-one 
map between a present-day dark-matter velocity $v$ within this distribution 
and a corresponding wavenumber $k$.  This map is also invertible in the 
sense that one may define a function $k_{\rm hor}^{-1}(k)$ which maps a particular
wavenumber $k$ to a value of $v$ --- in particular, to the value of $v$ for 
which this input value of $k$ is the threshold value for free-streaming 
suppression.  Given this one-to-one correspondence, we shall take an unorthodox 
approach in what follows and regard Eq.~(\ref{eq:khor}) as defining a functional 
map between $v$ and the variable $k$ itself~\cite{Dienes:2020bmn}.
We emphasize, however, that this interpretation of Eq.~(\ref{eq:khor})
is simply a reflection of the threshold relationship that exists between $v$ 
and $k$.  

Of course, for certain classes of $g_v(v)$ distributions --- in particular, 
distributions which are unimodal and sharply peaked around some particular value 
of $v$ --- the fact that different values of $v$ within the $g_v(v)$ distribution 
correspond to different particle horizons is relatively unimportant.
Indeed, for velocity distributions of this sort, it is common practice to define a 
single ``free-streaming scale'' $k_{\rm FSH}$ for the distribution as a whole 
by replacing the present-day velocity $v$ in Eq.~(\ref{eq:khor}) with the 
average velocity
\begin{equation}
    \langle v\rangle\! ~=~
      \frac{g_{\rm int}}{2\pi^2 N(t)}\int_{-\infty}^{0} 
      d\log v\, v\,g_v(v)~.~
\end{equation}
For velocity distributions of this sort,
including those associated with WDM models, this approximation provides a 
reasonably reliable way of assessing the effects of free-streaming on $P(k)$.
However, we emphasize that it {\it is}\/ an approximation,  Moreover, as 
we shall see, this approximation is often inadequate to characterize the 
effects of free-streaming on the growth of structure in dark-matter 
scenarios with more general velocity distributions outside its regime of 
validity.

In order to illustrate how this approximation can yield misleading results 
when applied to more complicated dark-matter velocity distributions, we 
consider a class of such distributions with a particular functional form and 
assess the impact of free-streaming on $P(k)$ for distributions within this class. 
In particular, we shall consider $g_v(v)$ distributions of the form
\begin{eqnarray}
  g_v(v) ~&=&~  \sum_{i=0}^{1}
    \frac{\mathcal{N}\Omega_i}{\sqrt{2\pi}
    \hat{\sigma}_i\Omega_{\rm DM}}
    \nonumber \\ & & ~~\times
    \exp\left\{\!-\frac{1}{2\hat{\sigma}^2_i}
    \left[\log\left(\!\frac{\hat{p}(v)}{\langle \hat{p} \rangle_i}\!\right) +
    \frac{1}{2}\hat{\sigma}^2_i\right]^2\!\right\} ~,~~~~~~~
  \label{eq:gsumraw}
\end{eqnarray}
where $\hat{p}(v)\equiv \gamma m v$.  In other words this choice for $g_v(v)$ 
is nothing but a log-normal distribution {\it in $p$-space}, with $\Omega_i$, 
$\langle \hat{p}\rangle_i$ and $\hat{\sigma}_i$ respectively representing the
abundance, average momentum, and width associated with the corresponding peak in 
$p$-space.  Here $\Omega_{\rm DM} = \Omega_0 + \Omega_1$ is the total present-day 
dark-matter abundance, $\mathcal{N}$ is an appropriate normalization factor, and
we adopt the convention that the index $i$ labels the peaks in order of decreasing 
average speed.  Of course, since $\hat{p}(v)$ is generally a non-linear function 
of $v$, a log-normal function in $p$-space is generally not a log-normal function 
in $v$-space.   However, in cases for which $\langle \hat p\rangle_i  \ll m$ for all $i$ and for which $\hat \sigma_i$ is not exceedingly large,
$g_v(v)$ mostly receives support for $v\ll 1$.    In such cases, 
this function reduces to 
\begin{eqnarray}
  g_v(v) &~\approx~& \sum_{i=0}^{1}
    \frac{\mathcal{N}\Omega_i}{\sqrt{2\pi}\sigma_i\Omega_{\rm DM}}
    \nonumber \\ & & ~~\times
    \exp\left\{\!-\frac{1}{2\sigma^2_i}
    \left[\log\left(\!\frac{v}{\langle v \rangle_i}\!\right) +
    \frac{1}{2}\sigma^2_i\right]^2\!\right\} ~,~~~~~~~
  \label{eq:gsum}
\end{eqnarray}
where $\Omega_i$, $\langle v \rangle_i$ and $\sigma_i$ respectively represent the
abundance, average speed, and widths associated with the corresponding peaks in $v$-space.
This provides a very good approximation for any 
phenomenologically viable present-day dark-matter velocity distribution, which 
necessarily receives non-negligible support only at velocities $v \ll 1$.
We emphasize that while we have chosen
this functional form for $g_v(v)$ for purposes of illustration,
bi-modal $g_v(v)$ distributions with approximately this form arise naturally 
in a variety of non-minimal dark-sector 
scenarios~\cite{Konig:2016dzg,Du:2021jcj,Decant:2021mhj,Heeck:2017xbu,Dienes:2020bmn}.

\begin{figure*}[ht!]
  \centering
  \includegraphics[clip, width=0.98\textwidth]{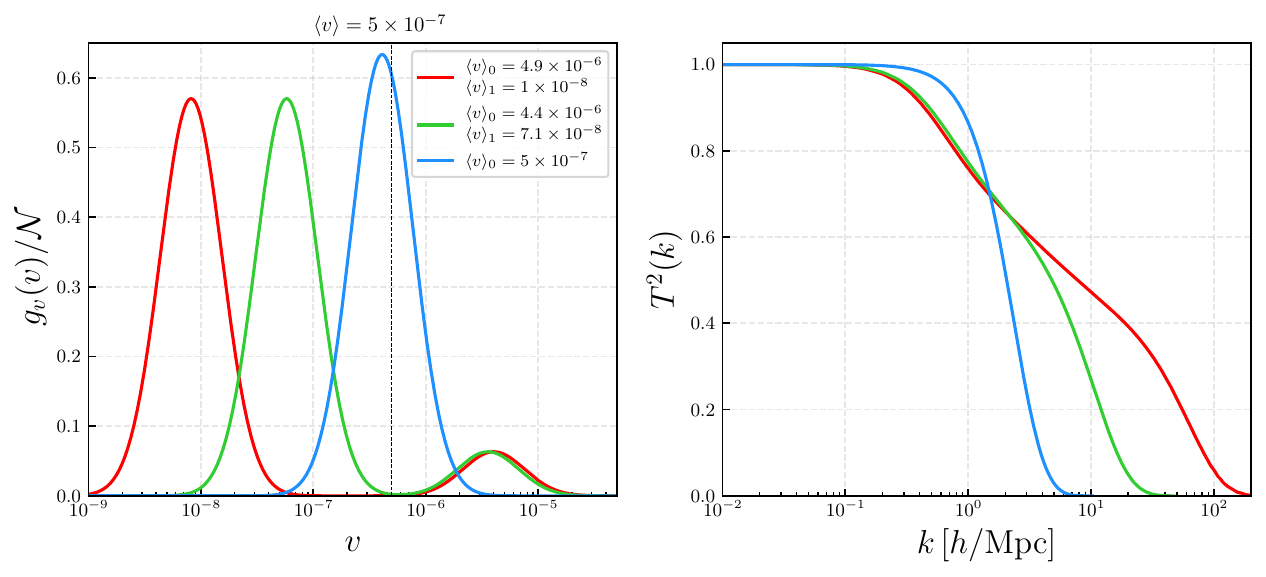}
  \caption{Three dark-matter velocity distributions $g_v(v)$ of the general form 
    specified in Eq.~(\protect\ref{eq:gsum}) (left panel), all of which 
    have the same average velocity $\langle v\rangle = 5\times 10^{-7}$ and 
    therefore the same nominal free-streaming scale $k_{\rm FSH}$, but lead to 
    very different matter power spectra $P(k)$ (right panel).  The blue curve
    in each panel corresponds to the case in which $\Omega_1 = 0$
    and $g_v(v)$ consists of a single Gaussian peak with 
    $\langle v\rangle_0 = 5\times 10^{-7}$.  The red curve 
    corresponds to the case of a bi-modal $g_v(v)$ distribution with 
    $\Omega_0/\Omega_{\rm DM} = 0.1$, $\langle v\rangle_0 = 4.9\times 10^{-6}$,
    and $\langle v\rangle_0 = 1.0\times 10^{-8}$.  The green curve 
    corresponds to the case of a bi-modal $g_v(v)$ distribution with 
    $\Omega_0/\Omega_{\rm DM} = 0.1$, $\langle v\rangle_0 = 4.4\times 10^{-6}$,
    and $\langle v\rangle_0 = 7.1\times 10^{-8}$.  For all of these distributions,
    we have taken $\sigma_0 = \sigma_1 = 0.63$.
    We observe from the right panel that the matter power spectra associated 
    these latter $g_v(v)$ distributions differ significantly from the 
    spectrum associated with the unimodal one.  
  \label{fig:gvsandPks}}
\end{figure*}

In the left panel of Fig.~\ref{fig:gvsandPks}, we show three 
$g_v(v)$ distributions of the form given in Eq.~(\ref{eq:gsum}).  
All of these distributions have the same average velocity 
$\langle v\rangle = 5\times 10^{-7}$ and therefore the same nominal 
free-streaming scale $k_{\rm FSH}$.  The blue curve corresponds to the case 
in which $\Omega_1 = 0$ and $g_v(v)$ consists of a single 
Gaussian peak with $\langle v\rangle_0 = 5\times 10^{-7}$.  
We note that the center of the peak does not coincide with $\langle v\rangle_0$
because $g_v(v)$ is a log-normal distribution with respect to $v$ itself, and the 
mean value for such a distribution is offset from the maximum.  
The red curve corresponds to the parameter choices $\Omega_0/\Omega_{\rm DM} = 0.1$, 
$\langle v\rangle_0 = 4.9\times 10^{-6}$, and $\langle v\rangle_0 = 1.0\times 10^{-8}$,
while the green curve corresponds to the parameter choices 
$\Omega_0/\Omega_{\rm DM} = 0.1$, $\langle v\rangle_0 = 4.4\times 10^{-6}$,
and $\langle v\rangle_0 = 7.1\times 10^{-8}$.   
For all of these distributions, we have taken $\sigma_0 = \sigma_1 = 0.63$, 
a value which corresponds to the standard deviation of $\log v$ obtained for 
any WDM distribution, regardless of the mass of the dark-matter particle.

In the right panel of Fig.~\ref{fig:gvsandPks}, we show the transfer function 
$T^2(k)$ obtained for each of these three $g_v(v)$ distributions.  We observe 
that the two transfer functions obtained for the bi-modal distributions differ 
dramatically from that obtained for the unimodal distribution with the same 
nominal free-streaming scale, despite the fact that the fractional abundance
$\Omega_0/\Omega_{\rm DM}$ associated with the higher-velocity peak is quite 
small.  These results, then, illustrate how $k_{\rm FSH}$ alone fails to provide 
a complete and accurate picture of how free-streaming affects the linear matter 
power spectrum in dark-matter scenarios with complicated, multi-modal $g_v(v)$ 
distributions and how the variation of $k_{\rm hor}(v)$ across the dark-matter 
velocity distribution must be taken into account in such scenarios in order
to obtain an accurate description of $P(k)$. 

While the results in Fig.~\ref{fig:gvsandPks} provide a qualitative 
picture of the extent to which the matter power spectra associated
with complicated, multi-modal $g_v(v)$ distributions differ from those 
associated with narrow, unimodal such distributions, it is also 
illuminating to investigate these differences in a more systematic, 
quantitative manner.  For example, it is interesting to assess the
degree to which the matter power spectrum that follows from a given 
$g_v(v)$ distribution differs not merely from the $P(k)$ curve
associated with one particular narrow, unimodal distribution, but from 
{\it any}\/ such distribution.

In order to perform such an analysis for
$g_v(v)$ distributions of the form given in Eq.~(\ref{eq:gsum}),
we begin by evaluating the transfer function $T^2(k)$ for the 
$g_v(v)$ distribution of interest using the {\rm CLASS} code package.
We then compare this distribution to a family of transfer functions 
obtained for a representative sample of narrow, unimodal $g_v(v)$
distributions.  Since any two narrow, unimodal $g_v(v)$ distributions with 
the same $\langle v\rangle$ yield very similar matter power spectra,
it is sufficient to include only $g_v(v)$ distributions associated with 
WDM models in this sample.  To a very good approximation, the transfer 
function for a WDM model depends only on the mass $m_{\rm WDM}$ of the 
dark-matter particle and takes the
form~\cite{Bode:2000gq,Hansen:2001zv,Viel:2005qj}
\begin{equation}
  T_{\rm WDM}^2(k) ~\approx~ \left[1 + (\alpha k)^{2\nu}\right]^{-10/\nu} \ ,
\label{eq:T2WDM}
\end{equation}
where \il{\nu = 1.12} and where 
\begin{equation}
\alpha ~=~ \frac{0.049~{\rm Mpc}}{h}
  \left(\frac{m_{\rm WDM}}{\rm keV}\right)^{\!\!-1.11} 
    \hskip -0.04truein 
      \left(\frac{\Omega_{\rm WDM}}{0.25}\right)^{\!\!0.11}
    \hskip -0.04truein 
      \left(\frac{h}{0.7}\right)^{\!\!1.22}~.
\end{equation}
Thus, the family of transfer functions to which we shall compare 
$T^2(k)$ for our $g_v(v)$ distribution of interest consists of 
a sample of WDM transfer functions $T_{\rm WDM}^2(k)$ corresponding 
to different values of $m_{\rm WDM}$.  In constructing this sample, 
we survey a broad range of $m_{\rm WDM}$ masses with a step size 
sufficiently small that further reducing that step size does not 
significantly impact our results.   

For each value of $m_{\rm WDM}$ in this survey,
we sample both the corresponding transfer function $T^2_{\rm WDM}(k)$ 
given by Eq.~(\ref{eq:T2WDM}) and the transfer function $T^2(k)$ obtained for 
our double-peak $g_v(v)$ distribution at a series of wavenumbers $k_j$ separated 
by regular intervals in $(\log k)$-space within the range 
$0.1~h{\rm Mpc}^{-1} \leq k \leq 2000~h{\rm Mpc}^{-1}$.  We assess the 
goodness of fit between the two curves using the chi-square statistic 
\begin{equation}
    \chi^2(m_{\rm WDM}) ~=~ \sum_j
      \frac{[T^2(k_j) - T^2_{\rm WDM}(k_j)]^2}{\sigma_{T^2}^2(k_j)}~,~
   \label{eq:chisq}
\end{equation}
where $\sigma_{T^2}(k_j)$ is the uncertainty in the transfer function at $k_j$.
For simplicity, since the choice of $\sigma_{T^2}(k_j)$ values for this 
theoretical comparison is somewhat arbitrary, we take $\sigma_{T^2}(k_j)$ 
to be equal to a common value $\sigma_{T^2}$ for all $k_j$.  Under this 
assumption, $\sigma_{T^2}$ may be viewed simply as a normalization factor.
We take the minimum value 
\begin{equation}
    \chi^2_{\rm min} ~=~ \underset{m_{\rm WDM}}{\min}
      \left\{\chi^2(m_{\rm WDM}) \right\}
   \label{eq:chisqsmin}
\end{equation}
from among all of the $\chi^2(m_{\rm WDM})$ values obtained in this way to be our
relative measure of the distinctiveness of $T^2(k)$.  

\begin{figure}[t!]
  \centering
  \includegraphics[clip, width=0.49\textwidth]{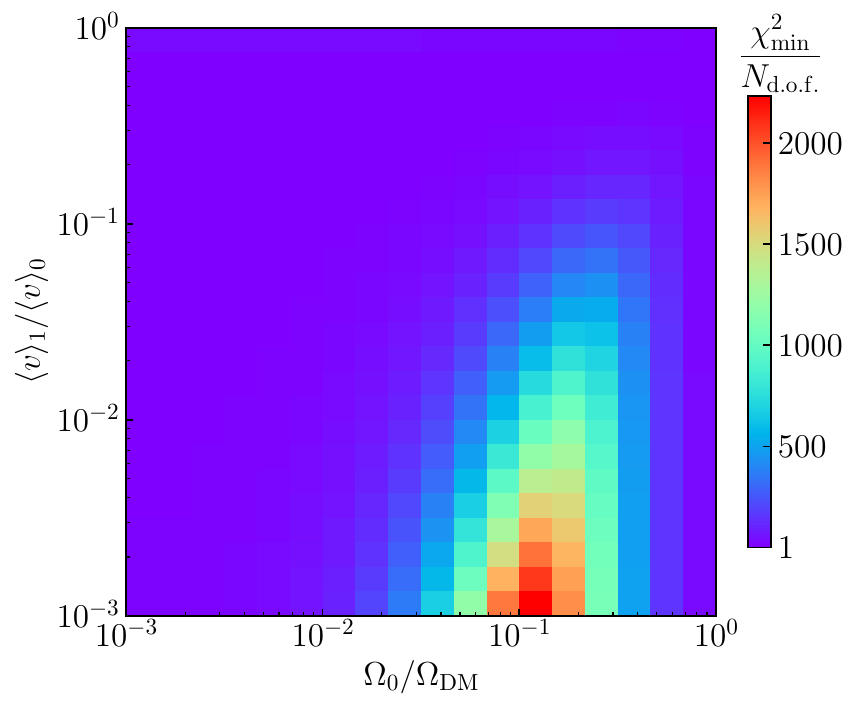}~
  \caption{The minimum value $\chi^2_{\rm min}$ of the chi-square statistic obtained 
    by fitting the linear matter power spectrum obtained for bi-modal dark-matter
    velocity distribution to the spectrum obtained for any WDM distribution,
    displayed in the $(\langle v\rangle_1/\langle v\rangle_0, 
    \Omega_1/\Omega_0)$-plane.  The results displayed here correspond to the 
    parameter choices $\sigma_0 = \sigma_1 = 0.63$ and 
    $\langle v\rangle_0 = 10^{-6}$.
  \label{fig:Chi2min}}
\end{figure}

We can endow this goodness-of-fit 
statistic with a meaningful interpretation by choosing the normalization factor 
$\sigma_{T^2}$ in accord with the usual statistical expectation that 
$\chi^2_{\rm min}/N_{\rm d.o.f.} \sim 1$ when the fit 
between $T^2(k)$ and the optimal $T^2_{\rm WDM}(k)$ distribution in our 
sample is good, where $N_{\rm d.o.f.}$ is the number of degrees of freedom.  Thus,  
we choose $\sigma_{T^2}$ such that $\chi^2_{\rm min}/N_{\rm d.o.f.} = 1$ when 
$\Omega_0 = 1$ and $\sigma_0 = 0.63$ and the $g_v(v)$ distribution reduces to a 
unimodal distribution with a standard deviation equal to that of a WDM distribution.
With this choice of normalization, $\chi^2_{\rm min}/N_{\rm d.o.f.} \sim \mathcal{O}(1)$ 
indicates that the matter power spectrum obtained from the corresponding double-peak 
$g_v(v)$ distribution does not differ significantly from that obtained in some WDM scenario, 
while $\chi^2_{\rm min}/N_{\rm d.o.f.} \gg 1$ indicates a significant difference.

In Fig.~\ref{fig:Chi2min}, we show the value of 
$\chi_{\rm min}^2/N_{\rm d.o.f.}$ within
the $(\langle v\rangle_1/\langle v\rangle_0,\Omega_1/\Omega_{\rm DM})$-plane 
for the $g_v(v)$ distribution in Eq.~(\ref{eq:gsum}) with 
$\sigma_0 = \sigma_1 = 0.63$ and $\langle v\rangle_0 = 10^{-6}$.
We observe that there exists a sizable region of parameter space within 
which $\chi_{\rm min}^2/N_{\rm d.o.f.} \gg 1$ and the 
matter power spectrum obtained from the double-peak $g_v(v)$ distribution
in Eq.~(\ref{eq:gsum}) indeed differs significantly from that obtained 
from any WDM distribution.  The largest values of 
$\chi_{\rm min}^2/N_{\rm d.o.f.}$ are obtained within regions
of the plane wherein $\Omega_0$ is smaller than but not negligible in comparison 
with $\Omega_1$ and there is s significant separation between the mean velocities
of the two peaks.

To summarize the results of this section, we have reviewed the physical 
principles behind the suppression of small-scale structure due to the 
free-streaming of dark-matter particles.  We have demonstrated that while
the variation of the horizon wavenumber $k_{\rm hor}(v)$ across the 
dark-matter velocity distribution has relatively little impact on $P(k)$ 
when $g_v(v)$ is unimodal and narrow, it can have an enormous impact on
$P(k)$ for more general $g_v(v)$ distributions.
Moreover, on the basis of the threshold relationship between $v$ and $k$
that exists by virtue of Eq.~(\ref{eq:khor}), we have been able to define
an invertible map between these two variables --- a map which can be 
exploited in order to extract information about $g_v(v)$ from 
$P(k)$~\cite{Dienes:2020bmn}.


\section{From Matter Power Spectra to Halo-Mass Functions\label{sec:HaloMassFunction}}


Having summarized the relationship between $g_v(v)$ and $P(k)$, we now discuss 
the relationship between $P(k)$ and $dn/d\log M$.  In relating these two quantities, 
we follow the analytic approach originally posited by Press and 
Schechter~\cite{Press:1973iz} and subsequently justified using the excursion-set 
formalism of Bond {\it et al.}~\cite{Bond:1990iw}.  

At late times, regions of 
space with sufficiently large average overdensity collapse under their own gravity 
and form compact, virialized objects --- \ie, matter halos.  The probability that 
a randomly chosen spherical region of space with radius $R$ will collapse prior 
to a given cosmological time $t$ depends on the statistical properties of
$\delta(\vec{\mathbf{x}},t)$.  The crucial quantity in this 
regard is the spatial average $\sigma^2(t,R)$ of the variance of 
$\delta(\vec{\mathbf{x}},t)$ within the same region.  This spatial average 
may be written as
\begin{equation}
  \sigma^2(t,R) ~\equiv~ \int_{-\infty}^{\infty}d\log k\,\, W^2(k,R) \frac{k^3 P(k,t)}{2\pi^2}~,
  \label{eq:VarianceOfDensPertAvgd}
\end{equation} 
where $W(k,R)$ is the Fourier transform in $k$-space of the position-space top-hat 
function {\il{W(r,R) \equiv \Theta(1-r/R)}}, where $\Theta(x)$ denotes the Heaviside 
function.  This enforces the condition that only points at distances {\il{r \leq R}} 
away from the center of the region are included in the average.  However, this 
definition of $\sigma^2(t,R)$ may also be generalized to include other functional 
forms for $W(k,R)$.  In this paper, we shall instead adopt a window function which 
is a top-hat function in $k$-space~\cite{Bertschinger:2006nq}: 
\begin{equation}
  W(k,R) ~=~ \Theta(1-kR)~.
  \label{eq:Windowfn}
\end{equation}

One well-known advantage of the window function in Eq.~(\ref{eq:Windowfn}) is that
its flatness in $k$-space allows $\sigma^2(t,R)$ to be sensitive to the natural shape of 
the matter power spectrum itself, rather than that of $W(k,R)$~\cite{Schneider:2013ria}.
This window function also has other advantages.  One of these 
is that only density perturbations with wavenumbers {\il{k \leq R^{-1}}} have any effect on 
$\sigma^2(t,R)$.  The primary drawback of this form 
of $W(k,R)$, however, is that the precise mathematical relationship between the value of $R$ 
associated with a halo and the corresponding halo mass $M$ is not well defined.
Nevertheless, since symmetry considerations dictate that {\il{M \propto R^3}}, the 
relationship between $M$ and $R$ may be parametrized as
\begin{equation}
  M ~\equiv~ \frac{4\pi}{3}\overline{\rho} (c_W R)^3~, 
  \label{eq:RtoMmap}
\end{equation}
where $\overline{\rho}$ is the present-day mass density of the universe and
where the value of the coefficient $c_W$ may be obtained from the results of numerical
simulations.  Following Ref.~\cite{Schneider:2014rda}, we take {\il{c_W\approx 2.5}}.
The value of $\overline{\rho} = \Omega_m\rhocrit$ can be determined from the total 
present-day matter abundance and critical density, which we take to be 
$\Omega_m \approx 0.315$ and 
$\rhocrit \approx 2.775 \times 10^{11} h^2~M_\odot 
{\rm Mpc}^{-3}$~\cite{Planck:2018vyg}, respectively.
Given that a well-defined one-to-one relationship exists between $M$ and $R$, the
spatially-averaged variance $\sigma^2(t,M)$ may also be viewed as a function of the 
halo mass $M$.

Within the Press-Schechter formalism, the present-day halo-mass function which 
follows from any particular $P(k)$ profile takes the form   
\begin{equation}
  \frac{dn}{d\log M} ~=~ \frac{\overline{\rho}}{2M}\eta(M)\frac{d\log\nu(M)}{d\log M}~,
  \label{eq:PressSchechter} 
\end{equation}
where {\il{\nu(M) \equiv \delta_c^2/\sigma^2(\tnow,M)}}, with {\il{\delta_c\approx 1.686}}  
denoting the critical overdensity, and where the function $\eta(M)$, which depends on
$M$ only through the function $\nu(M)$, represents the probability density of obtaining an 
averaged fractional overdensity at a given location.  For the window function in
Eq.~(\ref{eq:Windowfn}), regardless of the form of $\eta(M)$, the expression for 
$dn/d\log M$ in Eq.~(\ref{eq:PressSchechter}) simplifies to
\begin{eqnarray}
  \frac{dn}{d\log M} ~=~ \frac{\overline{\rho}}{12\pi^2 M} \nu(M) \eta(M) 
    \frac{P\big(1/R(M)\big)}{\delta_c^2 R^3(M)}~,~~~
  \label{eq:PressSchechterSimp} 
\end{eqnarray}
where $R(M)$ is the particular value of $R$ which corresponds to a given halo mass $M$
through Eq.~(\ref{eq:RtoMmap}).

A variety of possible forms for the function $\eta(M)$ have been proposed, based 
either on purely theoretical grounds or based on the results of $N$-body or hydrodynamic 
simulations~\cite{Press:1973iz,Sheth:1999mn,Sheth:1999su,Jenkins:2000bv,
Warren:2005ey,Tinker:2008ff,
Crocce:2009mg,Bhattacharya:2010wy,Watson:2012mt,Seppi:2020isf}.  
In what follows, we adopt the form~\cite{Sheth:1999mn,Sheth:1999su}     
\begin{equation}
  \eta(M) ~=~ \sqrt{\frac{2\nu(M)}{\pi}} 
    A\left[1 + \nu^{-\alpha}(M)\right]e^{-\nu(M)/2}~,
  \label{eq:ShethTormanf}
\end{equation}
where {\il{A \approx 0.3222}} and {\il{\alpha = 0.3}}.  This form for 
$\eta(M)$ is mathematically simple and accords reasonably well with the 
results of numerical simulations.  We shall discuss the way in which 
alternative functional forms for $\eta(M)$ could affect the results 
of our analysis in Sect.~\ref{sec:Conclusions}.  

In order to quantify the extent to which $dn/d\log M$ deviates from the 
corresponding result $(dn/d\log M)_{\rm CDM}$ that we would obtain for 
purely cold dark matter (CDM), we shall henceforth define the 
dimensionless {\it structure-suppression function}\/ according 
to the relation
\begin{equation}
  S(M) ~\equiv~ \sqrt{  \frac{dn/d\log M }{ (dn/d\log M)_{\rm CDM} }}~.
  \label{eq:StrucSuppFn}
\end{equation}
This function may be viewed as playing an analogous role with respect to the halo-mass
function that the transfer function $T(k)$ plays with respect to the linear matter 
power spectrum.  Moreover, for the Press-Schechter halo-mass function in
Eq.~(\ref{eq:PressSchechterSimp}), we find that $S^2(M)$ and 
$T^2(k)$ are related in any given dark-matter model by  
\begin{equation}
  S^2(M) ~=~ \frac{\nu(M) \eta(M)}{\nu_{\rm CDM}(M)\eta_{\rm CDM}(M)} T^2(1/R(M))~,
  \label{eq:StrucSuppFnSimp}
\end{equation}
where $\nu(M)$ and $\eta(M)$ are obtained from the corresponding matter power spectrum 
$P(k)$ for the dark-matter model in question.

By examining the mathematical relationship between $g_v(v)$ and 
$S^2(M)$, we may hope to develop some intuition about the manner in which the 
detailed shape of the dark-matter velocity distribution affects the 
statistical properties of dark-matter halos and subhalos.
As a first step toward developing that intuition, we begin by examining 
the relationship between $k$ and the halo mass $M$.
The underlying reason we were able to construct a map between $v$ and $k$ in
Sect.~\ref{sec:VelsMPS} is that $k_{\rm hor}(v)$ represents the threshold 
value of $k$ below which dark-matter particles with momentum $v$ cannot 
suppress structure by free-streaming.  Fortunately, a similar relationship 
exists between $k$ and $M$ by virtue of the window function $W(k,R)$.  
Indeed, we see from the relationship between $R$ and $M$ in 
Eq.~(\ref{eq:RtoMmap}) that $W(k,R)$ establishes a threshold 
value of $M$ for any given wavenumber $k$ above which density perturbations 
with that value of $k$ have no effect on $\sigma^2(\tnow,M)$ and therefore 
no effect on the halo-mass function.  In particular, we see from
Eqs.~(\ref{eq:Windowfn}) and~(\ref{eq:RtoMmap}) that 
this threshold value of $M$ is specified by the function        
\begin{equation}
  M_{\rm hal}(k) ~\equiv~ \frac{4\pi}{3}\overline{\rho} \left(\frac{c_W}{k}\right)^3~.
  \label{eq:Mhal}
\end{equation}
Just as we regarded Eq.~(\ref{eq:khor}) as defining a functional map between 
$v$ and $k$, we shall likewise regard Eq.~(\ref{eq:Mhal}) as defining a 
functional map between $k$ and $M$.  As with our map between $v$ and $k$, 
this map between $k$ and $M$ is likewise one-to-one and invertible.

We note that the threshold relationship between $k$ and $M$ which have used in
in constructing the functional map in Eq.~(\ref{eq:Mhal}) is only 
precisely defined for 
a window function with a sharp cutoff in $k$ which decreases with increasing $M$.  
Indeed, this is one of the reasons why we have adopted the form for $W(k,R)$ 
in Eq.~(\ref{eq:Windowfn}) in this analysis.  However, this does not mean that a 
heuristic functional map between $k$ and $M$ cannot be defined for other 
forms of $W(k,R)$ as well.  Indeed, by construction any sensible functional form 
for $W(k,R)$ must like serve to suppress the contribution to $\sigma^2(t,R)$ from 
small-scale fluctuations with wavenumbers $k \gg R^{-1}$.  
Thus, even in cases in which $W(k,R)$ does not have a sharp cutoff 
in $k$, a qualitative threshold relationship exists between $k$ and $M$ 
which can be used to formulated an invertible functional map between 
these variables.  In such cases, one could account for the ambiguity in 
the cutoff by incorporating an additional overall proportionality factor 
into Eq.~(\ref{eq:Mhal}) analogous to the parameter $\xi$ 
in Eq.~(\ref{eq:khor}), the optimal value of which would likewise be 
determined empirically.  We emphasize that this proportionality factor is 
not necessarily equivalent to the quantity $c_W^3$ that appears in 
Eq.~(\ref{eq:Mhal}) as a consequence of ambiguities in the 
relationship between $M$ and $R$, but that $c_W^3$ could    
of course be incorporated into this factor. 

Combining the two functional maps in Eqs.~(\ref{eq:khor}) and~(\ref{eq:Mhal}) and 
taking the {\il{a_{\rm prod}\rightarrow 0}} limit appropriate for particles 
which are relativistic at production, we can likewise define a direct 
functional map between $v$ and $M$, which takes the form
\begin{equation}
  M_{\rm hal}(v) ~\equiv~ \frac{4\pi\bar{\rho}c_W^3}{3\xi^3}
    \left[\int_0^1 \frac{da}{Ha^2}\frac{\gamma v}{\sqrt{\gamma^2 v^2 + a^2}}\right]^3~.~
  \label{eq:vtoMmap}
\end{equation}
As with the individual maps in Eqs.~(\ref{eq:khor}) and~(\ref{eq:Mhal}), this map is
one-to-one and invertible.

\begin{figure}[ht!]
  \centering
  \includegraphics[clip, width=0.48\textwidth]{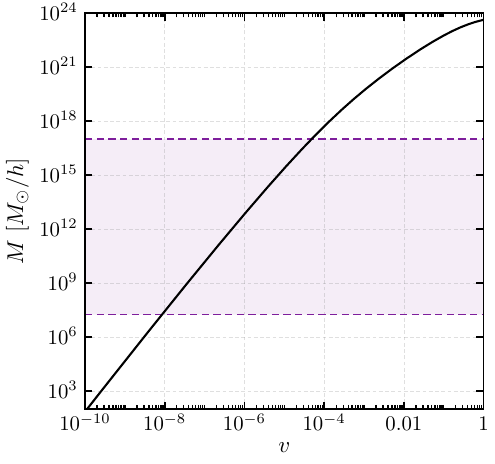}~
  \caption{The halo-mass variable $M_{\rm hal}(v)$, as defined in
    Eq.~(\protect\ref{eq:vtoMmap}), plotted as a function of $v$.  
    The purple shaded region corresponds to the range of 
    $M$ shown in Fig.~\protect\ref{fig:BradyBunch}.    
  \label{fig:Mvsv}}
\end{figure}

The physical motivation for defining $M_{\rm hal}(v)$, which we plot 
in Fig.~\ref{fig:Mvsv}, is that it conveys information about what parts of 
$g_v(v)$ are capable of affecting the halo-mass function at a given mass 
scale.  Dark-matter particles with velocities below a given value $v$ can 
only suppress structure at wavenumbers {\il{k > k_{\rm hor}^{-1}(v)}} and can 
therefore only affect the number density of halos with masses 
{\il{M < M_{\rm hal}(k_{\rm hor}^{-1}(v))}}.  Thus, the value of $M$ associated 
with a given value of $v$ through Eq.~(\ref{eq:vtoMmap}) represents the maximum 
halo mass for which dark-matter particles with that velocity can affect $S^2(M)$.  

Some comments about this functional map between $v$ and $M$ are in order.  
First, we may use this map in order to perform a change of variables and 
define a velocity distribution in $M$-space which corresponds to the 
distribution $g_v(v)$ in $v$-space.  In particular, by changing variables 
from $v$ to $M$ in Eq.~(\ref{eq:ComovingNumberDens}), we find that
\begin{equation}
  N(t) ~=~ \frac{g_{\rm int}}{2\pi^2}\int d\log M g_M(M)~,~    
\end{equation}
where
\begin{equation}
  g_M(M,t) ~\equiv~ g_v\big(M^{-1}_{\rm hal}(M),t\big)\left|
    \frac{d\log v}{d\log M_{\rm hal}(v)}\right|~, 
  \label{eq:gMMdefBody}
\end{equation}
where the factor $\left|d\log v/d\log M(v)\right|$ is simply the Jacobian
for this change of variables.  
This $M$-space velocity distribution in Eq.~(\ref{eq:gMMdefBody}) has 
a straightforward physical meaning.  In particular, up to an overall 
factor of $g_{\rm int}/2\pi$, this distribution is simply the differential 
number density of dark-matter particles per unit $\log M$ with velocities 
just barely sufficient to free-stream out of regions of space within which 
the dark matter would otherwise collapse into halos of mass $M$.

One advantage of defining $g_M(M)$ is that it can be used to facilitate 
a direct graphical comparison between the features of the the dark-matter 
velocity distribution and the features of $S^2(M)$.  Since $g_M(M)$ is also 
a function of $M$, it can be plotted on the same axes as the 
structure-suppression function.  As we shall see, juxtaposing $g_M(M)$ 
and $S^2(M)$ in this way can provide valuable insights into
the relationship between these two functions.

Another advantage of defining $g_M(M)$ in this way is that it allows us to
characterize the fraction of dark-matter particles which are ``hot'' relative to 
the mass scale $M$ --- \ie, capable of free-streaming out of regions which would 
collapse into halos of mass $M$ --- in a straightforward manner.
The threshold velocity above which dark-matter particles can free-stream out of
such regions is $v = M^{-1}_{\rm hal}(M)$.  The portion of $g_v(v')$ with 
velocities $v' > v$ above this threshold corresponds to the portion 
of $g_M(M')$ with $M' > M$.  Thus, up to an overall proportionality constant, 
the number density of dark-matter particles with velocities sufficient 
to free-stream out of halos of mass $M$ can be obtained by integrating 
$g_M(M')$ over $M'$ above this threshold.  Motivated by this consideration, 
we define the {\it hot-fraction function}\/
\begin{equation}
   F(M) ~\equiv~   \frac{\int_{\log M}^\infty d \log M'\, g_M(M')}
     {\int_{-\infty}^\infty d\log M'\, g_M(M')}~,~
  \label{eq:hotfrac}
\end{equation} 
which represents the fraction of the total dark-matter abundance associated with 
these particles.


\section{Halo-Mass Functions for Non-Trivial Dark-Matter Velocity Distributions\label{sec:ResultsHMF}}


We shall now examine how the detailed shape of $g_v(v)$ affects the shape of 
the halo-mass function within the context of the analytic formalism outlined
in Sect.~\ref{sec:HaloMassFunction}.  In doing so, we shall begin by focusing
on a comparatively simple form for $g_v(v)$ which can be modified in a 
controlled way and assessing the effect of these modifications on $S^2(M)$.  
In particular, we shall consider the limiting case of Eq.~(\ref{eq:gsum}) in which
and $\sigma_1$ is narrow and $\langle v\rangle_1$ is sufficiently
small that the dark-matter particles associated with the lower-velocity peak 
behave like CDM.  In this limit, the 
$g_v(v)$ distribution reduces to a single Gaussian peak accompanied 
by a purely cold component which makes up the remainder of $\Omega_{\rm DM}$.
This single peak is characterized by only three parameters: its
average velocity $\langle v \rangle_0$, width $\sigma_0$, and abundance 
$\Omega_0$.

\begin{figure*}[t!]
  \centering
  \includegraphics[clip, width=0.98\textwidth]{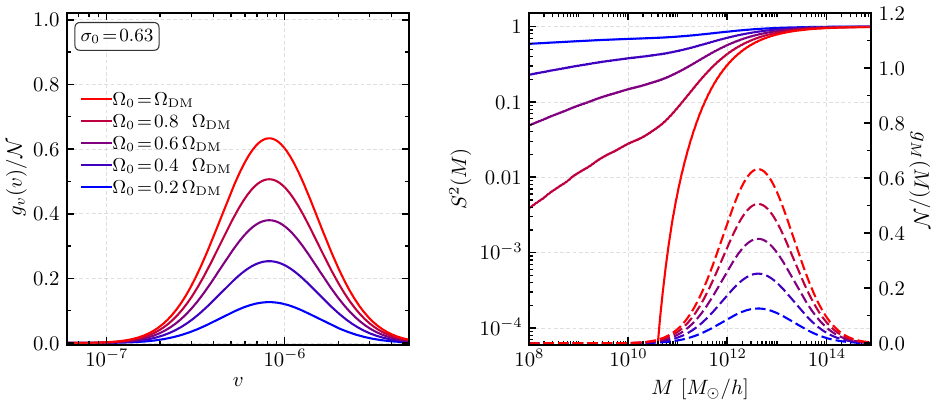}
  \caption{Several dark-matter velocity distributions, each consisting of a 
    single Gaussian peak with a different fractional abundance 
    $\Omega_0/\Omega_{\rm DM}$ and a purely cold component which contributes 
    the remainder of $\Omega_{\rm DM}$, along with the corresponding 
    structure-suppression functions $S^2(M)$.  The left panels shows the 
    $g_v(v)$ distributions as functions of $v$; the right panel 
    shows both the corresponding distributions $g_M(M)$ that we obtain 
    from our functional map between $v$ and $M$ as functions of $M$
    and the $S^2(M)$ curves that we obtain for these distributions. 
    For each of the distributions shown, we have taken $\sigma_0 = 0.63$ 
    and $\langle v\rangle_0 = 10^{-6}$.  We observe that the logarithmic slope
    $d\log S^2(M)/d\log M$ of the structure-suppression function at small $M$ 
    is correlated with the total abundance of dark-matter particles in the 
    $g_M(M')$ distribution with $M' > M$.
  \label{fig:HMFVaryOmegaSP}}
  \bigskip
\bigskip
  \includegraphics[clip, width=0.98\textwidth]{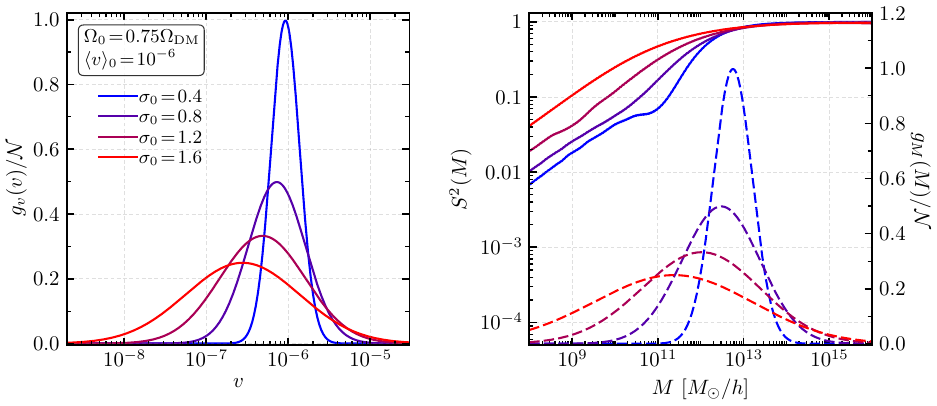}
  \caption{Similar to Fig.~\protect\ref{fig:HMFVaryOmegaSP}, but
    for a set of dark-matter velocity distributions in which the width
    $\sigma_0$ of the Gaussian peak is varied while the total abundance
    $\Omega_0 = \Omega_{\rm DM}$ and average velocity 
    $\langle v\rangle_0 = 10^{-6}$ are held fixed.  We observe that the 
    value of $S^2(M)$ in the region to the left of the peak in $g_M(M)$ 
    is sensitive to the width of the peak, the value of 
    $d\log S^2(M)/d\log M$ in this same region is not.
  \label{fig:HMFVarySigmaSP}}
\end{figure*}

In Fig.~\ref{fig:HMFVaryOmegaSP}, we illustrate the effect of varying
$\Omega_0$ while holding its 
width $\sigma_0$ and average velocity $\langle v\rangle_0$ fixed.  In the left
panel, we show the $g_v(v)$ distributions obtained for several such distributions
with different $\Omega_0$ values as functions of $v$.  In the right panel, we
show both the corresponding distributions $g_M(M)$ that we obtain from our 
functional map between $v$ and $M$ as functions of $M$ and the corresponding 
structure-suppression functions $S^2(M)$.  

Moving from right to left across the right panel of Fig.~\ref{fig:HMFVaryOmegaSP},
we observe that the $S^2(M)$ remains effectively unsuppressed in the region to 
the right of the peak in $g_M(M)$.
More importantly, we also observe that that the logarithmic slope 
$d\log S^2(M)/d\log M$ of the structure-suppression function in the region
immediately to the left of the peak appears to be correlated with the abundance 
$\Omega_0$.  

In Fig.~\ref{fig:HMFVarySigmaSP}, we illustrate the effect of varying the
width $\sigma_0$ of the Gaussian peak while holding its abundance 
$\Omega_0 = \Omega_{\rm DM}$ and average velocity $\langle v\rangle_0$ fixed.
We emphasize that since each Gaussian peak in Eq.~(\ref{eq:gsum}) is 
centered at $\log \langle v\rangle_i - \sigma_i^2/2$ in ($\log v$)-space, 
the location of the peak in $g_v(v)$ shown in the left panel of the figure 
shifts to the left as the corresponding $\sigma_0$ increases.  A corresponding
shift is of course also evident in the $g_M(M)$ distributions shown in the
right panel.

We observe that that increasing $\sigma_0$ induces to a more gradual suppression 
of $S^2(M)$ as we move from right to left across the right panel of 
Fig.~\ref{fig:HMFVarySigmaSP}, ultimately resulting in less net suppression in
$S^2(M)$ at small values of $M$.
However, we also observe that while the value of $S^2(M)$ in the region to the left 
of the peak in $g_M(M)$ is sensitive to the width of the peak, the logarithmic
slope $d\log S^2(M)/d\log M$ in that same region is not.

Taken together, the results shown in Figs.~\ref{fig:HMFVaryOmegaSP} 
and~\ref{fig:HMFVarySigmaSP}
suggest that the value of $d\log S^2(M)/d\log M$ at any given value
of $M$ is correlated with the total abundance contribution associated with 
the portion of $g_M(M')$ with $M' > M$.  This total abundance 
contribution is simply $\Omega_{\rm DM}F(M)$, where $F(M)$ is the hot-fraction
function in Eq.~(\ref{eq:hotfrac}).  Indeed, an equivalent statement of this 
conjecture, which follows from the invertible map in Eq.~(\ref{eq:vtoMmap}), is that
the value of $d\log S^2(M)/d\log M$ is correlated with the total 
abundance contributed by dark-matter particles within $g_v(v')$ with velocities
$v' > v$, where $v = M^{-1}_{\rm hal}(M)$ is the threshold velocity associated with
$M$ by the inverse of this map.

\begin{figure*}[t!]
  \centering
  \includegraphics[clip, width=0.98\textwidth]{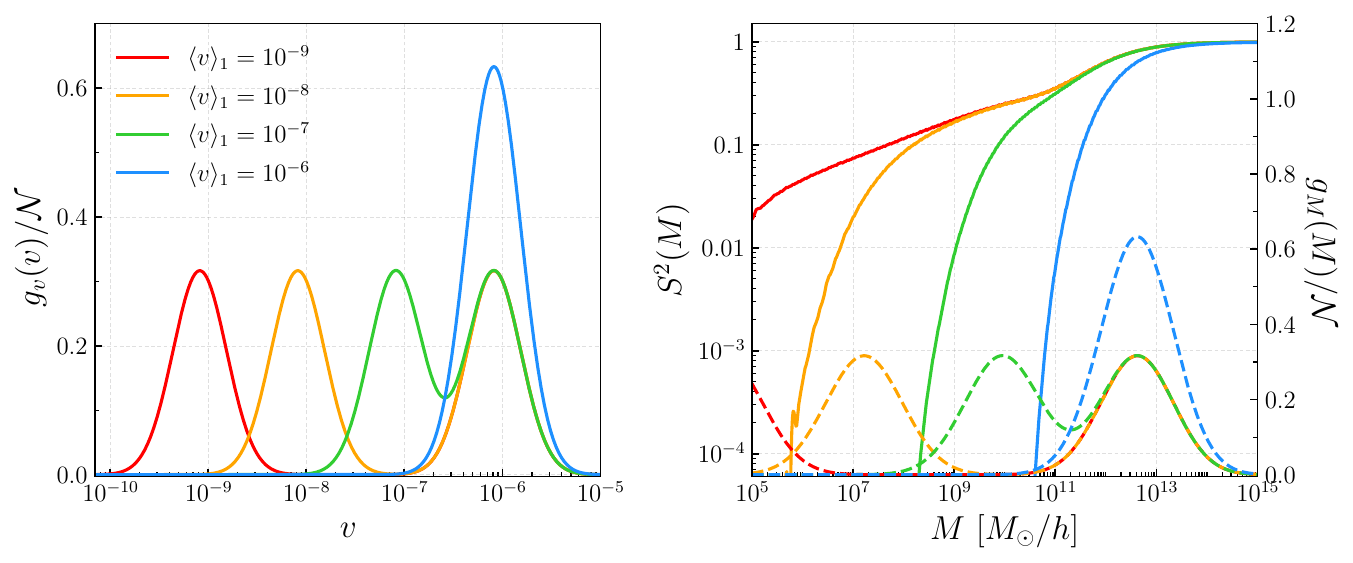}
  \caption{Similar to Fig.~\protect\ref{fig:HMFVaryOmegaSP}, but
    for a set of double-peak dark-matter velocity distributions 
    in which the average velocity $\langle v\rangle_1$ of the 
    lower-velocity peak is varied, while $\langle v\rangle_0 = 10^{-6}$, 
    the widths $\sigma_0 = \sigma_1 = 0.63$, and the abundances 
    $\Omega_1 = \Omega_2 = \Omega_{\rm DM}/2$ are all held fixed.
    We observe that as we scan from right to left across the right panel
    of the figure, any two $d\log S^2(M)/d\log M$ curves 
    coincide up until the point at which the corresponding $g_M(M)$ 
    distributions begin to differ.
  \label{fig:HMFVaryPeakLocs}}
\bigskip
\bigskip
  \centering
  \includegraphics[clip, width=0.98\textwidth]{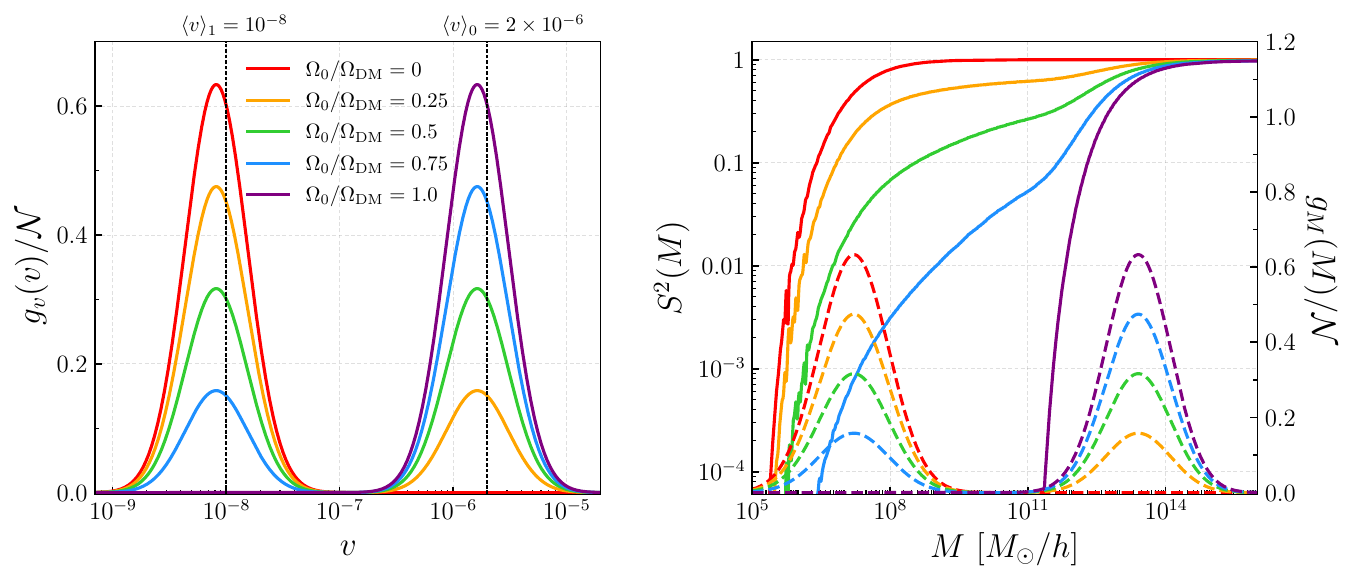}
  \caption{Similar to Fig.~\protect\ref{fig:HMFVaryOmegaSP}, but
    for a set of double-peak dark-matter velocity distributions of 
    in which the abundances $\Omega_0$ and $\Omega_1$ associated
    with the two peaks are varied, while 
    $\langle v\rangle_0 = 2\times 10^{-6}$,
    $\langle v\rangle_1 = 10^{-8}$, the 
    widths $\sigma_0 = \sigma_1 = 0.63$, and the sum 
    $\Omega_0 + \Omega_1 = \Omega_{\rm DM}$ of these abundances 
    are all held fixed.  We observe that $d\log S^2(M)/d\log M$ at a 
    location $M$ immediately to the left of any individual peak in 
    $g_M(M)$ is correlated with the value of $F(M)$ at that same location.
  \label{fig:HMFVaryPeakHeights}}
\end{figure*}

In order to test this conjecture further, we now consider more complicated
$g_v(v)$ distributions in which both Gaussian peaks in Eq.~(\ref{eq:gsum})
occur at velocities large enough to have a significant impact on structure.
In Fig.~\ref{fig:HMFVaryPeakLocs}, we illustrate the effect of varying 
$\langle v\rangle_1$ of the lower-velocity peak while holding 
$\langle v\rangle_0 = 10^{-6}$, the widths $\sigma_0 = \sigma_1 = 0.63$, 
and the abundances $\Omega_0 = \Omega_1$ fixed.  The blue curve in the left
panel represents the case in which $\langle v\rangle_1 = \langle v\rangle_0 $ 
and the two peaks in $g_v(v)$ coincide, effectively yielding a single Gaussian.
By contrast, the green, yellow, and red curves represent $g_v(v)$ distributions
with successively greater distances between $\langle v\rangle_1$ and 
$\langle v\rangle_0$.  The corresponding $S^2(M)$ curves and $g_M(M)$ 
distributions are shown in the right panel of the figure. 

We observe that
the $S^2(M)$ curves obtained for the the $g_v(v)$ distributions in which there 
is a significant separation between the peaks qualitatively differ from the curves 
obtained for narrow, unimodal $g_v(v)$ distributions.  Moreover, these curves
provide additional insight into the relationship between $S^2(M)$ and $g_M(M)$.
For example, we observe that $d\log S^2(M)/d\log M$ at a location $M$ immediately 
to the left of each peak in $g_M(M)$ is correlated with 
the value of $F(M)$ at such locations.  
Indeed, as we scan from larger to smaller values of $M$ across either of the 
peaks in a given $g_M(M)$ distribution, we see that $d\log S^2(M)/d\log M$ 
becomes increasingly steep as the cumulative abundance associated with 
dark-matter particles whose speeds $v$ map to halo masses above $M$ 
increases.  By contrast, 
as we scan from larger to smaller values of $M$ across the region
between each pair of peaks, where $g_M(M)$ is negligible, we see that 
$d\log S^2(M)/d\log M$ either remains constant or increases. Indeed, this is 
the case regardless of how much separation there is between the peaks.

The behavior of the curves in the right panel of 
Fig.~\ref{fig:HMFVaryPeakLocs} also illustrates the 
``locality'' inherent in the relationship between $g_M(M)$ and $S^2(M)$.
If two $g_v(v')$ distributions are identical above some velocity $v$, 
but differ for $v' < v$, the corresponding $S^2(M')$ distributions differ
only for masses $M' < M(v)$.  As a result, any two structure-suppression functions
shown in Fig.~\ref{fig:HMFVaryPeakLocs} coincide almost perfectly at
large $M$, and continue to track each other as $M$ decreases, all the way down 
to the point which the corresponding $g_M(M)$ distributions begin to diverge.  
We emphasize that this is true regardless of whether $d^2\log S^2(M)/(d\log M)^2$
is negative throughout the range of $M$ above which the two $g_M(M)$
distributions coincide.  Indeed, it still holds even if 
$d^2\log S^2(M)/(d\log M)^2 \geq 0$ across some or all of this range.

Finally, in Fig.~\ref{fig:HMFVaryPeakHeights}, we illustrate the effect on $S^2(M)$ 
of varying the relative abundances associated with the two peaks while holding 
$\langle v\rangle_0 = 2\times 10^6$, $\langle v\rangle_1 = 10^8$, 
and the widths $\sigma_0 = \sigma_1 = 0.63$ fixed.  As in 
Fig.~\ref{fig:HMFVaryPeakLocs}, we see that $d\log S^2(M)/d\log M$
decreases as we scan from larger to smaller values of $M$ across any particular 
peak in $g_M(M)$, but that it remains constant or increases as we scan across 
regions in which $g_M(M)$ is negligible.  However, we also see that the 
corresponding change in $d\log S^2(M)/d\log M$ is indeed correlated with $F(M)$.

Indeed, taken together, the results shown in Figs.~\ref{fig:HMFVaryOmegaSP}---\ref{fig:HMFVaryPeakHeights} 
suggest that there is a direct relationship 
between $d\log S^2(M)/d\log M$ and $F(M)$.  In particular, as we shall
demonstrate explicitly in Sect.~\ref{sec:ConjAll}, we find that within any 
interval in $M$-space at which $g_M(M)$ is non-negligible, this relationship 
is well described by a simple empirical relation of the form 
\begin{equation} 
  \frac{d \log S^2(M)}{d\log M} ~\approx~  \frac{7}{10} \, F^2(M)~.
  \label{eq:empir}
\end{equation}
Since $F(M)$ is by definition a monotonically decreasing function of $M$,
this relation implies that $d^2 \log\, S^2(M)/(d\log M)^2 \leq 0$ within any 
such interval.  We stress that this empirical relation is quite robust and holds 
regardless of how complicated the dark-matter velocity distribution might be. 

We also note that in cases in which the features in the dark-matter 
velocity distribution are well clustered in $M$-space --- \ie, in which there
are no extended intervals of $M$-space within which $g_M(M) \approx 0$ between 
these features --- we may obtain an approximate 
expression for $S^2(M)$ itself by integrating Eq.~(\ref{eq:empir})  directly:
\begin{eqnarray}
     && \log S^2(M) ~\approx~ \frac{7}{10}\int_{\log M}^{\infty} 
               d\log M' \, F^2(M')  \nonumber\\
     &&~~~=~ \frac{7}{10}\int_{\log M}^{\infty} 
           d\log M' \left[\frac{1}{\mathcal{N}}\int_{\log M'}^{\infty} 
            d\log M'' g_M(M'')\right]^2~.\nonumber\\ 
  \label{eq:intempir}
\end{eqnarray}
Moreover, as a consequence to the locality inherent in the relationship 
between $g_M(M)$ and $S^2(M)$, this procedure may also be applied to more     
general dark-matter velocity distributions in order to derive an 
an approximation for $S^2(M)$ at values of $M$ above all 
such extended intervals.  Such an approximation is potentially useful, as is  
would enable one to circumvent the rather complex procedure outlined in 
Sect.~\ref{sec:HaloMassFunction} through which the calculation of $S^2(M)$ is 
normally performed.  Of course, the usefulness of this approach will ultimately 
depend on the precision with which one might seek to evaluate $S^2(M)$.

The precise mathematical form of Eq.~(\ref{eq:empir}) and the
value of the numerical constant of course depend on the particular choice 
for the function $\eta(M)$ specified in Eq.~(\ref{eq:ShethTormanf}).  
Nevertheless, we emphasize that a similar mathematical relationship 
between $S^2(M)$ and 
$g_M(M)$ will exist for any alternative functional form of 
$\eta(M)$ which likewise respects the physical thresholds that served as 
the basis for our functional map between $v$ and $M$ in 
Eq.~(\ref{eq:vtoMmap}).  Indeed, this relationship is a consequence of 
{\it locality}\/ --- \ie, the fact that $d\log S^2(M)/d\log M$ is sensitive 
only to the portion of $g_M(M')$ for which $M' > M$.  As we have seen, this 
locality ultimately stems from the assumption inherent in
Eq.~(\ref{eq:PressSchechter}) that $\eta(M)$ depends on $M$ only 
through quantities which respect these physical thresholds. 

That said, given that a full $N$-body analysis of the class of
highly non-trivial
$g_v(v)$ distributions that we are considering in this paper has never 
been performed, one might wonder whether this property of $\eta(M)$ still 
holds for these $g_v(v)$ distributions.  While a conclusive answer 
to this question cannot be obtained without extensive numerical 
simulation, there are nevertheless indications that $\eta(M)$ indeed 
respects the same thresholds, even for more complicated forms of $g_v(v)$.
For example, an $N$-body analysis of the kinds of halo-mass functions which 
arise in so-called mixed-dark-matter scenarios --- \ie, scenarios in which 
the dark-matter velocity distribution includes both a WDM and a CDM 
contribution --- was performed in Ref.~\cite{Parimbelli:2021mtp}.
Despite the more complicated form of $g_v(v)$ which characterizes these
scenarios, the resulting $dn/d\log M$ were nevertheless 
found to be well modeled by an analytic function of the general form 
Eq.~(\ref{eq:PressSchechter}) with a universal $\eta(M)$ function which
respects the relevant physical thresholds.


\section{Connecting to Observables\label{sec:Observables}}


While the effect that the detailed shape of $g_v(v)$ on $S^2(M)$ has on the 
$S^2(M)$ is interesting in its own right, it is also interesting to extend 
this analysis a step further and consider how this detailed shape 
affects astrophysical observables which serve as probes of structure in the 
non-linear regime.  We focus here on two such observables: satellite 
counts within the halos of large galaxies and cluster-number counts.  
As we shall see, the detailed shape of the dark-matter velocity distribution 
can have a non-trivial impact on both of these observables.

\subsection{Satellite Counts\label{sec:SatelliteCounts}}

One observable which which provides information about the spatial 
distribution of matter within the universe is the number of satellite 
galaxies $N_{\rm SH}$ with masses $M$ above some observability threshold 
$M_{\rm min}$ which reside within a typical host halo of mass $M_0$.  Of
particular interest is the number of satellite galaxies within the 
halo of the Milky Way.

A theoretical prediction for $N_{\rm SH}$ can be derived analytically 
for a given $g_v(v)$ distribution and a given value of $M_0$.  
This prediction is derived
not from the halo-mass function, but rather from the conditional mass function 
$dN(M,z|M_0,z_0)/dM$.  This latter quantity represents the differential 
number of halos per unit mass $M$ present in the universe at redshift $z$ which, 
on average, will have been incorporated into a single host halo of mass $M_0$ 
by the time the universe reaches the redshift $z_0 < z$.  An 
approximate analytic expression for $dN(M,z|M_0,z_0)/dM$ can be derived from 
the same excursion-set formalism~\cite{Bond:1990iw} from which 
Eq.~(\ref{eq:PressSchechter}) can be obtained.  In particular, it can be shown
that~\cite{Lacey:1993iv}
\begin{equation}
  \frac{dN(M,z|M_0,z_0)}{dM} = - \frac{M_0}{M} \sigma^2(M) 
    \zeta(M,z|M_0,z_0) \frac{d \sigma^2(M)}{dM}\, ,~~
  \label{eq:ConditionalMF}  
\end{equation}
where $\sigma^2(M) \equiv \sigma^2(\tnow,M)$ and where 
$\zeta(M,z|M_0,z_0)$ represents the probability that a particle present in
a halo of mass $M$ at redshift $z$ would have been incorporated into a halo of 
mass $M_0$ by the time the universe reaches redshift $z_0$.  Under the 
simplifying assumption of spherical collapse, this conditional probability
takes the analytic form 
\begin{eqnarray}
  \zeta(M,t|M_0,t_0) &=& \frac{\delta(z)-\delta(z_0)}
    {(2\pi)^{1/2}[\sigma^2(M) - \sigma^2(M_0)]^{3/2}} \nonumber \\
    & & \times~
    \exp\left(-\frac{[\delta(z)-\delta(z_0)]^2}
    {2[\sigma^2(M) - \sigma^2(M_0)]}\right)\, ,~~~~~~
\end{eqnarray}
where $\delta(z) \equiv \delta_c/D(z)$ is defined in terms of the 
universal growth factor $D(z)$ for perturbations at redshift $z$.

In order to estimate the number of subhalos with masses above a
given threshold, we follow the procedure outlined in 
Refs.~\cite{Giocoli:2007gf,Schneider:2014rda}.  
We integrate the conditional mass function over $z$ in order to obtain the 
differential number $dN_{\rm SH}/dM$ of subhalos per unit $M$:
\begin{equation}
  \frac{dN_{\rm SH}}{dM} ~=~ \frac{1}{\mathcal{N}_{\rm SH}}
    \int_{0}^\infty \frac{dN(M,t|M_0,t_0)}{dM} \frac{d\delta(z)}{dz} dz ~,~
  \label{eq:SubhaloNumber}
\end{equation}
where $\mathcal{N}_{\rm SH}$ is a normalization factor which accounts for the 
fact that this integration overcounts the number of halos by including the
same progenitor at multiple redshifts.  Thus,  
the total expected number of subhalos with masses above a given mass 
threshold $M_{\rm min}$ within a host halo of mass $M_0$ is 
\begin{equation}
  N_{\rm SH} ~=~ \int_{M_{\rm min}}^{M_0} \frac{d N_{\rm SH}}{d M} dM~. 
  \label{eq:NSHAbove}
\end{equation}

In order to assess the degree to which $N_{\rm SH}$ depends on the detailed
shape of the dark-matter velocity distribution, we evaluate the predicted
number of Milky-Way satellites for a variety of $g_v(v)$ distributions of 
the form given in Eq.~(\ref{eq:gsum}) using Eq.~(\ref{eq:NSHAbove}).
For the window function in Eq.~(\ref{eq:Windowfn}), the expression
in Eq.~(\ref{eq:SubhaloNumber}) reduces to
\begin{eqnarray}
  \frac{d N_{\rm SH}}{d M} &~=~& 
    \frac{1}{6\pi^2\mathcal{N}_{\rm SH}}\left(\frac{M_0}{M^2}\right)
    \nonumber \\
    & & \times ~
    \frac{P(1/R(M))}{R^3(M)\sqrt{2\pi[\sigma^2(M) - \sigma^2(M_0)]}}~.~~~
  \label{eq:dNSHdM}
\end{eqnarray}
We choose this normalization factor $\mathcal{N}_{\rm SH}$ such 
that the value of $N_{\rm SH}$ obtained from Eq.~(\ref{eq:NSHAbove}) for a 
Milky-Way-sized galaxy in a purely CDM scenario accords with the value obtained 
from $N$-body simulations.  In particular, we base our value of 
$\mathcal{N}_{\rm SH}$ on the results obtained by 
Aquarius project~\cite{Lovell:2013ola}, which obtained 
a value $N_{\rm SH} = 157$ for the number of subhhalos of mass 
$M > 10^8\, h^{-1} M_\odot$ in a Milky-Way-like galaxy of mass 
$M_0 = 1.77 \times 10^{12}\, h^{-1} M_\odot$.  Accordingly, we adopt
$M_{\rm min} = 10^8\, h^{-1} M_\odot$ as our mass threshold.  For this
value of $M_{\rm min}$, we find numerically that $\mathcal{N} \approx 46.9$, 
which is roughly similar to the result $\mathcal{N}_{\rm SH} = 44.5$ 
obtained in Ref.~\cite{Schneider:2014rda}.  

In order to present the results of this analysis in an illustrative way,
we shall also define the quantity $N_{\rm SH}^{\rm WDM}(k_{\rm FSH})$ for a 
given $g_v(v)$ distribution to represent the value of $N_{\rm SH}$ that 
one would obtain for a WDM velocity distribution with the same average velocity 
$\langle v\rangle$ as $g(v)$ --- and hence also the same nominal free-streaming scale 
$k_{\rm FSH}$.  Since the dark-matter velocity distribution for a WDM 
model is completely specified by the single parameter $m_{\rm WDM}$, 
$N_{\rm SH}^{\rm WDM}(k_{\rm FSH})$ is uniquely defined.  The 
ratio\footnote{In keeping with our notation wherein subhalos are indicated 
through the subscript `SH', we have chosen to denote this ratio with the 
Hebrew letter \smallshin, pronounced ``shin'' and signifying the 
sound ``{\tt sh}''.}
\begin{equation}
    \shin ~\equiv~ \frac{N_{\rm SH}}{N_{\rm SH}^{\rm WDM}(k_{\rm FSH})}
    \label{eq:AlphaDef}
\end{equation}
of the actual value of $N_{\rm SH}$ obtained for this $g_v(v)$ to the value of
$N_{\rm SH}^{\rm WDM}(k_{\rm FSH})$ quantifies the degree to which $N_{\rm SH}$ 
departs from the WDM result due to the detailed shape of $g_v(v)$.

\begin{figure*}[t!]
  \centering
  \includegraphics[clip, width=0.48\textwidth]{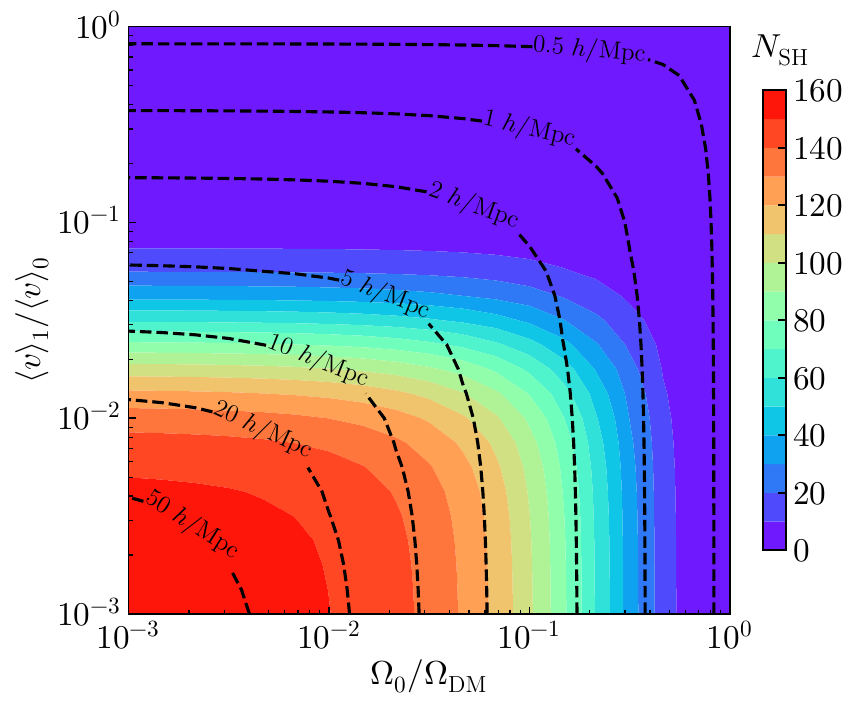}~~
  \includegraphics[clip, width=0.48\textwidth]{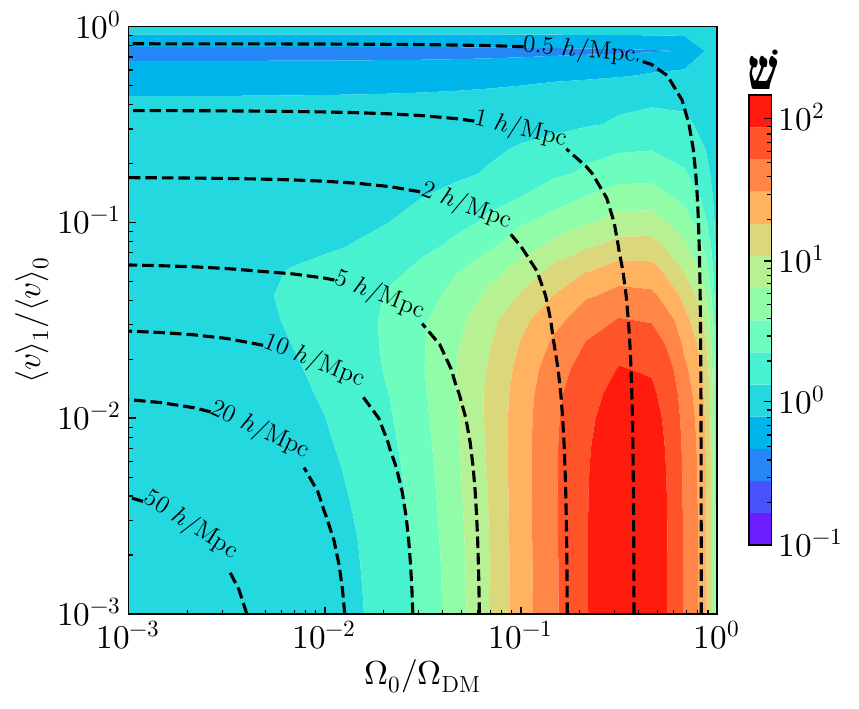}
  \caption{{\it Left panel}\/: The expected number $N_{\rm SH}$ 
    of satellites with masses $M > 10^8\, h^{-1} M_\odot$ 
    within the halo of a Milky-Way-sized galaxy for a dark-matter 
    scenario with a velocity distribution given by Eq.~(\ref{eq:gsum}), 
    displayed within the 
    $(\Omega_0/\Omega_{\rm DM}, \langle v\rangle_0/\langle v\rangle_1)$-plane.
    The results shown in the figure correspond to the parameter choices
    $\langle v\rangle_0 = 10^{-6}$ and $\sigma_0 = \sigma_1 = 0.63$.
    {\it Right panel}\/: The ratio \smallshin\ of $N_{\rm SH}$ to 
    the number of satellites obtained for the WDM model with the same 
    nominal free-streaming scale $k_{\rm FSH}$, likewise 
    displayed within the 
    $(\Omega_0/\Omega_{\rm DM}, \langle v\rangle_0/\langle v\rangle_1)$-plane.
    Contours of $k_{\rm FSH}$ (dashed black lines) are also provided in each 
    panel for reference.  
  \label{fig:MWSubs}}
\end{figure*}

In the left panel of Fig.~\ref{fig:MWSubs}, we show contours within the 
$(\Omega_0/\Omega_{\rm DM}, \langle v\rangle_0/\langle v\rangle_1)$-plane
of the expected number of satellite galaxies with masses 
$M > 10^8\, h^{-1} M_\odot$ contained within a Milky-Way-sized galaxy of mass 
$M_0 = 1.77 \times 10^{12}\, h^{-1} M_\odot$.  The results shown here 
correspond to the parameter choices $\langle v\rangle_0 = 10^{-6}$ and 
$\sigma_0 = \sigma_1 = 0.63$.  In the right panel, we show contours of 
the ratio \shin\  within the 
$(\Omega_0/\Omega_{\rm DM} \langle v\rangle_0/\langle v\rangle_1)$-plane 
for the same choice of parameters.  In each panel, we have also included 
contours (dashed black lines) of the nominal free-streaming scale 
$k_{\rm FSH}$. 

In interpreting these results displayed in Fig.~\ref{fig:MWSubs}, we begin
be noting that we may establish a rough bound on $g_v(v)$ by requiring that 
$N_{\rm SH}$ accord with the number $N_{\rm SH}^{\rm (obs)}$ actually observed 
in the halo of the Milky Way.  The known Milky-Way satellites with masses above 
our $M_{\rm min}$ threshold include 11 classical satellites and 15 
ultra-faint satellites discovered by the Sloan Digital Sky Survey (SDSS),
as well as a number of additional satellites identified by the Dark Energy 
Survey (DES)~\cite{Koposov:2015cua,DES:2015txk,DES:2015zwj,Fermi-LAT:2016uux}.
However, the current catalogue of $\sim 50$ such satellites is presumably
highly incomplete, given that SDSS and DES do not together cover the entire
sky and are subject to flux thresholds.  We therefore
estimate $N_{\rm SH}^{\rm (obs)}$ by multiplying the number of SDSS 
satellites by 3.5 in order to account for the limited sky coverage of 
the SDSS~\cite{Polisensky:2010rw,Schneider:2014rda,Schneider:2016uqi} and
adding to this the number of classical satellites.  This accounting yields
$N_{\rm SH}^{\rm (obs)} = 62$ for the total number of Milky-Way 
subhalos with masses $M > M_{\rm min}$.

The results shown in the right panel of Fig.~\ref{fig:MWSubs} reflect
the extent to which the detailed shape of $g_v(v)$ affects the value of 
$N_{\rm SH}$.  The ratio \shin\  differs significantly from unity within
the region wherein $\Omega_0/\Omega_{\rm DM}$ is large and 
$\langle v\rangle_1/\langle v\rangle_0 \leq 0.1$ --- in some places by
as much as two orders of magnitude.  These results make it 
clear that the detailed shape of $g_v(v)$ can have a significant impact 
on the substructure of galactic halos.  

In addition to the rough lower bound on $N_{\rm SH}$, other considerations 
related to structure formation likewise constrain on the form of $g_v(v)$.
For example, Lyman-$\alpha$-forest data impose stringent constraints on
the shape of the matter power spectrum at wavenumbers 
$1\, h/{\rm Mpc} \lesssim k \lesssim 50\, h/{\rm Mpc}$.  These constraints
likewise depend of the detailed shape of $g_v(v)$.  Constraints on 
$g_v(v)$ distributions of the form in Eq.~(\ref{eq:gsum}) were 
derived in Ref.~\cite{Dienes:2021cxp}.  For the values of 
$\langle v\rangle_0$, $\sigma_0$, and $\sigma_1$ we have adopted in 
Fig.~\ref{fig:MWSubs}, an analysis employing the area criterion
$\delta A$~\cite{Murgia:2017lwo} excludes the region of the
$(\Omega_0/\Omega_{\rm DM}, \langle v\rangle_0/\langle v\rangle_1)$-plane
wherein $\Omega_0/\Omega_{\rm DM} \gtrsim 0.1$ and 
$\langle v\rangle_1/\langle v\rangle_0 \gtrsim 0.25$.  While this 
excludes the region wherein \shin\ is significantly below 
unity, the allowed region of parameter space includes sizable 
regions wherein $\shin \sim \mathcal{O}(10-100)$.

\subsection{Cluster-Number Counts\label{sec:ClusterNumber}}

Another observable which provides information about the spatial distribution of 
matter within the universe is the number of galaxy clusters $N_{\rm C}$ observed 
within a given region of the sky.  However, unlike satellite counts, 
cluster-number counts of this sort are directly related to the halo-mass 
function and therefore serve as a probe of $S^2(M)$ --- and thus to the
results derived in Sect.~\ref{sec:ResultsHMF}.

Observationally speaking, the cluster-number count obtained from a given survey  
is simply the total number of galaxy clusters observed within a particular region
of the sky out to some maximum redshift $z_{\rm max}$ with masses which lie above
some threshold $M_{\rm min}(z)$ which may be redshift-dependent.  Thus, a 
theoretical prediction for $N_{\rm C}$ within a given dark-matter scenario 
may be obtained by evaluating~\cite{Lima:2005tt} 
\begin{equation}
  N_{\rm C} ~=~ \int_0^{z_{\rm max}} z \frac{dV}{dz} 
    \int_{\log M_{\rm min}(z)}^\infty d\log M\frac{dn}{d\log M}~,
\end{equation}
where $dV/dz$ is the comoving volume element per unit $z$.  This
comoving volume element may be written in the form
\begin{equation}
  \frac{dV}{dz} ~=~ 4\pi \Delta \Omega \frac{c \chi^2(z)}{H(z)}~,~ 
\end{equation}
where $\Delta \Omega$ is solid angle on the sky under observation,  
where $c$ is the speed of light, where $\chi(z)$ is the comoving
distance, and where $H(z)$ is the Hubble parameter.

In assessing the cluster-number count which follows from any particular 
$g_v(v)$ distribution, we adopt a set of parameters which allow us to 
compare our predictions with results predicted for
the Euclid survey~\cite{Sartoris:2015aga}.  In particular, we
take $\Delta \Omega \approx 4.57$~sr and $z_{\rm max} = 2$ and we
we adopt the redshift-dependent detection mass threshold presented in 
Ref.~\cite{Sartoris:2015aga}, in which $M_{\rm min}(z)$ varies between 
$\sim 10^{13.8}$ and $10^{14.1}~M_{\odot}$.

In order to examine the effect of the primordial
dark-matter velocity distribution on $N_{\rm C}$, we shall once again
consider $g_v(v)$ distributions of the form given in Eq.~(\ref{eq:gsum}).
Moreover, since our aim is to highlight the effect of varying the 
detailed shape of $g_v(v)$, rather than the effect of varying the nominal 
free-streaming scale $k_{\rm FSH}$, we shall proceed by examining 
how $N_{\rm C}$ varies along surfaces of fixed $k_{\rm FSH}$ within 
our parameter space.  In order that we may compare our results for $N_{\rm C}$ 
to those we have obtained for $N_{\rm SH}$ in a straightforward manner, we 
shall once again take $\langle v\rangle_0 = 10^{-6}$ and $\sigma_0 = \sigma_1 = 0.63$
and focus on the effect of varying $\Omega_0/\Omega_{\rm DM}$ and 
$\langle v_1\rangle/\langle v_0\rangle$ along contours of constant 
$k_{\rm FSH}$.

\begin{figure}[t!]
  \centering
  \includegraphics[clip, width=0.48\textwidth]{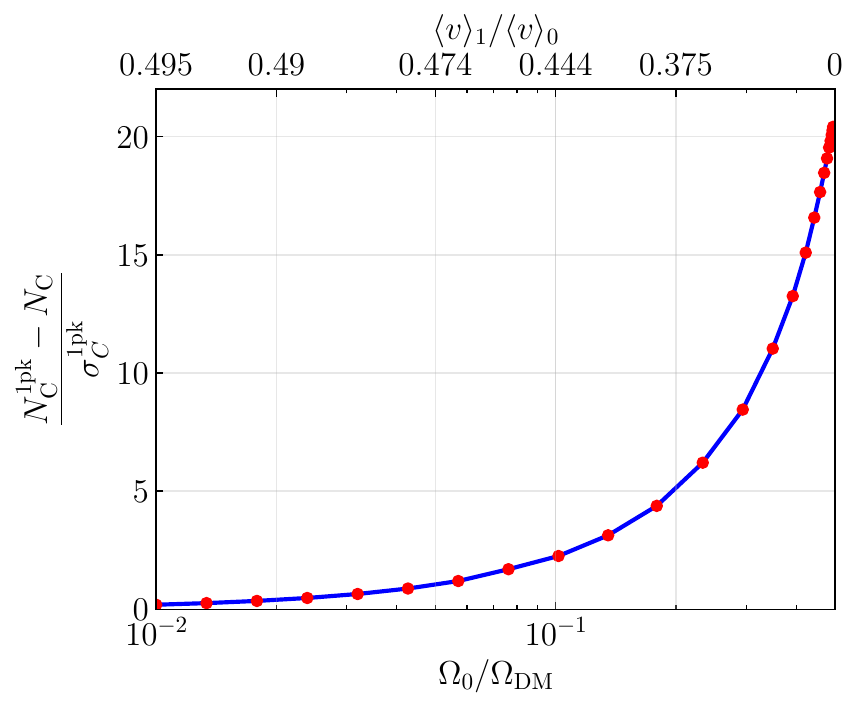}
  \caption{The difference between the cluster-number count 
    $N_{\rm C}^{1{\rm pk}}$ obtained for $g_v(v)$ distribution with a
    single Gaussian peak and the count $N_{\rm C}$ obtained for 
    a $g_v(v)$ distributions of the form given in given by Eq.~(\ref{eq:gsum}),
    normalized to the the Poisson uncertainty $\sigma_{\rm C}^{1{\rm pk}}$ 
    associated with the single-peak distribution, shown as a function of 
    $\Omega_0/\Omega_{\rm DM}$ for fixed $k_{\rm FSH} \approx 0.76~h/\mathrm{Mpc}$.  
    The value of $\langle v_1\rangle/\langle v_0\rangle$ which corresponds
    to a given value of $\Omega_0/\Omega_{\rm DM}$ along this $k_{\rm FSH}$ 
    contour is indicated along the top axis.  We note that 
    $N_{\rm C} < N_{\rm C}^{1{\rm pk}}$
  \label{fig:ClusterCount}}
\end{figure}

In Fig.~\ref{fig:ClusterCount}, we show the extent to which $N_{\rm C}$ 
differs from the cluster count $N_{\rm C}^{1{\rm pk}}$ obtained for 
the $g_v(v)$ distribution 
which has the same value of $k_{\rm FSH}$, but consists of a single Gaussian peak 
whose width is likewise set to the value $\sigma_0 = 0.63$ characteristic of 
a WDM distribution.  In particular, we fix $\langle v\rangle = 5\times 10^{-7}$, 
which fixes $k_{\rm FSH} \approx 0.76~h/\mathrm{Mpc}$, and show how the ratio
$(N_{\rm C}^{1{\rm pk}} - N_{\rm C})/\sigma_{\rm C}^{1{\rm pk}}$ varies as a 
function of $\Omega_0/\Omega_{\rm DM}$ for this fixed value of $k_{\rm FSH}$, 
where $\sigma_{\rm C}^{1{\rm pk}}$ is the Poisson uncertainty associated with 
the single-peak distribution.  
This ratio provides an estimate of the statistical 
significance of the impact of the detailed shape of $g_v(v)$ on $N_{\rm C}$.
The value of $\langle v\rangle_1/\langle v\rangle_0$ which which corresponds 
to a given value of $\Omega_0/\Omega_{\rm DM}$ along this contour is indicated along 
the top axis of the figure.  We emphasize that in moving from left to right across
Fig.~\ref{fig:ClusterCount} corresponds to following a contour in the 
$(\Omega_0/\Omega_{\rm DM}, \langle v\rangle_0/\langle v\rangle_1)$-plane 
which lies nearby the $k_{\rm FSH} = 1~h/\mathrm{Mpc}$ contour 
indicated Fig.~\ref{fig:MWSubs}.

The results displayed in Fig.~\ref{fig:ClusterCount} clearly indicate that 
the detailed shape of $g_v(v)$ can have a significant impact on cluster-number 
counts.  The most significant deviation occurs along the portion of the 
$k_{\rm FSH}$ contour where $\Omega_0/\Omega_{\rm DM}$ is large and 
$\langle v_1\rangle/\langle v_0\rangle$ is small.  Indeed, this portion of
the contour corresponds to the region of parameter space in 
Fig.~\ref{fig:MWSubs} within which the greatest deviations likewise arose 
between the value of $N_{\rm SH}$ for our double-peak $g_v(v)$ distributions and
the expected value for a WDM distribution with the same nominal free-streaming
scale. 

To summarize the results of this section, we find that the detailed shape of 
the primordial dark-matter velocity distribution not only affects the halo-mass
function, but also has an impact on observables such as $N_{\rm SH}$ and 
$N_{\rm C}$.  Thus, these observables can play a role in probing and 
constraining $g_v(v)$ and the halo- or subhalo-mass functions which follow
from it.


\section{A Reconstruction Conjecture\label{sec:ConjAll}}


In Sect.~\ref{sec:HaloMassFunction}, we examined the mathematical 
relationship between $g_M(M)$ and $S^2(M)$ in the extended Press-Schechter
formalism.  In doing so, we identified a pair of physical thresholds
which underpin this relationship.  The presence of these thresholds 
permits us to construct a map between $v$ and $M$ analogous to 
the expression in Eq.~(\ref{eq:vtoMmap}) for any well-behaved 
functional forms for $W(k,R)$ and $\eta(M)$.  Moreover, in 
Sect.~\ref{sec:Observables}, we also saw that this relationship between 
$g_M(M)$ and $S^2(M)$ is such that the detailed shape of the dark-matter 
velocity distribution has a potentially measurable impact on astrophysical
observables such as cluster-number counts.

Motivated by these findings, we propose a method for inverting the 
procedure outlined in Sect.~\ref{sec:HaloMassFunction} and reconstructing 
the detailed shape of $g_v(v)$ directly from information contained in 
$S^2(M)$.  This conjecture is clearly more speculative than the results 
we have presented thus far.  Indeed, it is more sensitive to the 
particular functional forms one chooses for $W(k,R)$ and $\eta(M)$.
Moreover, the fact that the halo-mass function is not directly measurable, 
but rather must itself be inferred indirectly from observational data, 
imposes some restrictions on its practical applicability.  
Nevertheless, as we shall argue below, this 
reconstruction conjecture can provide a potentially useful method of 
extracting information about the primordial velocity distribution of 
the dark matter, and by extension the underlying particle-physics processes
which produced it in the early universe.

\subsection{Formulating the Conjecture\label{sec:ConjFormal}}

Our stated goal, then, is to invert Eq.~(\ref{eq:empir}) and obtain 
information about $g_M(M)$ from $S^2(M)$.
We begin by considering the case in which the features in $g_M(M)$ are well 
clustered in $M$-space.  
In this case,
a statement about the functional form of $g_M(M)$ may then be obtained in a 
straightforward manner from Eq.~(\ref{eq:empir}).  Taking the logarithmic 
derivative of both sides of this relation and using the definition of the 
hot-fraction function in Eq.~(\ref{eq:hotfrac}) to relate $dF(M)/dM$ to $g_M(M)$, 
we find that
\begin{equation}
  \frac{ g_M(M)}{\calN} ~\approx~ 
     \sqrt{ \frac{5}{14}} \left( \frac{d \log S^2(M)}{d\log M}\right)^{-1/2}\,
     \left| \frac{d^2 \log S^2(M)}{(d\log M)^2}\right|~.
  \label{eq:recconjrough}
\end{equation}
This is the basic form of our conjecture which allows us to reconstruct 
the salient features of the dark-matter velocity distribution $g_M(M)$ 
directly from the first and second logarithmic derivatives of the 
structure-suppression function. 

As stated above, our conjecture in Eq.~(\ref{eq:recconjrough}) holds under the 
assumption that $d^2 \log\, S^2(M)/(d\log M)^2$ is always either zero or 
negative --- \ie, that $\log \,S^2(M)$ is always either a straight line or 
concave-down when plotted versus $\log\, M$.  However, this conjecture may 
also be extended to the more general case in which this condition is not always
satisfied.  Indeed, while we have seen in Sect.~\ref{sec:ResultsHMF} that 
$\log \,S^2(M)$ can in fact be concave-up, we have also seen that this behavior 
only arises within intervals of $M$-space within which $g_M(M)$ is negligible.  
Thus, in order to generalize our reconstruction conjecture to account for this 
possibility, we need only to posit that {\il{g_M(M)\approx 0}} whenever 
{\il{d^2 \log\, S^2(M)/(d\log\, M)^2 > 0}}.  In other words, we posit that
\begin{eqnarray}
  \frac{ g_M(M)}{\calN} &~\approx~&  
    \sqrt{ \frac{5}{14}} \left( \frac{d \log S^2(M)}{d\log M}\right)^{-1/2}
    \nonumber \\ && ~~~\times \left| 
    \min\left(0, \frac{d^2 \log S^2(M)}{(d\log M)^2}\right) \right|~.~~~
\label{eq:recconj}
\end{eqnarray}
This, then, is the complete statement of our reconstruction conjecture. 
 
Several important caveats must be borne in mind regarding this conjecture.
First, we emphasize that it is not meant to be a precise mathematical 
statement.  Indeed, given the rather complicated nature of the 
Einstein-Boltzmann evolution equations which connect $g_M(M)$ to $S(M)$, 
we do not expect a relation of the simple form in Eq.~(\ref{eq:recconj}) 
to provide a precise inverse (except perhaps under some limiting 
approximations and simplifications). Rather, this conjecture is intended merely 
as an approximate practical guide --- a back-of-the-envelope method
for reproducing the rough characteristics of $g_M(M)$ given a particular 
structure-suppression function $S(M)$.

Second, as discussed in more detail in Sect.~\ref{sec:VelsMPS}, our map
between $v$ and $M$ in Eq.~(\ref{eq:vtoMmap}) has been formulated under the 
assumption that the dark matter is relativistic at the time at which it is 
produced.  When this is not the case, we expect that a more appropriate map 
between these two variables will depend on further details such as the time 
at which the dark matter is produced, and hence will carry a sensitivity to 
the particular dark-matter production scenario envisaged.  However, in the 
vast majority of situations in which this assumption is violated and a 
significant population of dark-matter particles is non-relativistic at the 
time of production, this population of non-relativistic dark-matter 
particles is typically sufficiently cold that it has no effect on $S^2(M)$
for $M$ within our regime of interest.  While it is possible to engineer 
situations in which the map in Eq.~(\ref{eq:vtoMmap}) might require 
modification while free-streaming effects on $S^2(M)$ are non-negligible, 
such situations require a somewhat unusual dark-matter cosmology --- a 
cosmology in which a significant non-relativistic yet ``lukewarm'' population 
of dark-matter particles is generated at exceedingly late times 
by some dynamics that contributes significantly to $f(v)$ within a 
particular range of velocities.  

Third, our procedure for calculating $P(k)$ from a given $g_v(v)$ implicitly 
incorporates certain assumptions.  One of these assumptions is that the background 
cosmology does not deviate significantly from that of the standard cosmology. 
For example, it is assumed that the time $\tMRE$ of matter-radiation equality is 
the same as in the standard cosmology and that the universe is effectively 
radiation-dominated at all times from the end of the reheating epoch 
until $\tMRE$.  It is also assumed that the primordial spectrum of density 
perturbations produced after inflation is Gaussian-random. Another of these 
assumptions is that the velocity distribution of dark-matter particles has 
ceased evolving, except as a consequence of redshifting effects, by some very 
early time deep within the radiation-dominated epoch. This implies not only that 
the production of the dark matter is effectively complete by that time, but also 
that the effect of scattering and decay processes involving dark-matter particles 
is negligible thereafter.  Of course, the above assumptions do not necessarily 
imply limitations on our conjecture {\it per se}\/ in these regimes. While it is 
possible that our conjecture ceases to provide accurate results for cosmological 
histories wherein the above assumptions are relaxed, it is also possible that 
our conjecture remains robust even in the presence of these deviations.

The restrictions implied by these caveats are not severe.  Indeed, as we 
shall demonstrate in Sect.~\ref{sec:Results}, our conjecture as stated here 
will still allow us to resurrect the salient features of $g_M(M)$ --- and 
hence also of $f(v)$ --- for a wide variety of different dark-sector scenarios.
Clearly, details such as the proportionality constants in 
Eq.~(\ref{eq:vtoMmap}) and the precise functional relationship between 
$d\log S^2(M)/d\ln M$ and $F(M)$ depend on the particular functional 
forms for $W(k,R)$ and $\eta(M)$.  However, the qualitative picture that 
we have developed for reconstructing $g_M(M)$ from $S^2(M)$ is predicated only 
on one crucial assumption --- the assumption of locality which follows from
the physical thresholds which have allowed us to formulate the map between 
$v$ to $M$ in Eq.~(\ref{eq:vtoMmap}).  Thus, a reconstruction conjecture
qualitatively similar to Eq.~(\ref{eq:recconj}) can likewise be formulated 
for any alternative functional forms that one might adopt for $W(k,R)$ and 
$\eta(M)$, provided that these functions respect the same thresholds.
It is also worth emphasizing that, as a consequence of this locality, 
our reconstruction procedure permits us to reconstruct the value of $g_M(M)$ 
at any particular $M$ solely based on information about $S^2(M)$ and its 
derivatives at that same value of $M$ without any additional information 
about the global properties of this structure-suppression function.
Thus, even if the functional form of $S^2(M)$ is known only across a 
limited range of halo masses, our conjecture can still be applied in order to 
reconstruct $g_M(M)$ across that same range of $M$.
 
It is also important to note the similarities and differences between our 
conjecture for reconstructing $f(v)$ from the shape of $S^2(M)$ and the similar 
proposal that we advanced in Ref.~\cite{Dienes:2020bmn} for extracting 
information about the dark-matter phase-space distribution from the linear 
matter power spectrum~\cite{Dienes:2020bmn}.  First, as emphasized in the
Introduction, the conjecture in Eq.~(\ref{eq:recconj}) does not rely on this 
previous proposal in any way.  Moreover, in principle, the conjecture in 
Eq.~(\ref{eq:recconj}) permits one to extract information about $f(v)$ at 
much lower velocities.  Measurements of $P(k)$ based on data obtained at 
low redshifts are currently reliable up to around 
{\il{k\lesssim 0.05}} -- $0.1\mbox{~Mpc}^{-1}$.  Information from 
Lyman-$\alpha$-forest measurements can provide additional information about 
$P(k)$ at wavenumbers up to around {\il{k \sim 1\mbox{~Mpc}^{-1}}}.  While 
future measurements of the 21-cm line of neutral hydrogen could in principle 
yield information about $P(k)$ at significantly higher redshifts, the present 
state of our knowledge of $P(k)$ permits us to reconstruct $f(v)$ only 
down to {\il{v \sim 5\times 10^{-7}}} using the methods of
Ref.~\cite{Dienes:2020bmn}.  

By contrast, our reconstruction conjecture in Eq.~(\ref{eq:recconj}) relies 
solely on information contained within the halo-mass function in order to 
reconstruct $f(v)$.  Thus, one could in principle use this conjecture to 
probe $f(v)$ down to {\il{v \sim 10^{-9}}} or lower.  In practice, 
exploiting this property of the conjecture is somewhat challenging, given
that astrophysical observables from which meaningful information about 
the structure-suppression function can currently be extracted, such as 
differential cluster-number counts, typically provide information about
$S^2(M)$ at halo-mass scales $M \gtrsim 10^{13}~M_{\odot}$ --- and thus to
regions of $g_v(v)$ wherein $v \gtrsim 10^{-6}$.  Nevertheless,
if observational techniques can be developed which permit one to probe 
$S^2(M)$ in a meaningful way at lower scales, our reconstruction 
conjecture can provide a method of extracting meaningful information 
about the dark-matter velocity distribution from those measurements.

\subsection{Testing the Conjecture\label{sec:Results}}

Having stated our conjecture, we now assess the extent to which it enables 
us to reconstruct the underlying dark-matter velocity distribution from the 
halo-mass function.  In 
particular, we shall test this conjecture within the context of an 
illustrative dark-matter model in which $g_M(M)$ deviates significantly 
from that of purely cold dark matter in a variety of ways within different 
regions of model-parameter space.  For a set of illustrative points in 
that parameter space, we will then reconstruct $g_M(M)$ using our conjecture 
and compare it with the corresponding ``true'' $g_M(M)$ distribution.
  
The model which we shall adopt for purposes of illustration --- a model which was 
introduced in Ref.~\cite{Dienes:2020bmn} --- is one in which the cosmological 
abundance of dark matter is generated non-thermally, via the decays of unstable 
dark-sector particles.  The dark sector in this scenario consists of an ensemble 
of $N$ real scalar fields $\phi_\ell$ with $\ell = 0, 1, 2, \ldots, N-1$ whose decay 
widths are dominated by two-body processes associated with trilinear terms
in the interaction Lagrangian of the form
\begin{equation}
  \mathcal{L}_{\rm int} ~\ni~ \sum_{\ell = 0}^N\sum_{i=0}^{\ell}\sum_{j=0}^{i} 
    c_{\ell i j} \phi_\ell \phi_i\phi_j~,~
  \label{eq:Lint}
\end{equation}
where the $c_{\ell i j}$ are coupling constants with dimensions of mass.  
The masses of the $\phi_\ell$ are given by $m_\ell = (2\ell + 1)m_0$, where 
$m_0$ is the mass of $\phi_0$, while these coupling constants are given by 
\begin{eqnarray}
  c_{\ell i j}~ &=&~ \mu R_{\ell i j} 
    \left(\frac{m_\ell - m_i - m_j}{2m_0}\right)^r
    \left(1+\frac{|m_i - m_j|}{2m_0}\right)^s \nonumber \\
    & & ~\times~ \Theta(m_\ell - m_i - m_j)~,~
  \label{eq:clij}
\end{eqnarray}
where $\mu$ is a parameter with dimensions of mass, where $\Theta(x)$ denotes 
the Heaviside function, and where 
\begin{equation}
  R_{\ell i j} ~=~ \begin{cases} 
    6 & \mbox{all indices different} \\
    3 & \mbox{only two indices equal} \\
    1 & \mbox{all indices equal} 
    \end{cases}
\end{equation}
is a combinatorial factor.  In what follows, for concreteness, we 
take $\mu = m_0/10$.  

Physically, the parameter $r$ appearing in Eq.~(\ref{eq:clij}) governs the 
manner in which the $c_{\ell i j}$ scale with the overall kinetic energy released 
during the decay process.  Taking $r > 0$ establishes a preference for highly 
exothermic decays in which a substantial fraction of the initial mass energy 
$m_\ell$ of the decaying particle is converted into kinetic energy, while 
taking $r < 0$ establishes a preference for decays in which a comparatively 
small fraction of $m_\ell$ is converted to kinetic energy.  By contrast, the 
parameter $s$ governs the manner in which the $c_{\ell i j}$ scale with the 
difference in mass $|m_i - m_j|$ between the two daughter particles.  Taking 
$s > 0$ establishes a preference for decays in which $m_i$ and $m_j$ are very 
similar, while taking $s < 0$ establishes a preference for decays with a 
significant difference between these two daughter-particle masses.
Thus, by varying the parameters of this model --- and in particular, by varying
$r$ and $s$ --- we are able to realize a variety of qualitatively different 
dark-matter velocity distributions in a straightforward way.

For any given choice of model parameters, we evaluate the resulting 
dark-matter velocity distribution $g_v(v)$ by numerically solving the 
coupled system of Boltzmann equations for the $\phi_\ell$.  We then
determine the linear matter power spectrum $P(k)$ for this
$g_v(v)$ distribution and the linear matter power spectrum
$P_{\rm CDM}(k)$ for purely cold dark matter numerically 
using \texttt{CLASS}.~  We obtain $dn/d\log M$ and 
$(dn/d\log M)_{\rm CDM}$ from the corresponding matter power 
spectra using the Press-Schechter formalism, as encapsulated 
in Eq.~(\ref{eq:PressSchechter}), and use these results to 
construct the structure-suppression function $S^2(M)$.
We then test our conjecture by using it to reconstruct $g_M(M)$ from 
$S^2(M)$, and assess how well this reconstructed $g_M(M)$ matches the 
$g_M(M)$ test function that we obtain from our original $g_v(v)$ 
function through use of the functional map in Eq.~(\ref{eq:vtoMmap}).

\begin{figure*}[h!]
  \centering
  \includegraphics[clip, width=0.99\textwidth]{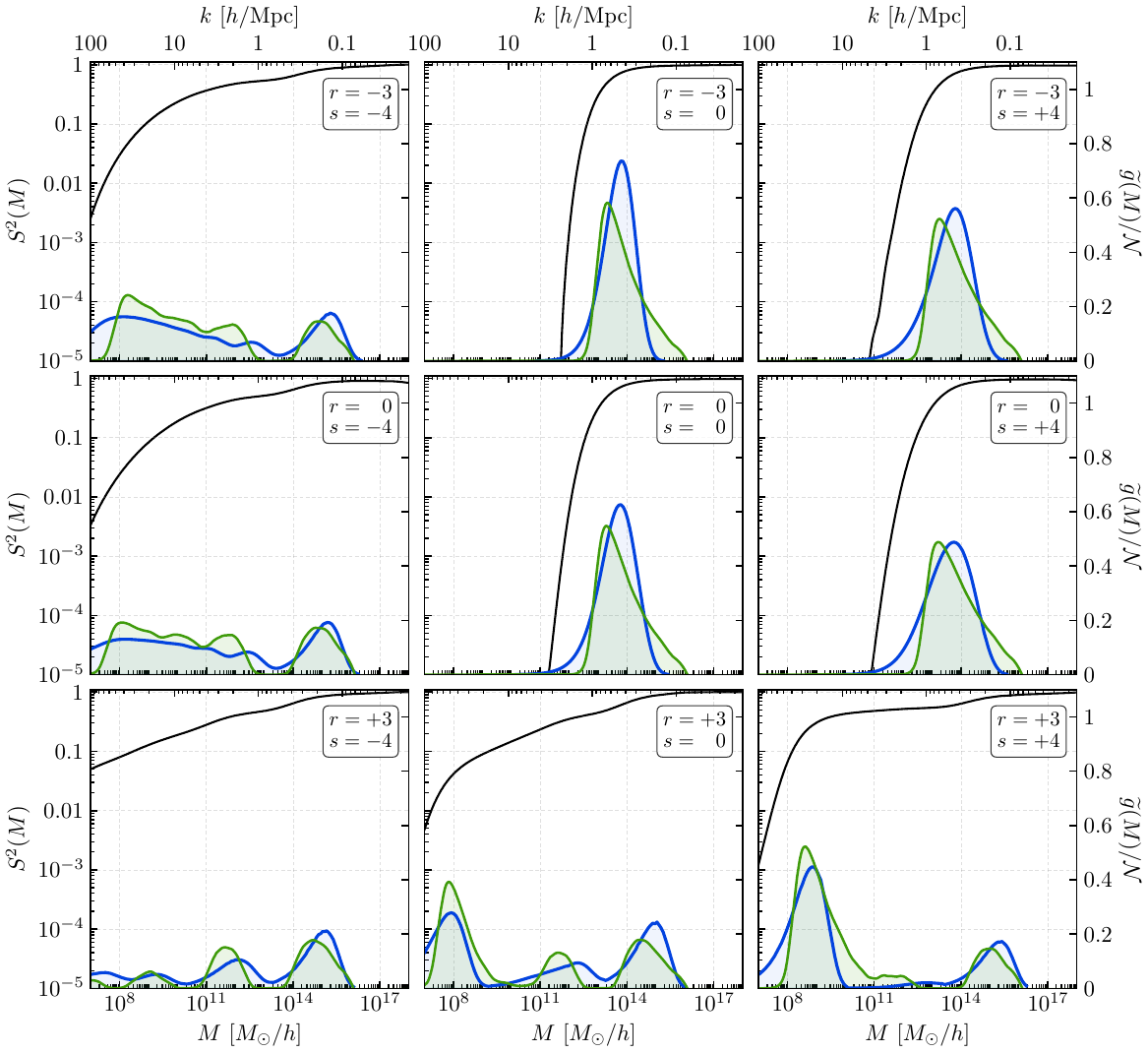}~
  \caption{An explicit test of our reconstruction conjecture for a variety 
    of different dark-matter phase-space distributions $g_M(M)$ which 
    arise in the context of our illustrative dark-matter model --- a 
    model in which the dark matter is produced non-thermally through decay 
    cascades within an extended dark sector.  These distributions 
    correspond to different choices of the parameters $r$ and $s$ in
    Eq.~(\protect\ref{eq:clij}).  Some of these distributions are unimodal, 
    some are bi-modal, and some exhibit even more 
    complex configurations of peaks and troughs.  The blue curve shown 
    in each panel represents the original dark-matter velocity 
    distribution $g_M(M)$.  The black curve represents the corresponding
    structure-suppression function $S^2(M)$ to which it gives rise.  
    The green curve represents the reconstruction of $g_M(M)$ from 
    $S^2(M)$ using in Eq.~(\ref{eq:recconj}).  In all cases, we see that 
    our conjecture successfully reproduces the salient features of the 
    original velocity distribution.
  \label{fig:BradyBunch}}
\end{figure*}

In Fig.~\ref{fig:BradyBunch}, we display the results of our analysis for nine 
different combinations of the model parameters $r$ and $s$.  
These parameter combinations have been chosen such that the corresponding 
dark-matter velocity distributions exhibit a wide variety of profiles.  
The blue curve displayed in each panel represents the ``true'' velocity 
distribution $g_M(M)$ for the corresponding choice of model parameters.  
Indeed, we see that the set of $g_M(M)$ functions obtained for this set of 
parameter combinations includes unimodal distributions as well as a variety 
of multi-modal distributions.  Thus, the velocity distributions shown 
in Fig.~\ref{fig:BradyBunch} collectively provide a thorough ``stress test'' 
of how well our conjecture performs when applied to qualitatively 
different kinds of dark-matter scenarios.   

The black curve appearing in each panel of Fig.~\ref{fig:BradyBunch} represents 
the structure-suppression function $S^2(M)$ which corresponds to the velocity 
distribution $g_M(M)$.  The green curve, on the other hand, represents the 
reconstructed $g_M(M)$ obtained solely from information contained in $S^2(M)$ 
using Eq.~(\ref{eq:recconj}).  In performing this test, we have chosen the 
value of the proportionality constant in Eq.~(\ref{eq:vtoMmap}) to be 
{\il{\xi = 9/4}}, as this tends to horizontally align the original and 
reconstructed dark-matter velocity distributions with each other.

We observe that in each case shown, our reconstruction 
conjecture indeed reproduces the salient features of the original velocity
distribution.  In particular, we see that our conjecture allows us to 
reconstruct not only the approximate locations of the peaks in $g_M(M)$, 
but also the relative areas under those peaks to an impressive degree of 
accuracy across the entire range of $M$ shown.  Thus, while our 
conjecture of course does not reproduce the detailed shapes of the 
features in $g_M(M)$ with perfect fidelity, the results in 
Fig.~\ref{fig:BradyBunch} attest that the simple relation in 
Eq.~(\ref{eq:recconj}) nevertheless provides a versatile tool for 
extracting meaningful information about the properties of the dark matter 
directly from the shape of the halo-mass function alone.


\section{Conclusions\label{sec:Conclusions}}


The dark-matter velocity distributions which arise within the context of 
non-minimal dark-sector scenarios can be complicated and even multi-modal.
Even within the linear regime, the small-scale-structure predictions of
such scenarios differ significantly from the predictions of scenarios
in which $f(v)$ is relatively narrow and unimodal, such as WDM.
In this paper, we have investigated how the detailed shape of the dark-matter
velocity distribution impacts structure in the non-linear regime.  In particular,
through use of the analytic Press-Schechter formalism, we have developed 
an intuition as to how features present in $f(v)$ affect the halo-mass and 
subhalo-mass functions.  We have also studied the implications of these 
results for observables such as the number counts of galaxy clusters and the 
expected number of satellites for a Milky-Way-sized galaxy.  Finally,    
we have proposed a conjecture which can be used to reconstruct the salient 
features of the primordial dark-matter velocity distribution $f(v)$ 
directly from the shape of the halo-mass function $dn/d\log M$.  
This reconstruction conjecture requires essentially no 
additional information about the properties of the dark matter beyond what is 
imprinted on $dn/d\log M$ itself.  Moreover, we have shown that our conjecture 
successfully reproduces the salient features of the underlying dark-matter velocity 
distribution even in situations in which that distribution is complicated and even 
multi-modal.

Several comments are in order.  First of all, our results are predicated on 
a number of theoretical assumptions concerning the form of the halo-mass 
function, the window function $W(k,R)$, \etc~  For example, in our analysis, 
we have adopted the functional form for $\eta(M)$ in 
Eq.~(\ref{eq:ShethTormanf}).  However, as discussed in
Sect.~\ref{sec:HaloMassFunction}, there exist a number of alternatives we 
could have chosen for $\eta(M)$.~  Likewise, while our choice of the window 
function in Eq.~(\ref{eq:Windowfn}) allows us to formulate the map between 
$k$ and $M$ in an unambiguous way, it is certainly possible to consider 
alternatives for $W(k,R)$. Such modifications would of course have an 
impact on our quantitative results for $S^2(M)$, $N_{\rm SH}$, and $N_{\rm C}$, 
as well as the precise form of the empirical relation between 
$d\log S^2(M)/d\log M$ and $F(M)$ in Eq.~(\ref{eq:empir}).  
However, for well-behaved window functions --- \ie, 
functions which have sufficiently flat tops and which decay sufficiently 
rapidly for {\il{k \gtrsim R^{-1}}} --- we expect that these modifications 
will not alter the qualitative nature of our results, including the 
fundamental relationship between $d\log S^2(M)/d\log M$ and $F(M)$ which 
underlies the reconstruction conjecture in Eq.~(\ref{eq:recconj}).  
This issue merits further exploration.

One interesting feature of our analysis is that it makes no particular 
assumption about the masses of the individual dark-matter particles 
themselves.  As such, this analysis is applicable not only to the case in 
which $g_v(v)$ receives contributions from a single dark-matter species 
with a well-defined mass, but also to the case in which multiple particle 
species contribute to the present-day dark-matter abundance (an extreme 
example of which occurs in the Dynamical Dark Matter
framework~{\mbox{\cite{Dienes:2011ja,Dienes:2011sa}}}).  The corresponding
distribution $g_v(v)$ in this latter case represents the aggregate 
velocity distribution for all particle species which contribute to the present-day 
dark-matter abundance.  Of course, this also means that while observables such as
$N_{\rm SH}$ and $N_{\rm C}$ are sensitive to the detailed shape of $g_v(v)$, they
are not capable of distinguishing between single-particle and multi-particle 
dark-matter scenarios which yield the same $g_v(v)$ distribution.  

We also note that we have assumed throughout this paper that the dark matter has 
negligible self-interactions.  Indeed, if dark-matter self-interactions were 
sufficiently strong, the the shape of $g_v(v)$ would continue to evolve at times 
$t > t_{\rm MRE}$.  It would be interesting to explore how the presence of 
appreciable dark-matter self-interactions would affect our results.

While we have focused in this paper primarily on the halo-mass function 
$dN/d\log M$, we also note that the subhalo-mass function --- \ie, the 
differential number of subhalos $dN_{\rm SH}/d\log M$ within a host halo 
of mass $M_0$ --- also provides an observational handle on the 
detailed shape of $g_v(v)$.  Indeed, we have already seen in 
Sect.~\ref{sec:SatelliteCounts} that even the integral number of subhalos 
$N_{\rm SH}$ within a given host halo can provide such a handle.  
Strong gravitational lensing provides a tool which can be used to probe 
this subhalo-mass function on mass scales 
{\il{M \sim 10^{6}}} -- $10^{8}$~$M_\odot$.  Analyses of small existing 
samples of strongly-lensed objects have already yielded meaningful 
constraints on the subhalo-mass 
function~{\mbox{\cite{Vegetti:2014lqa,Hsueh:2019ynk,Gilman:2019nap}}}.  
Moreover, a significant number of additional strong-lensing candidates 
have been identified within SDSS
data~\cite{Talbot:2020arv}.  It would be interesting to consider how
$dN_{\rm SH}/d\log M$ depends on $g_v(v)$ for different $M_0$.  It
would also be interesting to consider whether a
procedure could be developed for reconstructing $g_M(M)$ from 
the shape of $dN_{\rm SH}/d\log M$ at scales $M \lesssim 10^{10}~M_\odot$.

As noted in Sect.~\ref{sec:ConjAll}, the reconstruction conjecture we 
have formulated in Eq.~(\ref{eq:recconj}) has two distinct components:
first, the assertion that the hot-fraction function $F(k)$ is connected to the
{\it slope}\/ of the structure-suppression function, and second, that this 
relation takes the explicit form in Eq.~(\ref{eq:empir}).
Indeed, these two assertions together yield our final conjecture in 
Eq.~(\ref{eq:recconj}).  As evident in Fig.~\ref{fig:BradyBunch},
our conjecture is remarkably successful in reproducing the salient 
features of $g_M(M)$.
However, just as with the conjecture presented in Ref.~\cite{Dienes:2020bmn}, 
we regard this conjecture as at best purely empirical.  It is therefore possible 
that one or both aspects of this conjecture might be further refined.
For example, it is possible that the hot-fraction function $F(M)$ might also 
carry a weak dependence on other (higher) derivatives of the structure-suppression 
function, or on the value of the structure-suppression function itself.
Likewise, even with this assumption, it is possible
that the relation in Eq.~(\ref{eq:empir}) might carry higher-order corrections.
Although it is not possible to rigorously invert the mathematical procedure
discussed in Sect.~\ref{sec:HaloMassFunction} through which a given dark-matter 
velocity distribution $g_v(v)$ produces a corresponding structure-suppression 
function $S(M)$, it may be possible to trace through such a calculation 
analytically to leading order and thereby determine which features of $g_v(v)$ 
might dominate the resulting $S^2(M)$ and  vice versa.  In this way, one might 
hope to eventually derive our conjecture analytically, along with possible 
correction terms.

On the surface, the problem of reconstructing of $g_M(M)$ from $S^2(M)$ bears 
similarity to another well-known inverse problem in kinetic theory, namely the 
determination of the phase-space distribution $f(p)$ of a gas of particles from 
quantities which depend on $f(p)$ only through the moments of this distribution.  
While it is generally possible to solve this inverse problem, the procedure for doing 
so typically yields a significant number of mathematically consistent 
solutions --- solutions which may or may not be physically sensible.  The reason 
why our reconstruction conjecture does not give rise to a similarly large 
multiplicity of solutions is ultimately that the relationship between the $g_M(M)$ 
and $S^2(M)$ is local in the sense that the value of $g_M(M)$ at any particular 
value of $M$ depends only on the value of $S^2(M)$ and its derivatives at that 
same value of $M$.  As a result of this locality, our reconstruction conjecture 
gives us access to $g_M(M)$ directly, rather than requiring us to infer the shape 
of this distribution from its moments.

Our reconstruction conjecture in
principle provides a method of obtaining information about the dark-matter
velocity distribution from the shape of the halo-mass function.  Of course, 
the practical utility of this procedure depends on our ability to
determine the shape of the halo-mass function, which is itself not a directly
measurable quantity.  Doing this presents its own set of challenges.
Significant theoretical uncertainties exist in the relationship between the 
relevant astrophysical observables and halo mass.  Moreover, statistical 
fluctuations in the measured values of these observables can introduce a 
so-called Eddington bias~\cite{Eddington:1913}.  Furthermore, the accuracy 
to which the mass of an individual halo can be measured is limited both by 
the number density of background source images and by uncertainties in the 
shapes of foreground halos.

Nevertheless, despite these challenges, strides have been made toward 
extracting information about the halo-mass function from observation.  In
particular, methods have been proposed for obtaining information about 
$dn/d\log M$ from cosmic-microwave-background (CMB) 
data~\cite{Murray:2013sna}, galaxy-cluster 
number counts~\cite{Castro:2016jmw}, weak lensing of the
CMB~\cite{Madhavacheril:2014slf} and other background 
sources~\cite{Dong:2019mch,Sonnenfeld:2019}, the line widths of 
neutral hydrogen emitted by galaxies~\cite{Li:2019zvm},
and the foreground-background galaxy angular cross-correlation function 
in conjunction with with background samples of sub-millimeter
galaxies~\cite{Cueli:2021dai}.
Such methods make no assumptions about the primordial dark-matter velocity 
distribution $f(v)$, which it is the aim of our conjecture to extract.
Moreover, while many of these methods yield information about $dn/d\log M$
only at scales $M \gtrsim 10^{13}~M_\odot$, others others are in principle 
capable of probing the halo-mass function down to $M \sim 10^{10}~M_\odot$.
Empirical models relating the luminosities of galaxies to the masses of their 
host halos~\cite{Persic:1995ru,Vale:2004yt,Shankar:2006xz,Salucci:2007tm} 
can also potentially be used in conjunction with observation 
in order to provide meaningful information about $dn/d\log M$.  However,
since these models are predicated on the assumption that the dark-matter
velocity distribution takes a particular, simple form, significant numerical 
work would be required in order to assess their applicability to scenarios 
with significantly more complicated $g_v(v)$ distributions.
Thus, despite the challenges involved in determining the halo-mass function from 
observation, larger data sets and an improved understanding of the theoretical
relationship between astrophysical observables and halo mass could 
significantly reduce the uncertainties in $dn/d\log M$ in the near future.  
As such, a calculational tool of the sort we have proposed in this paper 
could potentially be a valuable addition to the toolbox of the 
dark-matter cosmologist.

\begin{acknowledgments}

We would like to thank K.~Abazajian, K.~Boddy, and Z.~Zhai for discussions.
KM wishes to thank the EXCEL Scholars Program for Undergraduate Research at Lafayette 
College, which helped to facilitate this research. 
The research activities of KRD are supported in part by the Department of Energy 
under Grant DE-FG02-13ER41976 (DE-SC0009913) and by the National Science Foundation 
through its employee IR/D program. 
FH is supported in part by the National Natural Science Foundation of China (NSFC) 
under Grants 11690022, 12047503, 11875003, and 12022514 and is also supported by the 
Strategic Priority Research Program and Key Research Program of Frontier Science of the 
Chinese Academy of Sciences under Grants XDB21010200, XDB23000000, and ZDBS-LY-7003.
The research activities of JK are supported in part by the Science and Technology 
Research Council (STFC) under the Consolidated Grant ST/T00102X/1.   
The research activities of KM and BT are supported in 
part by the National Science Foundation under Grants PHY-1720430 and PHY-2014104.  
The opinions and conclusions expressed herein are those of the authors, and do not 
represent any funding agencies. 

\end{acknowledgments}

\appendix

\bibliographystyle{apsrev4-2} 
\bibliography{references}

\begin{thebibliography}{60}%
\makeatletter
\providecommand \@ifxundefined [1]{%
 \@ifx{#1\undefined}
}%
\providecommand \@ifnum [1]{%
 \ifnum #1\expandafter \@firstoftwo
 \else \expandafter \@secondoftwo
 \fi
}%
\providecommand \@ifx [1]{%
 \ifx #1\expandafter \@firstoftwo
 \else \expandafter \@secondoftwo
 \fi
}%
\providecommand \natexlab [1]{#1}%
\providecommand \enquote  [1]{``#1''}%
\providecommand \bibnamefont  [1]{#1}%
\providecommand \bibfnamefont [1]{#1}%
\providecommand \citenamefont [1]{#1}%
\providecommand \href@noop [0]{\@secondoftwo}%
\providecommand \href [0]{\begingroup \@sanitize@url \@href}%
\providecommand \@href[1]{\@@startlink{#1}\@@href}%
\providecommand \@@href[1]{\endgroup#1\@@endlink}%
\providecommand \@sanitize@url [0]{\catcode `\\12\catcode `\$12\catcode
  `\&12\catcode `\#12\catcode `\^12\catcode `\_12\catcode `\%12\relax}%
\providecommand \@@startlink[1]{}%
\providecommand \@@endlink[0]{}%
\providecommand \url  [0]{\begingroup\@sanitize@url \@url }%
\providecommand \@url [1]{\endgroup\@href {#1}{\urlprefix }}%
\providecommand \urlprefix  [0]{URL }%
\providecommand \Eprint [0]{\href }%
\providecommand \doibase [0]{https://doi.org/}%
\providecommand \selectlanguage [0]{\@gobble}%
\providecommand \bibinfo  [0]{\@secondoftwo}%
\providecommand \bibfield  [0]{\@secondoftwo}%
\providecommand \translation [1]{[#1]}%
\providecommand \BibitemOpen [0]{}%
\providecommand \bibitemStop [0]{}%
\providecommand \bibitemNoStop [0]{.\EOS\space}%
\providecommand \EOS [0]{\spacefactor3000\relax}%
\providecommand \BibitemShut  [1]{\csname bibitem#1\endcsname}%
\let\auto@bib@innerbib\@empty
\bibitem [{\citenamefont {König}\ \emph {et~al.}(2016)\citenamefont {König},
  \citenamefont {Merle},\ and\ \citenamefont {Totzauer}}]{Konig:2016dzg}%
  \BibitemOpen
  \bibfield  {author} {\bibinfo {author} {\bibfnamefont {J.}~\bibnamefont
  {König}}, \bibinfo {author} {\bibfnamefont {A.}~\bibnamefont {Merle}},\ and\
  \bibinfo {author} {\bibfnamefont {M.}~\bibnamefont {Totzauer}},\ }\href
  {https://doi.org/10.1088/1475-7516/2016/11/038} {\bibfield  {journal}
  {\bibinfo  {journal} {JCAP}\ }\textbf {\bibinfo {volume} {11}},\ \bibinfo
  {pages} {038}},\ \Eprint {https://arxiv.org/abs/1609.01289} {arXiv:1609.01289
  [hep-ph]} \BibitemShut {NoStop}%
\bibitem [{\citenamefont {Du}\ \emph {et~al.}(2022)\citenamefont {Du},
  \citenamefont {Huang}, \citenamefont {Li}, \citenamefont {Li},\ and\
  \citenamefont {Yu}}]{Du:2021jcj}%
  \BibitemOpen
  \bibfield  {author} {\bibinfo {author} {\bibfnamefont {Y.}~\bibnamefont
  {Du}}, \bibinfo {author} {\bibfnamefont {F.}~\bibnamefont {Huang}}, \bibinfo
  {author} {\bibfnamefont {H.-L.}\ \bibnamefont {Li}}, \bibinfo {author}
  {\bibfnamefont {Y.-Z.}\ \bibnamefont {Li}},\ and\ \bibinfo {author}
  {\bibfnamefont {J.-H.}\ \bibnamefont {Yu}},\ }\href
  {https://doi.org/10.1088/1475-7516/2022/04/012} {\bibfield  {journal}
  {\bibinfo  {journal} {JCAP}\ }\textbf {\bibinfo {volume} {04}}\bibfield
  {number} {\bibinfo  {number} { (04)},\ \bibinfo {pages} {012}},\ }\Eprint
  {https://arxiv.org/abs/2111.01267} {arXiv:2111.01267 [hep-ph]} \BibitemShut
  {NoStop}%
\bibitem [{\citenamefont {Decant}\ \emph {et~al.}(2022)\citenamefont {Decant},
  \citenamefont {Heisig}, \citenamefont {Hooper},\ and\ \citenamefont
  {Lopez-Honorez}}]{Decant:2021mhj}%
  \BibitemOpen
  \bibfield  {author} {\bibinfo {author} {\bibfnamefont {Q.}~\bibnamefont
  {Decant}}, \bibinfo {author} {\bibfnamefont {J.}~\bibnamefont {Heisig}},
  \bibinfo {author} {\bibfnamefont {D.~C.}\ \bibnamefont {Hooper}},\ and\
  \bibinfo {author} {\bibfnamefont {L.}~\bibnamefont {Lopez-Honorez}},\ }\href
  {https://doi.org/10.1088/1475-7516/2022/03/041} {\bibfield  {journal}
  {\bibinfo  {journal} {JCAP}\ }\textbf {\bibinfo {volume} {03}}\bibfield
  {number} {\bibinfo  {number} { (03)},\ \bibinfo {pages} {041}},\ }\Eprint
  {https://arxiv.org/abs/2111.09321} {arXiv:2111.09321 [astro-ph.CO]}
  \BibitemShut {NoStop}%
\bibitem [{\citenamefont {Heeck}\ and\ \citenamefont
  {Teresi}(2017)}]{Heeck:2017xbu}%
  \BibitemOpen
  \bibfield  {author} {\bibinfo {author} {\bibfnamefont {J.}~\bibnamefont
  {Heeck}}\ and\ \bibinfo {author} {\bibfnamefont {D.}~\bibnamefont {Teresi}},\
  }\href {https://doi.org/10.1103/PhysRevD.96.035018} {\bibfield  {journal}
  {\bibinfo  {journal} {Phys. Rev. D}\ }\textbf {\bibinfo {volume} {96}},\
  \bibinfo {pages} {035018} (\bibinfo {year} {2017})},\ \Eprint
  {https://arxiv.org/abs/1706.09909} {arXiv:1706.09909 [hep-ph]} \BibitemShut
  {NoStop}%
\bibitem [{\citenamefont {Dienes}\ \emph {et~al.}(2020)\citenamefont {Dienes},
  \citenamefont {Huang}, \citenamefont {Kost}, \citenamefont {Su},\ and\
  \citenamefont {Thomas}}]{Dienes:2020bmn}%
  \BibitemOpen
  \bibfield  {author} {\bibinfo {author} {\bibfnamefont {K.~R.}\ \bibnamefont
  {Dienes}}, \bibinfo {author} {\bibfnamefont {F.}~\bibnamefont {Huang}},
  \bibinfo {author} {\bibfnamefont {J.}~\bibnamefont {Kost}}, \bibinfo {author}
  {\bibfnamefont {S.}~\bibnamefont {Su}},\ and\ \bibinfo {author}
  {\bibfnamefont {B.}~\bibnamefont {Thomas}},\ }\href
  {https://doi.org/10.1103/PhysRevD.101.123511} {\bibfield  {journal} {\bibinfo
   {journal} {Phys. Rev. D}\ }\textbf {\bibinfo {volume} {101}},\ \bibinfo
  {pages} {123511} (\bibinfo {year} {2020})},\ \Eprint
  {https://arxiv.org/abs/2001.02193} {arXiv:2001.02193 [astro-ph.CO]}
  \BibitemShut {NoStop}%
\bibitem [{\citenamefont {Press}\ and\ \citenamefont
  {Schechter}(1974)}]{Press:1973iz}%
  \BibitemOpen
  \bibfield  {author} {\bibinfo {author} {\bibfnamefont {W.~H.}\ \bibnamefont
  {Press}}\ and\ \bibinfo {author} {\bibfnamefont {P.}~\bibnamefont
  {Schechter}},\ }\href {https://doi.org/10.1086/152650} {\bibfield  {journal}
  {\bibinfo  {journal} {Astrophys. J.}\ }\textbf {\bibinfo {volume} {187}},\
  \bibinfo {pages} {425} (\bibinfo {year} {1974})}\BibitemShut {NoStop}%
\bibitem [{\citenamefont {Bardeen}\ \emph {et~al.}(1986)\citenamefont
  {Bardeen}, \citenamefont {Bond}, \citenamefont {Kaiser},\ and\ \citenamefont
  {Szalay}}]{Bardeen:1985tr}%
  \BibitemOpen
  \bibfield  {author} {\bibinfo {author} {\bibfnamefont {J.~M.}\ \bibnamefont
  {Bardeen}}, \bibinfo {author} {\bibfnamefont {J.}~\bibnamefont {Bond}},
  \bibinfo {author} {\bibfnamefont {N.}~\bibnamefont {Kaiser}},\ and\ \bibinfo
  {author} {\bibfnamefont {A.}~\bibnamefont {Szalay}},\ }\href
  {https://doi.org/10.1086/164143} {\bibfield  {journal} {\bibinfo  {journal}
  {Astrophys. J.}\ }\textbf {\bibinfo {volume} {304}},\ \bibinfo {pages} {15}
  (\bibinfo {year} {1986})}\BibitemShut {NoStop}%
\bibitem [{\citenamefont {Bond}\ \emph {et~al.}(1991)\citenamefont {Bond},
  \citenamefont {Cole}, \citenamefont {Efstathiou},\ and\ \citenamefont
  {Kaiser}}]{Bond:1990iw}%
  \BibitemOpen
  \bibfield  {author} {\bibinfo {author} {\bibfnamefont {J.}~\bibnamefont
  {Bond}}, \bibinfo {author} {\bibfnamefont {S.}~\bibnamefont {Cole}}, \bibinfo
  {author} {\bibfnamefont {G.}~\bibnamefont {Efstathiou}},\ and\ \bibinfo
  {author} {\bibfnamefont {N.}~\bibnamefont {Kaiser}},\ }\href
  {https://doi.org/10.1086/170520} {\bibfield  {journal} {\bibinfo  {journal}
  {Astrophys. J.}\ }\textbf {\bibinfo {volume} {379}},\ \bibinfo {pages} {440}
  (\bibinfo {year} {1991})}\BibitemShut {NoStop}%
\bibitem [{\citenamefont {Sheth}\ and\ \citenamefont
  {Tormen}(1999)}]{Sheth:1999mn}%
  \BibitemOpen
  \bibfield  {author} {\bibinfo {author} {\bibfnamefont {R.~K.}\ \bibnamefont
  {Sheth}}\ and\ \bibinfo {author} {\bibfnamefont {G.}~\bibnamefont {Tormen}},\
  }\href {https://doi.org/10.1046/j.1365-8711.1999.02692.x} {\bibfield
  {journal} {\bibinfo  {journal} {Mon. Not. Roy. Astron. Soc.}\ }\textbf
  {\bibinfo {volume} {308}},\ \bibinfo {pages} {119} (\bibinfo {year}
  {1999})},\ \Eprint {https://arxiv.org/abs/astro-ph/9901122}
  {arXiv:astro-ph/9901122} \BibitemShut {NoStop}%
\bibitem [{\citenamefont {Sheth}\ \emph {et~al.}(2001)\citenamefont {Sheth},
  \citenamefont {Mo},\ and\ \citenamefont {Tormen}}]{Sheth:1999su}%
  \BibitemOpen
  \bibfield  {author} {\bibinfo {author} {\bibfnamefont {R.~K.}\ \bibnamefont
  {Sheth}}, \bibinfo {author} {\bibfnamefont {H.~J.}\ \bibnamefont {Mo}},\ and\
  \bibinfo {author} {\bibfnamefont {G.}~\bibnamefont {Tormen}},\ }\href
  {https://doi.org/10.1046/j.1365-8711.2001.04006.x} {\bibfield  {journal}
  {\bibinfo  {journal} {Mon. Not. Roy. Astron. Soc.}\ }\textbf {\bibinfo
  {volume} {323}},\ \bibinfo {pages} {1} (\bibinfo {year} {2001})},\ \Eprint
  {https://arxiv.org/abs/astro-ph/9907024} {arXiv:astro-ph/9907024}
  \BibitemShut {NoStop}%
\bibitem [{\citenamefont {Castro}\ \emph {et~al.}(2016)\citenamefont {Castro},
  \citenamefont {Marra},\ and\ \citenamefont {Quartin}}]{Castro:2016jmw}%
  \BibitemOpen
  \bibfield  {author} {\bibinfo {author} {\bibfnamefont {T.}~\bibnamefont
  {Castro}}, \bibinfo {author} {\bibfnamefont {V.}~\bibnamefont {Marra}},\ and\
  \bibinfo {author} {\bibfnamefont {M.}~\bibnamefont {Quartin}},\ }\href
  {https://doi.org/10.1093/mnras/stw2072} {\bibfield  {journal} {\bibinfo
  {journal} {Mon. Not. Roy. Astron. Soc.}\ }\textbf {\bibinfo {volume} {463}},\
  \bibinfo {pages} {1666} (\bibinfo {year} {2016})},\ \Eprint
  {https://arxiv.org/abs/1605.07548} {arXiv:1605.07548 [astro-ph.CO]}
  \BibitemShut {NoStop}%
\bibitem [{\citenamefont {Dong}\ \emph {et~al.}(2019)\citenamefont {Dong},
  \citenamefont {Zhang}, \citenamefont {Yang}, \citenamefont {Zhang},\ and\
  \citenamefont {Luo}}]{Dong:2019mch}%
  \BibitemOpen
  \bibfield  {author} {\bibinfo {author} {\bibfnamefont {F.}~\bibnamefont
  {Dong}}, \bibinfo {author} {\bibfnamefont {J.}~\bibnamefont {Zhang}},
  \bibinfo {author} {\bibfnamefont {X.}~\bibnamefont {Yang}}, \bibinfo {author}
  {\bibfnamefont {J.}~\bibnamefont {Zhang}},\ and\ \bibinfo {author}
  {\bibfnamefont {W.}~\bibnamefont {Luo}},\ }\href
  {https://doi.org/10.3847/1538-4357/ab3a9d} {\bibfield  {journal} {\bibinfo
  {journal} {Astrophys. J.}\ }\textbf {\bibinfo {volume} {883}},\ \bibinfo
  {pages} {155} (\bibinfo {year} {2019})},\ \Eprint
  {https://arxiv.org/abs/1905.11886} {arXiv:1905.11886 [astro-ph.CO]}
  \BibitemShut {NoStop}%
\bibitem [{\citenamefont {Sonnenfeld}\ \emph {et~al.}(2019)\citenamefont
  {Sonnenfeld}, \citenamefont {Wang},\ and\ \citenamefont
  {Bahcall}}]{Sonnenfeld:2019}%
  \BibitemOpen
  \bibfield  {author} {\bibinfo {author} {\bibfnamefont {A.}~\bibnamefont
  {Sonnenfeld}}, \bibinfo {author} {\bibfnamefont {W.}~\bibnamefont {Wang}},\
  and\ \bibinfo {author} {\bibfnamefont {N.}~\bibnamefont {Bahcall}},\ }\href
  {https://doi.org/10.1051/0004-6361/201834260} {\bibfield  {journal} {\bibinfo
   {journal} {Astronomy \& Astrophysics}\ }\textbf {\bibinfo {volume} {622}},\
  \bibinfo {pages} {A30} (\bibinfo {year} {2019})},\ \Eprint
  {https://arxiv.org/abs/1811.04934} {arXiv:1811.04934 [astro-ph.GA]}
  \BibitemShut {NoStop}%
\bibitem [{\citenamefont {Li}\ \emph {et~al.}(2019)\citenamefont {Li},
  \citenamefont {Lelli}, \citenamefont {McGaugh}, \citenamefont {Pawlowski},
  \citenamefont {Zwaan},\ and\ \citenamefont {Schombert}}]{Li:2019zvm}%
  \BibitemOpen
  \bibfield  {author} {\bibinfo {author} {\bibfnamefont {P.}~\bibnamefont
  {Li}}, \bibinfo {author} {\bibfnamefont {F.}~\bibnamefont {Lelli}}, \bibinfo
  {author} {\bibfnamefont {S.}~\bibnamefont {McGaugh}}, \bibinfo {author}
  {\bibfnamefont {M.~S.}\ \bibnamefont {Pawlowski}}, \bibinfo {author}
  {\bibfnamefont {M.~A.}\ \bibnamefont {Zwaan}},\ and\ \bibinfo {author}
  {\bibfnamefont {J.}~\bibnamefont {Schombert}},\ }\href
  {https://doi.org/10.3847/2041-8213/ab53e6} {\bibfield  {journal} {\bibinfo
  {journal} {Astrophys. J. Lett.}\ }\textbf {\bibinfo {volume} {886}},\
  \bibinfo {pages} {L11} (\bibinfo {year} {2019})},\ \Eprint
  {https://arxiv.org/abs/1911.00517} {arXiv:1911.00517 [astro-ph.GA]}
  \BibitemShut {NoStop}%
\bibitem [{\citenamefont {Cueli}\ \emph {et~al.}(2021)\citenamefont {Cueli},
  \citenamefont {Bonavera}, \citenamefont {Gonz\'alez-Nuevo},\ and\
  \citenamefont {Lapi}}]{Cueli:2021dai}%
  \BibitemOpen
  \bibfield  {author} {\bibinfo {author} {\bibfnamefont {M.~M.}\ \bibnamefont
  {Cueli}}, \bibinfo {author} {\bibfnamefont {L.}~\bibnamefont {Bonavera}},
  \bibinfo {author} {\bibfnamefont {J.}~\bibnamefont {Gonz\'alez-Nuevo}},\ and\
  \bibinfo {author} {\bibfnamefont {A.}~\bibnamefont {Lapi}},\ }\href
  {https://doi.org/10.1051/0004-6361/202039326} {\bibfield  {journal} {\bibinfo
   {journal} {Astron. Astrophys.}\ }\textbf {\bibinfo {volume} {645}},\
  \bibinfo {pages} {A126} (\bibinfo {year} {2021})},\ \Eprint
  {https://arxiv.org/abs/2102.03890} {arXiv:2102.03890 [astro-ph.CO]}
  \BibitemShut {NoStop}%
\bibitem [{\citenamefont
  {Lesgourgues}(2011{\natexlab{a}})}]{Lesgourgues:2011re}%
  \BibitemOpen
  \bibfield  {author} {\bibinfo {author} {\bibfnamefont {J.}~\bibnamefont
  {Lesgourgues}},\ }\href@noop {} {\  (\bibinfo {year} {2011}{\natexlab{a}})},\
  \Eprint {https://arxiv.org/abs/1104.2932} {arXiv:1104.2932 [astro-ph.IM]}
  \BibitemShut {NoStop}%
\bibitem [{\citenamefont {Blas}\ \emph {et~al.}(2011)\citenamefont {Blas},
  \citenamefont {Lesgourgues},\ and\ \citenamefont {Tram}}]{Blas:2011rf}%
  \BibitemOpen
  \bibfield  {author} {\bibinfo {author} {\bibfnamefont {D.}~\bibnamefont
  {Blas}}, \bibinfo {author} {\bibfnamefont {J.}~\bibnamefont {Lesgourgues}},\
  and\ \bibinfo {author} {\bibfnamefont {T.}~\bibnamefont {Tram}},\ }\href
  {https://doi.org/10.1088/1475-7516/2011/07/034} {\bibfield  {journal}
  {\bibinfo  {journal} {JCAP}\ }\textbf {\bibinfo {volume} {07}},\ \bibinfo
  {pages} {034}},\ \Eprint {https://arxiv.org/abs/1104.2933} {arXiv:1104.2933
  [astro-ph.CO]} \BibitemShut {NoStop}%
\bibitem [{\citenamefont
  {Lesgourgues}(2011{\natexlab{b}})}]{Lesgourgues:2011rg}%
  \BibitemOpen
  \bibfield  {author} {\bibinfo {author} {\bibfnamefont {J.}~\bibnamefont
  {Lesgourgues}},\ }\href@noop {} {\  (\bibinfo {year} {2011}{\natexlab{b}})},\
  \Eprint {https://arxiv.org/abs/1104.2934} {arXiv:1104.2934 [astro-ph.CO]}
  \BibitemShut {NoStop}%
\bibitem [{\citenamefont {Lesgourgues}\ and\ \citenamefont
  {Tram}(2011)}]{Lesgourgues:2011rh}%
  \BibitemOpen
  \bibfield  {author} {\bibinfo {author} {\bibfnamefont {J.}~\bibnamefont
  {Lesgourgues}}\ and\ \bibinfo {author} {\bibfnamefont {T.}~\bibnamefont
  {Tram}},\ }\href {https://doi.org/10.1088/1475-7516/2011/09/032} {\bibfield
  {journal} {\bibinfo  {journal} {JCAP}\ }\textbf {\bibinfo {volume} {09}},\
  \bibinfo {pages} {032}},\ \Eprint {https://arxiv.org/abs/1104.2935}
  {arXiv:1104.2935 [astro-ph.CO]} \BibitemShut {NoStop}%
\bibitem [{\citenamefont {Bode}\ \emph {et~al.}(2001)\citenamefont {Bode},
  \citenamefont {Ostriker},\ and\ \citenamefont {Turok}}]{Bode:2000gq}%
  \BibitemOpen
  \bibfield  {author} {\bibinfo {author} {\bibfnamefont {P.}~\bibnamefont
  {Bode}}, \bibinfo {author} {\bibfnamefont {J.~P.}\ \bibnamefont {Ostriker}},\
  and\ \bibinfo {author} {\bibfnamefont {N.}~\bibnamefont {Turok}},\ }\href
  {https://doi.org/10.1086/321541} {\bibfield  {journal} {\bibinfo  {journal}
  {Astrophys. J.}\ }\textbf {\bibinfo {volume} {556}},\ \bibinfo {pages} {93}
  (\bibinfo {year} {2001})},\ \Eprint {https://arxiv.org/abs/astro-ph/0010389}
  {arXiv:astro-ph/0010389} \BibitemShut {NoStop}%
\bibitem [{\citenamefont {Hansen}\ \emph {et~al.}(2002)\citenamefont {Hansen},
  \citenamefont {Lesgourgues}, \citenamefont {Pastor},\ and\ \citenamefont
  {Silk}}]{Hansen:2001zv}%
  \BibitemOpen
  \bibfield  {author} {\bibinfo {author} {\bibfnamefont {S.~H.}\ \bibnamefont
  {Hansen}}, \bibinfo {author} {\bibfnamefont {J.}~\bibnamefont {Lesgourgues}},
  \bibinfo {author} {\bibfnamefont {S.}~\bibnamefont {Pastor}},\ and\ \bibinfo
  {author} {\bibfnamefont {J.}~\bibnamefont {Silk}},\ }\href
  {https://doi.org/10.1046/j.1365-8711.2002.05410.x} {\bibfield  {journal}
  {\bibinfo  {journal} {Mon. Not. Roy. Astron. Soc.}\ }\textbf {\bibinfo
  {volume} {333}},\ \bibinfo {pages} {544} (\bibinfo {year} {2002})},\ \Eprint
  {https://arxiv.org/abs/astro-ph/0106108} {arXiv:astro-ph/0106108}
  \BibitemShut {NoStop}%
\bibitem [{\citenamefont {Viel}\ \emph {et~al.}(2005)\citenamefont {Viel},
  \citenamefont {Lesgourgues}, \citenamefont {Haehnelt}, \citenamefont
  {Matarrese},\ and\ \citenamefont {Riotto}}]{Viel:2005qj}%
  \BibitemOpen
  \bibfield  {author} {\bibinfo {author} {\bibfnamefont {M.}~\bibnamefont
  {Viel}}, \bibinfo {author} {\bibfnamefont {J.}~\bibnamefont {Lesgourgues}},
  \bibinfo {author} {\bibfnamefont {M.~G.}\ \bibnamefont {Haehnelt}}, \bibinfo
  {author} {\bibfnamefont {S.}~\bibnamefont {Matarrese}},\ and\ \bibinfo
  {author} {\bibfnamefont {A.}~\bibnamefont {Riotto}},\ }\href
  {https://doi.org/10.1103/PhysRevD.71.063534} {\bibfield  {journal} {\bibinfo
  {journal} {Phys.\ Rev.\ D}\ }\textbf {\bibinfo {volume} {71}},\ \bibinfo
  {pages} {063534} (\bibinfo {year} {2005})},\ \Eprint
  {https://arxiv.org/abs/astro-ph/0501562} {arXiv:astro-ph/0501562}
  \BibitemShut {NoStop}%
\bibitem [{\citenamefont {Bertschinger}(2006)}]{Bertschinger:2006nq}%
  \BibitemOpen
  \bibfield  {author} {\bibinfo {author} {\bibfnamefont {E.}~\bibnamefont
  {Bertschinger}},\ }\href {https://doi.org/10.1103/PhysRevD.74.063509}
  {\bibfield  {journal} {\bibinfo  {journal} {Phys. Rev. D}\ }\textbf {\bibinfo
  {volume} {74}},\ \bibinfo {pages} {063509} (\bibinfo {year} {2006})},\
  \Eprint {https://arxiv.org/abs/astro-ph/0607319} {arXiv:astro-ph/0607319}
  \BibitemShut {NoStop}%
\bibitem [{\citenamefont {Schneider}\ \emph {et~al.}(2013)\citenamefont
  {Schneider}, \citenamefont {Smith},\ and\ \citenamefont
  {Reed}}]{Schneider:2013ria}%
  \BibitemOpen
  \bibfield  {author} {\bibinfo {author} {\bibfnamefont {A.}~\bibnamefont
  {Schneider}}, \bibinfo {author} {\bibfnamefont {R.~E.}\ \bibnamefont
  {Smith}},\ and\ \bibinfo {author} {\bibfnamefont {D.}~\bibnamefont {Reed}},\
  }\href {https://doi.org/10.1093/mnras/stt829} {\bibfield  {journal} {\bibinfo
   {journal} {Mon.\ Not.\ Roy.\ Astron.\ Soc.}\ }\textbf {\bibinfo {volume}
  {433}},\ \bibinfo {pages} {1573} (\bibinfo {year} {2013})},\ \Eprint
  {https://arxiv.org/abs/1303.0839} {arXiv:1303.0839 [astro-ph.CO]}
  \BibitemShut {NoStop}%
\bibitem [{\citenamefont {Schneider}(2015)}]{Schneider:2014rda}%
  \BibitemOpen
  \bibfield  {author} {\bibinfo {author} {\bibfnamefont {A.}~\bibnamefont
  {Schneider}},\ }\href {https://doi.org/10.1093/mnras/stv1169} {\bibfield
  {journal} {\bibinfo  {journal} {Mon. Not. Roy. Astron. Soc.}\ }\textbf
  {\bibinfo {volume} {451}},\ \bibinfo {pages} {3117} (\bibinfo {year}
  {2015})},\ \Eprint {https://arxiv.org/abs/1412.2133} {arXiv:1412.2133
  [astro-ph.CO]} \BibitemShut {NoStop}%
\bibitem [{\citenamefont {Aghanim}\ \emph {et~al.}(2020)\citenamefont {Aghanim}
  \emph {et~al.}}]{Planck:2018vyg}%
  \BibitemOpen
  \bibfield  {author} {\bibinfo {author} {\bibfnamefont {N.}~\bibnamefont
  {Aghanim}} \emph {et~al.} (\bibinfo {collaboration} {Planck}),\ }\href
  {https://doi.org/10.1051/0004-6361/201833910} {\bibfield  {journal} {\bibinfo
   {journal} {Astron. Astrophys.}\ }\textbf {\bibinfo {volume} {641}},\
  \bibinfo {pages} {A6} (\bibinfo {year} {2020})},\ \bibinfo {note} {[Erratum:
  Astron.Astrophys. 652, C4 (2021)]},\ \Eprint
  {https://arxiv.org/abs/1807.06209} {arXiv:1807.06209 [astro-ph.CO]}
  \BibitemShut {NoStop}%
\bibitem [{\citenamefont {Jenkins}\ \emph {et~al.}(2001)\citenamefont
  {Jenkins}, \citenamefont {Frenk}, \citenamefont {White}, \citenamefont
  {Colberg}, \citenamefont {Cole}, \citenamefont {Evrard}, \citenamefont
  {Couchman},\ and\ \citenamefont {Yoshida}}]{Jenkins:2000bv}%
  \BibitemOpen
  \bibfield  {author} {\bibinfo {author} {\bibfnamefont {A.}~\bibnamefont
  {Jenkins}}, \bibinfo {author} {\bibfnamefont {C.}~\bibnamefont {Frenk}},
  \bibinfo {author} {\bibfnamefont {S.~D.}\ \bibnamefont {White}}, \bibinfo
  {author} {\bibfnamefont {J.}~\bibnamefont {Colberg}}, \bibinfo {author}
  {\bibfnamefont {S.}~\bibnamefont {Cole}}, \bibinfo {author} {\bibfnamefont
  {A.~E.}\ \bibnamefont {Evrard}}, \bibinfo {author} {\bibfnamefont
  {H.}~\bibnamefont {Couchman}},\ and\ \bibinfo {author} {\bibfnamefont
  {N.}~\bibnamefont {Yoshida}},\ }\href
  {https://doi.org/10.1046/j.1365-8711.2001.04029.x} {\bibfield  {journal}
  {\bibinfo  {journal} {Mon. Not. Roy. Astron. Soc.}\ }\textbf {\bibinfo
  {volume} {321}},\ \bibinfo {pages} {372} (\bibinfo {year} {2001})},\ \Eprint
  {https://arxiv.org/abs/astro-ph/0005260} {arXiv:astro-ph/0005260}
  \BibitemShut {NoStop}%
\bibitem [{\citenamefont {Warren}\ \emph {et~al.}(2006)\citenamefont {Warren},
  \citenamefont {Abazajian}, \citenamefont {Holz},\ and\ \citenamefont
  {Teodoro}}]{Warren:2005ey}%
  \BibitemOpen
  \bibfield  {author} {\bibinfo {author} {\bibfnamefont {M.~S.}\ \bibnamefont
  {Warren}}, \bibinfo {author} {\bibfnamefont {K.}~\bibnamefont {Abazajian}},
  \bibinfo {author} {\bibfnamefont {D.~E.}\ \bibnamefont {Holz}},\ and\
  \bibinfo {author} {\bibfnamefont {L.}~\bibnamefont {Teodoro}},\ }\href
  {https://doi.org/10.1086/504962} {\bibfield  {journal} {\bibinfo  {journal}
  {Astrophys. J.}\ }\textbf {\bibinfo {volume} {646}},\ \bibinfo {pages} {881}
  (\bibinfo {year} {2006})},\ \Eprint {https://arxiv.org/abs/astro-ph/0506395}
  {arXiv:astro-ph/0506395} \BibitemShut {NoStop}%
\bibitem [{\citenamefont {Tinker}\ \emph {et~al.}(2008)\citenamefont {Tinker},
  \citenamefont {Kravtsov}, \citenamefont {Klypin}, \citenamefont {Abazajian},
  \citenamefont {Warren}, \citenamefont {Yepes}, \citenamefont {Gottlober},\
  and\ \citenamefont {Holz}}]{Tinker:2008ff}%
  \BibitemOpen
  \bibfield  {author} {\bibinfo {author} {\bibfnamefont {J.~L.}\ \bibnamefont
  {Tinker}}, \bibinfo {author} {\bibfnamefont {A.~V.}\ \bibnamefont
  {Kravtsov}}, \bibinfo {author} {\bibfnamefont {A.}~\bibnamefont {Klypin}},
  \bibinfo {author} {\bibfnamefont {K.}~\bibnamefont {Abazajian}}, \bibinfo
  {author} {\bibfnamefont {M.~S.}\ \bibnamefont {Warren}}, \bibinfo {author}
  {\bibfnamefont {G.}~\bibnamefont {Yepes}}, \bibinfo {author} {\bibfnamefont
  {S.}~\bibnamefont {Gottlober}},\ and\ \bibinfo {author} {\bibfnamefont
  {D.~E.}\ \bibnamefont {Holz}},\ }\href {https://doi.org/10.1086/591439}
  {\bibfield  {journal} {\bibinfo  {journal} {Astrophys. J.}\ }\textbf
  {\bibinfo {volume} {688}},\ \bibinfo {pages} {709} (\bibinfo {year}
  {2008})},\ \Eprint {https://arxiv.org/abs/0803.2706} {arXiv:0803.2706
  [astro-ph]} \BibitemShut {NoStop}%
\bibitem [{\citenamefont {Crocce}\ \emph {et~al.}(2010)\citenamefont {Crocce},
  \citenamefont {Fosalba}, \citenamefont {Castander},\ and\ \citenamefont
  {Gaztanaga}}]{Crocce:2009mg}%
  \BibitemOpen
  \bibfield  {author} {\bibinfo {author} {\bibfnamefont {M.}~\bibnamefont
  {Crocce}}, \bibinfo {author} {\bibfnamefont {P.}~\bibnamefont {Fosalba}},
  \bibinfo {author} {\bibfnamefont {F.~J.}\ \bibnamefont {Castander}},\ and\
  \bibinfo {author} {\bibfnamefont {E.}~\bibnamefont {Gaztanaga}},\ }\href
  {https://doi.org/10.1111/j.1365-2966.2009.16194.x} {\bibfield  {journal}
  {\bibinfo  {journal} {Mon. Not. Roy. Astron. Soc.}\ }\textbf {\bibinfo
  {volume} {403}},\ \bibinfo {pages} {1353} (\bibinfo {year} {2010})},\ \Eprint
  {https://arxiv.org/abs/0907.0019} {arXiv:0907.0019 [astro-ph.CO]}
  \BibitemShut {NoStop}%
\bibitem [{\citenamefont {Bhattacharya}\ \emph {et~al.}(2011)\citenamefont
  {Bhattacharya}, \citenamefont {Heitmann}, \citenamefont {White},
  \citenamefont {Lukic}, \citenamefont {Wagner},\ and\ \citenamefont
  {Habib}}]{Bhattacharya:2010wy}%
  \BibitemOpen
  \bibfield  {author} {\bibinfo {author} {\bibfnamefont {S.}~\bibnamefont
  {Bhattacharya}}, \bibinfo {author} {\bibfnamefont {K.}~\bibnamefont
  {Heitmann}}, \bibinfo {author} {\bibfnamefont {M.}~\bibnamefont {White}},
  \bibinfo {author} {\bibfnamefont {Z.}~\bibnamefont {Lukic}}, \bibinfo
  {author} {\bibfnamefont {C.}~\bibnamefont {Wagner}},\ and\ \bibinfo {author}
  {\bibfnamefont {S.}~\bibnamefont {Habib}},\ }\href
  {https://doi.org/10.1088/0004-637X/732/2/122} {\bibfield  {journal} {\bibinfo
   {journal} {Astrophys. J.}\ }\textbf {\bibinfo {volume} {732}},\ \bibinfo
  {pages} {122} (\bibinfo {year} {2011})},\ \Eprint
  {https://arxiv.org/abs/1005.2239} {arXiv:1005.2239 [astro-ph.CO]}
  \BibitemShut {NoStop}%
\bibitem [{\citenamefont {Watson}\ \emph {et~al.}(2013)\citenamefont {Watson},
  \citenamefont {Iliev}, \citenamefont {D'Aloisio}, \citenamefont {Knebe},
  \citenamefont {Shapiro},\ and\ \citenamefont {Yepes}}]{Watson:2012mt}%
  \BibitemOpen
  \bibfield  {author} {\bibinfo {author} {\bibfnamefont {W.~A.}\ \bibnamefont
  {Watson}}, \bibinfo {author} {\bibfnamefont {I.~T.}\ \bibnamefont {Iliev}},
  \bibinfo {author} {\bibfnamefont {A.}~\bibnamefont {D'Aloisio}}, \bibinfo
  {author} {\bibfnamefont {A.}~\bibnamefont {Knebe}}, \bibinfo {author}
  {\bibfnamefont {P.~R.}\ \bibnamefont {Shapiro}},\ and\ \bibinfo {author}
  {\bibfnamefont {G.}~\bibnamefont {Yepes}},\ }\href
  {https://doi.org/10.1093/mnras/stt791} {\bibfield  {journal} {\bibinfo
  {journal} {Mon. Not. Roy. Astron. Soc.}\ }\textbf {\bibinfo {volume} {433}},\
  \bibinfo {pages} {1230} (\bibinfo {year} {2013})},\ \Eprint
  {https://arxiv.org/abs/1212.0095} {arXiv:1212.0095 [astro-ph.CO]}
  \BibitemShut {NoStop}%
\bibitem [{\citenamefont {Seppi}\ \emph {et~al.}(2021)\citenamefont {Seppi},
  \citenamefont {Comparat}, \citenamefont {Nandra}, \citenamefont {Bulbul},
  \citenamefont {Prada}, \citenamefont {Klypin}, \citenamefont {Merloni},
  \citenamefont {Predehl},\ and\ \citenamefont {Chitham}}]{Seppi:2020isf}%
  \BibitemOpen
  \bibfield  {author} {\bibinfo {author} {\bibfnamefont {R.}~\bibnamefont
  {Seppi}}, \bibinfo {author} {\bibfnamefont {J.}~\bibnamefont {Comparat}},
  \bibinfo {author} {\bibfnamefont {K.}~\bibnamefont {Nandra}}, \bibinfo
  {author} {\bibfnamefont {E.}~\bibnamefont {Bulbul}}, \bibinfo {author}
  {\bibfnamefont {F.}~\bibnamefont {Prada}}, \bibinfo {author} {\bibfnamefont
  {A.}~\bibnamefont {Klypin}}, \bibinfo {author} {\bibfnamefont
  {A.}~\bibnamefont {Merloni}}, \bibinfo {author} {\bibfnamefont
  {P.}~\bibnamefont {Predehl}},\ and\ \bibinfo {author} {\bibfnamefont {J.~I.}\
  \bibnamefont {Chitham}},\ }\href
  {https://doi.org/10.1051/0004-6361/202039123} {\bibfield  {journal} {\bibinfo
   {journal} {Astron. Astrophys.}\ }\textbf {\bibinfo {volume} {652}},\
  \bibinfo {pages} {A155} (\bibinfo {year} {2021})},\ \Eprint
  {https://arxiv.org/abs/2008.03179} {arXiv:2008.03179 [astro-ph.CO]}
  \BibitemShut {NoStop}%
\bibitem [{\citenamefont {Parimbelli}\ \emph {et~al.}(2021)\citenamefont
  {Parimbelli}, \citenamefont {Scelfo}, \citenamefont {Giri}, \citenamefont
  {Schneider}, \citenamefont {Archidiacono}, \citenamefont {Camera},\ and\
  \citenamefont {Viel}}]{Parimbelli:2021mtp}%
  \BibitemOpen
  \bibfield  {author} {\bibinfo {author} {\bibfnamefont {G.}~\bibnamefont
  {Parimbelli}}, \bibinfo {author} {\bibfnamefont {G.}~\bibnamefont {Scelfo}},
  \bibinfo {author} {\bibfnamefont {S.~K.}\ \bibnamefont {Giri}}, \bibinfo
  {author} {\bibfnamefont {A.}~\bibnamefont {Schneider}}, \bibinfo {author}
  {\bibfnamefont {M.}~\bibnamefont {Archidiacono}}, \bibinfo {author}
  {\bibfnamefont {S.}~\bibnamefont {Camera}},\ and\ \bibinfo {author}
  {\bibfnamefont {M.}~\bibnamefont {Viel}},\ }\href
  {https://doi.org/10.1088/1475-7516/2021/12/044} {\bibfield  {journal}
  {\bibinfo  {journal} {JCAP}\ }\textbf {\bibinfo {volume} {12}}\bibfield
  {number} {\bibinfo  {number} { (12)},\ \bibinfo {pages} {044}},\ }\Eprint
  {https://arxiv.org/abs/2106.04588} {arXiv:2106.04588 [astro-ph.CO]}
  \BibitemShut {NoStop}%
\bibitem [{\citenamefont {Lacey}\ and\ \citenamefont
  {Cole}(1993)}]{Lacey:1993iv}%
  \BibitemOpen
  \bibfield  {author} {\bibinfo {author} {\bibfnamefont {C.~G.}\ \bibnamefont
  {Lacey}}\ and\ \bibinfo {author} {\bibfnamefont {S.}~\bibnamefont {Cole}},\
  }\href@noop {} {\bibfield  {journal} {\bibinfo  {journal} {Mon. Not. Roy.
  Astron. Soc.}\ }\textbf {\bibinfo {volume} {262}},\ \bibinfo {pages} {627}
  (\bibinfo {year} {1993})}\BibitemShut {NoStop}%
\bibitem [{\citenamefont {Giocoli}\ \emph {et~al.}(2008)\citenamefont
  {Giocoli}, \citenamefont {Pieri},\ and\ \citenamefont
  {Tormen}}]{Giocoli:2007gf}%
  \BibitemOpen
  \bibfield  {author} {\bibinfo {author} {\bibfnamefont {C.}~\bibnamefont
  {Giocoli}}, \bibinfo {author} {\bibfnamefont {L.}~\bibnamefont {Pieri}},\
  and\ \bibinfo {author} {\bibfnamefont {G.}~\bibnamefont {Tormen}},\ }\href
  {https://doi.org/10.1111/j.1365-2966.2008.13283.x} {\bibfield  {journal}
  {\bibinfo  {journal} {Mon. Not. Roy. Astron. Soc.}\ }\textbf {\bibinfo
  {volume} {387}},\ \bibinfo {pages} {689} (\bibinfo {year} {2008})},\ \Eprint
  {https://arxiv.org/abs/0712.1476} {arXiv:0712.1476 [astro-ph]} \BibitemShut
  {NoStop}%
\bibitem [{\citenamefont {Lovell}\ \emph {et~al.}(2014)\citenamefont {Lovell},
  \citenamefont {Frenk}, \citenamefont {Eke}, \citenamefont {Jenkins},
  \citenamefont {Gao},\ and\ \citenamefont {Theuns}}]{Lovell:2013ola}%
  \BibitemOpen
  \bibfield  {author} {\bibinfo {author} {\bibfnamefont {M.~R.}\ \bibnamefont
  {Lovell}}, \bibinfo {author} {\bibfnamefont {C.~S.}\ \bibnamefont {Frenk}},
  \bibinfo {author} {\bibfnamefont {V.~R.}\ \bibnamefont {Eke}}, \bibinfo
  {author} {\bibfnamefont {A.}~\bibnamefont {Jenkins}}, \bibinfo {author}
  {\bibfnamefont {L.}~\bibnamefont {Gao}},\ and\ \bibinfo {author}
  {\bibfnamefont {T.}~\bibnamefont {Theuns}},\ }\href
  {https://doi.org/10.1093/mnras/stt2431} {\bibfield  {journal} {\bibinfo
  {journal} {Mon. Not. Roy. Astron. Soc.}\ }\textbf {\bibinfo {volume} {439}},\
  \bibinfo {pages} {300} (\bibinfo {year} {2014})},\ \Eprint
  {https://arxiv.org/abs/1308.1399} {arXiv:1308.1399 [astro-ph.CO]}
  \BibitemShut {NoStop}%
\bibitem [{\citenamefont {Koposov}\ \emph {et~al.}(2015)\citenamefont
  {Koposov}, \citenamefont {Belokurov}, \citenamefont {Torrealba},\ and\
  \citenamefont {Evans}}]{Koposov:2015cua}%
  \BibitemOpen
  \bibfield  {author} {\bibinfo {author} {\bibfnamefont {S.~E.}\ \bibnamefont
  {Koposov}}, \bibinfo {author} {\bibfnamefont {V.}~\bibnamefont {Belokurov}},
  \bibinfo {author} {\bibfnamefont {G.}~\bibnamefont {Torrealba}},\ and\
  \bibinfo {author} {\bibfnamefont {N.~W.}\ \bibnamefont {Evans}},\ }\href
  {https://doi.org/10.1088/0004-637X/805/2/130} {\bibfield  {journal} {\bibinfo
   {journal} {Astrophys. J.}\ }\textbf {\bibinfo {volume} {805}},\ \bibinfo
  {pages} {130} (\bibinfo {year} {2015})},\ \Eprint
  {https://arxiv.org/abs/1503.02079} {arXiv:1503.02079 [astro-ph.GA]}
  \BibitemShut {NoStop}%
\bibitem [{\citenamefont {Bechtol}\ \emph {et~al.}(2015)\citenamefont {Bechtol}
  \emph {et~al.}}]{DES:2015txk}%
  \BibitemOpen
  \bibfield  {author} {\bibinfo {author} {\bibfnamefont {K.}~\bibnamefont
  {Bechtol}} \emph {et~al.} (\bibinfo {collaboration} {DES}),\ }\href
  {https://doi.org/10.1088/0004-637X/807/1/50} {\bibfield  {journal} {\bibinfo
  {journal} {Astrophys. J.}\ }\textbf {\bibinfo {volume} {807}},\ \bibinfo
  {pages} {50} (\bibinfo {year} {2015})},\ \Eprint
  {https://arxiv.org/abs/1503.02584} {arXiv:1503.02584 [astro-ph.GA]}
  \BibitemShut {NoStop}%
\bibitem [{\citenamefont {Drlica-Wagner}\ \emph {et~al.}(2015)\citenamefont
  {Drlica-Wagner} \emph {et~al.}}]{DES:2015zwj}%
  \BibitemOpen
  \bibfield  {author} {\bibinfo {author} {\bibfnamefont {A.}~\bibnamefont
  {Drlica-Wagner}} \emph {et~al.} (\bibinfo {collaboration} {DES}),\ }\href
  {https://doi.org/10.1088/0004-637X/813/2/109} {\bibfield  {journal} {\bibinfo
   {journal} {Astrophys. J.}\ }\textbf {\bibinfo {volume} {813}},\ \bibinfo
  {pages} {109} (\bibinfo {year} {2015})},\ \Eprint
  {https://arxiv.org/abs/1508.03622} {arXiv:1508.03622 [astro-ph.GA]}
  \BibitemShut {NoStop}%
\bibitem [{\citenamefont {Albert}\ \emph {et~al.}(2017)\citenamefont {Albert}
  \emph {et~al.}}]{Fermi-LAT:2016uux}%
  \BibitemOpen
  \bibfield  {author} {\bibinfo {author} {\bibfnamefont {A.}~\bibnamefont
  {Albert}} \emph {et~al.} (\bibinfo {collaboration} {Fermi-LAT, DES}),\ }\href
  {https://doi.org/10.3847/1538-4357/834/2/110} {\bibfield  {journal} {\bibinfo
   {journal} {Astrophys. J.}\ }\textbf {\bibinfo {volume} {834}},\ \bibinfo
  {pages} {110} (\bibinfo {year} {2017})},\ \Eprint
  {https://arxiv.org/abs/1611.03184} {arXiv:1611.03184 [astro-ph.HE]}
  \BibitemShut {NoStop}%
\bibitem [{\citenamefont {Polisensky}\ and\ \citenamefont
  {Ricotti}(2011)}]{Polisensky:2010rw}%
  \BibitemOpen
  \bibfield  {author} {\bibinfo {author} {\bibfnamefont {E.}~\bibnamefont
  {Polisensky}}\ and\ \bibinfo {author} {\bibfnamefont {M.}~\bibnamefont
  {Ricotti}},\ }\href {https://doi.org/10.1103/PhysRevD.83.043506} {\bibfield
  {journal} {\bibinfo  {journal} {Phys.\ Rev.\ D}\ }\textbf {\bibinfo {volume}
  {83}},\ \bibinfo {pages} {043506} (\bibinfo {year} {2011})},\ \Eprint
  {https://arxiv.org/abs/1004.1459} {arXiv:1004.1459 [astro-ph.CO]}
  \BibitemShut {NoStop}%
\bibitem [{\citenamefont {Schneider}(2016)}]{Schneider:2016uqi}%
  \BibitemOpen
  \bibfield  {author} {\bibinfo {author} {\bibfnamefont {A.}~\bibnamefont
  {Schneider}},\ }\href {https://doi.org/10.1088/1475-7516/2016/04/059}
  {\bibfield  {journal} {\bibinfo  {journal} {JCAP}\ }\textbf {\bibinfo
  {volume} {04}},\ \bibinfo {pages} {059}},\ \Eprint
  {https://arxiv.org/abs/1601.07553} {arXiv:1601.07553 [astro-ph.CO]}
  \BibitemShut {NoStop}%
\bibitem [{\citenamefont {Dienes}\ \emph {et~al.}(2021)\citenamefont {Dienes},
  \citenamefont {Huang}, \citenamefont {Kost}, \citenamefont {Thomas},\ and\
  \citenamefont {Yu}}]{Dienes:2021cxp}%
  \BibitemOpen
  \bibfield  {author} {\bibinfo {author} {\bibfnamefont {K.~R.}\ \bibnamefont
  {Dienes}}, \bibinfo {author} {\bibfnamefont {F.}~\bibnamefont {Huang}},
  \bibinfo {author} {\bibfnamefont {J.}~\bibnamefont {Kost}}, \bibinfo {author}
  {\bibfnamefont {B.}~\bibnamefont {Thomas}},\ and\ \bibinfo {author}
  {\bibfnamefont {H.-B.}\ \bibnamefont {Yu}},\ }\href@noop {} {\  (\bibinfo
  {year} {2021})},\ \Eprint {https://arxiv.org/abs/2112.09105}
  {arXiv:2112.09105 [astro-ph.CO]} \BibitemShut {NoStop}%
\bibitem [{\citenamefont {Murgia}\ \emph {et~al.}(2017)\citenamefont {Murgia},
  \citenamefont {Merle}, \citenamefont {Viel}, \citenamefont {Totzauer},\ and\
  \citenamefont {Schneider}}]{Murgia:2017lwo}%
  \BibitemOpen
  \bibfield  {author} {\bibinfo {author} {\bibfnamefont {R.}~\bibnamefont
  {Murgia}}, \bibinfo {author} {\bibfnamefont {A.}~\bibnamefont {Merle}},
  \bibinfo {author} {\bibfnamefont {M.}~\bibnamefont {Viel}}, \bibinfo {author}
  {\bibfnamefont {M.}~\bibnamefont {Totzauer}},\ and\ \bibinfo {author}
  {\bibfnamefont {A.}~\bibnamefont {Schneider}},\ }\href
  {https://doi.org/10.1088/1475-7516/2017/11/046} {\bibfield  {journal}
  {\bibinfo  {journal} {JCAP}\ }\textbf {\bibinfo {volume} {11}},\ \bibinfo
  {pages} {046}},\ \Eprint {https://arxiv.org/abs/1704.07838} {arXiv:1704.07838
  [astro-ph.CO]} \BibitemShut {NoStop}%
\bibitem [{\citenamefont {Lima}\ and\ \citenamefont {Hu}(2005)}]{Lima:2005tt}%
  \BibitemOpen
  \bibfield  {author} {\bibinfo {author} {\bibfnamefont {M.}~\bibnamefont
  {Lima}}\ and\ \bibinfo {author} {\bibfnamefont {W.}~\bibnamefont {Hu}},\
  }\href {https://doi.org/10.1103/PhysRevD.72.043006} {\bibfield  {journal}
  {\bibinfo  {journal} {Phys. Rev. D}\ }\textbf {\bibinfo {volume} {72}},\
  \bibinfo {pages} {043006} (\bibinfo {year} {2005})},\ \Eprint
  {https://arxiv.org/abs/astro-ph/0503363} {arXiv:astro-ph/0503363}
  \BibitemShut {NoStop}%
\bibitem [{\citenamefont {Sartoris}\ \emph {et~al.}(2016)\citenamefont
  {Sartoris} \emph {et~al.}}]{Sartoris:2015aga}%
  \BibitemOpen
  \bibfield  {author} {\bibinfo {author} {\bibfnamefont {B.}~\bibnamefont
  {Sartoris}} \emph {et~al.},\ }\href {https://doi.org/10.1093/mnras/stw630}
  {\bibfield  {journal} {\bibinfo  {journal} {Mon. Not. Roy. Astron. Soc.}\
  }\textbf {\bibinfo {volume} {459}},\ \bibinfo {pages} {1764} (\bibinfo {year}
  {2016})},\ \Eprint {https://arxiv.org/abs/1505.02165} {arXiv:1505.02165
  [astro-ph.CO]} \BibitemShut {NoStop}%
\bibitem [{\citenamefont {Dienes}\ and\ \citenamefont
  {Thomas}(2012{\natexlab{a}})}]{Dienes:2011ja}%
  \BibitemOpen
  \bibfield  {author} {\bibinfo {author} {\bibfnamefont {K.~R.}\ \bibnamefont
  {Dienes}}\ and\ \bibinfo {author} {\bibfnamefont {B.}~\bibnamefont
  {Thomas}},\ }\href {https://doi.org/10.1103/PhysRevD.85.083523} {\bibfield
  {journal} {\bibinfo  {journal} {Phys. Rev. D}\ }\textbf {\bibinfo {volume}
  {85}},\ \bibinfo {pages} {083523} (\bibinfo {year} {2012}{\natexlab{a}})},\
  \Eprint {https://arxiv.org/abs/1106.4546} {arXiv:1106.4546 [hep-ph]}
  \BibitemShut {NoStop}%
\bibitem [{\citenamefont {Dienes}\ and\ \citenamefont
  {Thomas}(2012{\natexlab{b}})}]{Dienes:2011sa}%
  \BibitemOpen
  \bibfield  {author} {\bibinfo {author} {\bibfnamefont {K.~R.}\ \bibnamefont
  {Dienes}}\ and\ \bibinfo {author} {\bibfnamefont {B.}~\bibnamefont
  {Thomas}},\ }\href {https://doi.org/10.1103/PhysRevD.85.083524} {\bibfield
  {journal} {\bibinfo  {journal} {Phys. Rev. D}\ }\textbf {\bibinfo {volume}
  {85}},\ \bibinfo {pages} {083524} (\bibinfo {year} {2012}{\natexlab{b}})},\
  \Eprint {https://arxiv.org/abs/1107.0721} {arXiv:1107.0721 [hep-ph]}
  \BibitemShut {NoStop}%
\bibitem [{\citenamefont {Vegetti}\ \emph {et~al.}(2014)\citenamefont
  {Vegetti}, \citenamefont {Koopmans}, \citenamefont {Auger}, \citenamefont
  {Treu},\ and\ \citenamefont {Bolton}}]{Vegetti:2014lqa}%
  \BibitemOpen
  \bibfield  {author} {\bibinfo {author} {\bibfnamefont {S.}~\bibnamefont
  {Vegetti}}, \bibinfo {author} {\bibfnamefont {L.}~\bibnamefont {Koopmans}},
  \bibinfo {author} {\bibfnamefont {M.}~\bibnamefont {Auger}}, \bibinfo
  {author} {\bibfnamefont {T.}~\bibnamefont {Treu}},\ and\ \bibinfo {author}
  {\bibfnamefont {A.}~\bibnamefont {Bolton}},\ }\href
  {https://doi.org/10.1093/mnras/stu943} {\bibfield  {journal} {\bibinfo
  {journal} {Mon. Not. Roy. Astron. Soc.}\ }\textbf {\bibinfo {volume} {442}},\
  \bibinfo {pages} {2017} (\bibinfo {year} {2014})},\ \Eprint
  {https://arxiv.org/abs/1405.3666} {arXiv:1405.3666 [astro-ph.GA]}
  \BibitemShut {NoStop}%
\bibitem [{\citenamefont {Hsueh}\ \emph {et~al.}(2020)\citenamefont {Hsueh},
  \citenamefont {Enzi}, \citenamefont {Vegetti}, \citenamefont {Auger},
  \citenamefont {Fassnacht}, \citenamefont {Despali}, \citenamefont
  {Koopmans},\ and\ \citenamefont {McKean}}]{Hsueh:2019ynk}%
  \BibitemOpen
  \bibfield  {author} {\bibinfo {author} {\bibfnamefont {J.-W.}\ \bibnamefont
  {Hsueh}}, \bibinfo {author} {\bibfnamefont {W.}~\bibnamefont {Enzi}},
  \bibinfo {author} {\bibfnamefont {S.}~\bibnamefont {Vegetti}}, \bibinfo
  {author} {\bibfnamefont {M.}~\bibnamefont {Auger}}, \bibinfo {author}
  {\bibfnamefont {C.~D.}\ \bibnamefont {Fassnacht}}, \bibinfo {author}
  {\bibfnamefont {G.}~\bibnamefont {Despali}}, \bibinfo {author} {\bibfnamefont
  {L.~V.}\ \bibnamefont {Koopmans}},\ and\ \bibinfo {author} {\bibfnamefont
  {J.~P.}\ \bibnamefont {McKean}},\ }\href
  {https://doi.org/10.1093/mnras/stz3177} {\bibfield  {journal} {\bibinfo
  {journal} {Mon. Not. Roy. Astron. Soc.}\ }\textbf {\bibinfo {volume} {492}},\
  \bibinfo {pages} {3047} (\bibinfo {year} {2020})},\ \Eprint
  {https://arxiv.org/abs/1905.04182} {arXiv:1905.04182 [astro-ph.CO]}
  \BibitemShut {NoStop}%
\bibitem [{\citenamefont {Gilman}\ \emph {et~al.}(2020)\citenamefont {Gilman},
  \citenamefont {Birrer}, \citenamefont {Nierenberg}, \citenamefont {Treu},
  \citenamefont {Du},\ and\ \citenamefont {Benson}}]{Gilman:2019nap}%
  \BibitemOpen
  \bibfield  {author} {\bibinfo {author} {\bibfnamefont {D.}~\bibnamefont
  {Gilman}}, \bibinfo {author} {\bibfnamefont {S.}~\bibnamefont {Birrer}},
  \bibinfo {author} {\bibfnamefont {A.}~\bibnamefont {Nierenberg}}, \bibinfo
  {author} {\bibfnamefont {T.}~\bibnamefont {Treu}}, \bibinfo {author}
  {\bibfnamefont {X.}~\bibnamefont {Du}},\ and\ \bibinfo {author}
  {\bibfnamefont {A.}~\bibnamefont {Benson}},\ }\href
  {https://doi.org/10.1093/mnras/stz3480} {\bibfield  {journal} {\bibinfo
  {journal} {Mon. Not. Roy. Astron. Soc.}\ }\textbf {\bibinfo {volume} {491}},\
  \bibinfo {pages} {6077} (\bibinfo {year} {2020})},\ \Eprint
  {https://arxiv.org/abs/1908.06983} {arXiv:1908.06983 [astro-ph.CO]}
  \BibitemShut {NoStop}%
\bibitem [{\citenamefont {Talbot}\ \emph {et~al.}(2021)\citenamefont {Talbot},
  \citenamefont {Brownstein}, \citenamefont {Dawson}, \citenamefont {Kneib},\
  and\ \citenamefont {Bautista}}]{Talbot:2020arv}%
  \BibitemOpen
  \bibfield  {author} {\bibinfo {author} {\bibfnamefont {M.~S.}\ \bibnamefont
  {Talbot}}, \bibinfo {author} {\bibfnamefont {J.~R.}\ \bibnamefont
  {Brownstein}}, \bibinfo {author} {\bibfnamefont {K.~S.}\ \bibnamefont
  {Dawson}}, \bibinfo {author} {\bibfnamefont {J.-P.}\ \bibnamefont {Kneib}},\
  and\ \bibinfo {author} {\bibfnamefont {J.}~\bibnamefont {Bautista}},\ }\href
  {https://doi.org/10.1093/mnras/stab267} {\bibfield  {journal} {\bibinfo
  {journal} {Mon. Not. Roy. Astron. Soc.}\ }\textbf {\bibinfo {volume} {502}},\
  \bibinfo {pages} {4617} (\bibinfo {year} {2021})},\ \Eprint
  {https://arxiv.org/abs/2007.09006} {arXiv:2007.09006 [astro-ph.GA]}
  \BibitemShut {NoStop}%
\bibitem [{\citenamefont {Eddignton}(1913)}]{Eddington:1913}%
  \BibitemOpen
  \bibfield  {author} {\bibinfo {author} {\bibfnamefont {A.}~\bibnamefont
  {Eddignton}},\ }\href {https://doi.org/10.1093/mnras/73.5.359} {\bibfield
  {journal} {\bibinfo  {journal} {Mon. Not. Roy. Astron. Soc.}\ }\textbf
  {\bibinfo {volume} {73}},\ \bibinfo {pages} {359} (\bibinfo {year}
  {1913})}\BibitemShut {NoStop}%
\bibitem [{\citenamefont {Murray}\ \emph {et~al.}(2013)\citenamefont {Murray},
  \citenamefont {Power},\ and\ \citenamefont {Robotham}}]{Murray:2013sna}%
  \BibitemOpen
  \bibfield  {author} {\bibinfo {author} {\bibfnamefont {S.}~\bibnamefont
  {Murray}}, \bibinfo {author} {\bibfnamefont {C.}~\bibnamefont {Power}},\ and\
  \bibinfo {author} {\bibfnamefont {A.}~\bibnamefont {Robotham}},\ }\href
  {https://doi.org/10.1093/mnrasl/slt079} {\bibfield  {journal} {\bibinfo
  {journal} {Mon. Not. Roy. Astron. Soc.}\ }\textbf {\bibinfo {volume} {434}},\
  \bibinfo {pages} {L61} (\bibinfo {year} {2013})},\ \Eprint
  {https://arxiv.org/abs/1306.5140} {arXiv:1306.5140 [astro-ph.CO]}
  \BibitemShut {NoStop}%
\bibitem [{\citenamefont {Madhavacheril}\ \emph {et~al.}(2015)\citenamefont
  {Madhavacheril} \emph {et~al.}}]{Madhavacheril:2014slf}%
  \BibitemOpen
  \bibfield  {author} {\bibinfo {author} {\bibfnamefont {M.}~\bibnamefont
  {Madhavacheril}} \emph {et~al.} (\bibinfo {collaboration} {ACT}),\ }\href
  {https://doi.org/10.1103/PhysRevLett.114.151302} {\bibfield  {journal}
  {\bibinfo  {journal} {Phys. Rev. Lett.}\ }\textbf {\bibinfo {volume} {114}},\
  \bibinfo {pages} {151302} (\bibinfo {year} {2015})},\ \bibinfo {note}
  {[Addendum: Phys.Rev.Lett. 114, 189901 (2015)]},\ \Eprint
  {https://arxiv.org/abs/1411.7999} {arXiv:1411.7999 [astro-ph.CO]}
  \BibitemShut {NoStop}%
\bibitem [{\citenamefont {Persic}\ \emph {et~al.}(1996)\citenamefont {Persic},
  \citenamefont {Salucci},\ and\ \citenamefont {Stel}}]{Persic:1995ru}%
  \BibitemOpen
  \bibfield  {author} {\bibinfo {author} {\bibfnamefont {M.}~\bibnamefont
  {Persic}}, \bibinfo {author} {\bibfnamefont {P.}~\bibnamefont {Salucci}},\
  and\ \bibinfo {author} {\bibfnamefont {F.}~\bibnamefont {Stel}},\ }\href
  {https://doi.org/10.1093/mnras/278.1.27} {\bibfield  {journal} {\bibinfo
  {journal} {Mon. Not. Roy. Astron. Soc.}\ }\textbf {\bibinfo {volume} {281}},\
  \bibinfo {pages} {27} (\bibinfo {year} {1996})},\ \Eprint
  {https://arxiv.org/abs/astro-ph/9506004} {arXiv:astro-ph/9506004}
  \BibitemShut {NoStop}%
\bibitem [{\citenamefont {Vale}\ and\ \citenamefont
  {Ostriker}(2004)}]{Vale:2004yt}%
  \BibitemOpen
  \bibfield  {author} {\bibinfo {author} {\bibfnamefont {A.}~\bibnamefont
  {Vale}}\ and\ \bibinfo {author} {\bibfnamefont {J.~P.}\ \bibnamefont
  {Ostriker}},\ }\href {https://doi.org/10.1111/j.1365-2966.2004.08059.x}
  {\bibfield  {journal} {\bibinfo  {journal} {Mon. Not. Roy. Astron. Soc.}\
  }\textbf {\bibinfo {volume} {353}},\ \bibinfo {pages} {189} (\bibinfo {year}
  {2004})},\ \Eprint {https://arxiv.org/abs/astro-ph/0402500}
  {arXiv:astro-ph/0402500} \BibitemShut {NoStop}%
\bibitem [{\citenamefont {Shankar}\ \emph {et~al.}(2006)\citenamefont
  {Shankar}, \citenamefont {Lapi}, \citenamefont {Salucci}, \citenamefont
  {De~Zotti},\ and\ \citenamefont {Danese}}]{Shankar:2006xz}%
  \BibitemOpen
  \bibfield  {author} {\bibinfo {author} {\bibfnamefont {F.}~\bibnamefont
  {Shankar}}, \bibinfo {author} {\bibfnamefont {A.}~\bibnamefont {Lapi}},
  \bibinfo {author} {\bibfnamefont {P.}~\bibnamefont {Salucci}}, \bibinfo
  {author} {\bibfnamefont {G.}~\bibnamefont {De~Zotti}},\ and\ \bibinfo
  {author} {\bibfnamefont {L.}~\bibnamefont {Danese}},\ }\href
  {https://doi.org/10.1086/502794} {\bibfield  {journal} {\bibinfo  {journal}
  {Astrophys. J.}\ }\textbf {\bibinfo {volume} {643}},\ \bibinfo {pages} {14}
  (\bibinfo {year} {2006})},\ \Eprint {https://arxiv.org/abs/astro-ph/0601577}
  {arXiv:astro-ph/0601577} \BibitemShut {NoStop}%
\bibitem [{\citenamefont {Salucci}\ \emph {et~al.}(2007)\citenamefont
  {Salucci}, \citenamefont {Lapi}, \citenamefont {Tonini}, \citenamefont
  {Gentile}, \citenamefont {Yegorova},\ and\ \citenamefont
  {Klein}}]{Salucci:2007tm}%
  \BibitemOpen
  \bibfield  {author} {\bibinfo {author} {\bibfnamefont {P.}~\bibnamefont
  {Salucci}}, \bibinfo {author} {\bibfnamefont {A.}~\bibnamefont {Lapi}},
  \bibinfo {author} {\bibfnamefont {C.}~\bibnamefont {Tonini}}, \bibinfo
  {author} {\bibfnamefont {G.}~\bibnamefont {Gentile}}, \bibinfo {author}
  {\bibfnamefont {I.}~\bibnamefont {Yegorova}},\ and\ \bibinfo {author}
  {\bibfnamefont {U.}~\bibnamefont {Klein}},\ }\href
  {https://doi.org/10.1111/j.1365-2966.2007.11696.x} {\bibfield  {journal}
  {\bibinfo  {journal} {Mon. Not. Roy. Astron. Soc.}\ }\textbf {\bibinfo
  {volume} {378}},\ \bibinfo {pages} {41} (\bibinfo {year} {2007})},\ \Eprint
  {https://arxiv.org/abs/astro-ph/0703115} {arXiv:astro-ph/0703115}
  \BibitemShut {NoStop}%
\end{thebibliography}%

\end{document}